\documentclass[aip,rsi,amsmath,amssymb,reprint,floatfix,10pt]{revtex4-2}

\usepackage{times}
\usepackage{amssymb,amsmath}
\usepackage{graphicx}
\usepackage{natbib}
\usepackage{multirow}
\usepackage{float}
\usepackage{braket}
\usepackage{silence}
\DeclareMathOperator{\sgn}{sgn}
\renewcommand\Re{\operatorname{Re}}

\usepackage[svgnames]{xcolor}

\begin{document}

\title{Unprecedented Spin-Lifetime of Itinerant Electrons in Natural Graphite Crystals}

\author{Bence~G.~M\'{a}rkus}
\affiliation{{Stavropoulos Center for Complex Quantum Matter, Department of Physics and Astronomy, University of Notre Dame, Notre Dame, Indiana 46556, USA}}

\author{D\'{a}vid~Beke}
\affiliation{{Stavropoulos Center for Complex Quantum Matter, Department of Physics and Astronomy, University of Notre Dame, Notre Dame, Indiana 46556, USA}}
\affiliation{{Institute for Solid State Physics and Optics, HUN-REN Wigner Research Centre for Physics, PO. Box 49, H-1525, Hungary}}
\affiliation{Kand\'o K\'alm\'an Faculty of Electrical Engineering, \'Obuda University, Tavaszmez\H{o} u. 17., H-1084 Budapest, Hungary}

\author{Lili~Vajtai}
\affiliation{{Stavropoulos Center for Complex Quantum Matter, Department of Physics and Astronomy, University of Notre Dame, Notre Dame, Indiana 46556, USA}}

\author{Andr\'{a}s~J\'{a}nossy}
\affiliation{{HUN-REN–BME Condensed Matter Research Group, Budapest University of 
Technology and Economics, M\H{u}egyetem rkp. 3., H-1111 Budapest, Hungary}}
\affiliation{{Department of Physics, Institute of Physics, Budapest University of Technology and Economics, M\H{u}egyetem rkp. 3., H-1111 Budapest, Hungary}}

\author{L\'{a}szl\'{o}~Forr\'{o}}
\affiliation{{Stavropoulos Center for Complex Quantum Matter, Department of Physics and Astronomy, University of Notre Dame, Notre Dame, Indiana 46556, USA}}

\author{Ferenc~Simon\email{simon.ferenc@ttk.bme.hu}}
\affiliation{{Stavropoulos Center for Complex Quantum Matter, Department of Physics and Astronomy, University of Notre Dame, Notre Dame, Indiana 46556, USA}}
\affiliation{{Department of Physics, Institute of Physics, Budapest University of Technology and Economics, M\H{u}egyetem rkp. 3., H-1111 Budapest, Hungary}}
\affiliation{{Institute for Solid State Physics and Optics, HUN-REN Wigner Research Centre for Physics, PO. Box 49, H-1525, Hungary}}

\date{\today}

\begin{abstract}
    A long spin-lifetime of electrons is the holy grail of spintronics, a field exploiting the electron angular momentum as information carrier and storage unit. Previous reports indicated a spin lifetime, $\tau_{\text{s}}$ near $10$ ns at best in graphene-based devices at low temperatures. 
    We detail the observation of $\tau_{\text{s}}$ approaching the ultralong $1{,}000$~ns at room temperature in natural graphite crystals using magnetic resonance spectroscopy. The relaxation time shows a giant anisotropy: the lifetime of spins, polarized perpendicular to the graphite plane, is more than $50$ times longer than for the in-plane polarization. The temperature dependence of $\tau_{\text{s}}$ proves that diffusion of spins to the crystallite edges, where relaxation occurs, limits the lifetime. This suggests that graphite is an excellent candidate for spintronic applications, seamlessly integrating with emerging 2D van der Waals technologies.
\end{abstract}



\maketitle

Spintronics exploits the intrinsic spin of electrons in addition to their charge, enabling faster, more energy-efficient memory and logic devices compared to conventional electronics \cite{FabianRMP}. This technology has revolutionized data storage with applications such as magnetic random-access memory (MRAM) and is poised to impact quantum computing and neuromorphic engineering \cite{WolfSCI}. 
Spintronic devices \cite{WolfSCI, FabianRMP, WuReview} require high-mobility materials with long spin relaxation times to ensure long-distance spin diffusion. The large mobility \cite{CastroNetoRMP2009} and the expected long $\tau_{\text{s}}$ in graphene (Refs. \onlinecite{DoraEPL2010, KawakamiFabianNatNanotechn2014}) are particularly appealing. In recent attempts to fabricate graphene-based spintronic devices \cite{Graphene_Spintronics_Review}, compatibility with two-dimensional heterostructures \cite{GeimNat2013} allowed to adjust the spin-orbit coupling (SOC) via the proximity effect \cite{OzyilmazNatComm2011, MorpurgoNatComm2015, MorpurgoPRX2016, Graphene_WS2_hetero_1, Graphene_WS2_hetero_2, Valenzuela2DMat2017, FerreiraPRL2017, DashNatComm2017, ValenzuelaNatMat2020, FabianGmitraPRB2015, FabianGmitraPRL2017, ZuticMatToday2019, FabianPRL2020, ValenzuelaNatMater2025}. Theory predicted a giant spin-relaxation anisotropy in graphene sandwiched with high SOC materials \cite{FabianRochePRL2017}, later confirmed in mono- and bilayer graphene \cite{ValenzuelaNatPhys2018, Fabian_vanWeesPRL2018, MakkPRB2018, KawakamiPRL2018, BouchiatAnis_PRL2018, vanWeesPRB2019}, contrasting with isotropic spin relaxation on SOC-free substrates \cite{TombrosPRL2008, ValenzuelaNatComm2016, KawakamiAnisPRB2018, GraphAnisPRB2018}. 

At present, the fabrication of graphene with a spin lifetime sufficiently long for device applications appears elusive. The value of $\tau_{\text{s}}$ in graphene is controversial; reported values range from $100$ ps to $12.6$~ns \cite{TombrosNAT2007, KawakamiPRL2011, OzyilmazPRL2011, ZomerPRB2012, RocheValenzuela2014, KamalakarNatComm2015, BeschotenNL2016, GrapheneSpintronicsReview2024, ChenACSAEM2025}. Theory suggests, extrinsic effects like adatoms \cite{FabianPRL2015} or ripples \cite{UpperLimit_T1_Cummings2025} limit the spin lifetime to these short values. Contact electrodes can also significantly affect $\tau_{\text{s}}$ (Refs. \onlinecite{Kawakami_electrode1, Kawakami_electrode2}). The lack of itinerant electrons in charge-neutral graphene requires gate biasing that induces a disturbing Bychkov--Rashba-type SOC \cite{GmitraPRB2009}.
 
Interestingly, graphite possesses a hitherto neglected symmetry, first described for graphene in the Kane--Mele Hamiltonian \cite{KaneMelePRL2005}. We recently observed in graphite a long $\tau_{\text{s}}$ over $100$ ns and a large relaxation anisotropy \cite{MarkusNatComm}. Following the experiments, a first-principle calculation taking into account this symmetry for graphite predicted a long $\tau_{\text{s}}$ intrinsic lifetime perpendicular to the graphite planes and a large lifetime anisotropy. A recent theoretical work predicted\cite{XuTheory} an intrinsic perpendicular spin lifetime of $\tau_{\text{s}}\approx 600~\text{ns}$, at $300$ K, considerably longer than observed at the time\cite{MarkusNatComm}. 
 
Here, we observe in pure, natural graphite crystals i) exceptionally long spin lifetimes up to $1{,}000$ ns, which predicts millimeter-long spin diffusion lengths at ambient temperatures ii) a giant intrinsic spin relaxation anisotropy of over $50$. These properties, together with the appreciable charge-carrier density and high in-plane mobility, make graphite an excellent candidate for the nonmagnetic conduction layer of spintronic devices. In some materials, localized spins have much longer lifetimes (exceeding $10$ ms)\cite{Gachter2022, Garreis2024, Denisov2024, NafradiNatComm, GrapheneNanoRibbon} but these systems, without mobile spins, are inappropriate for spintronics purposes.



\begin{figure*}[!ht]
	\centering
    \includegraphics[width=1\textwidth]{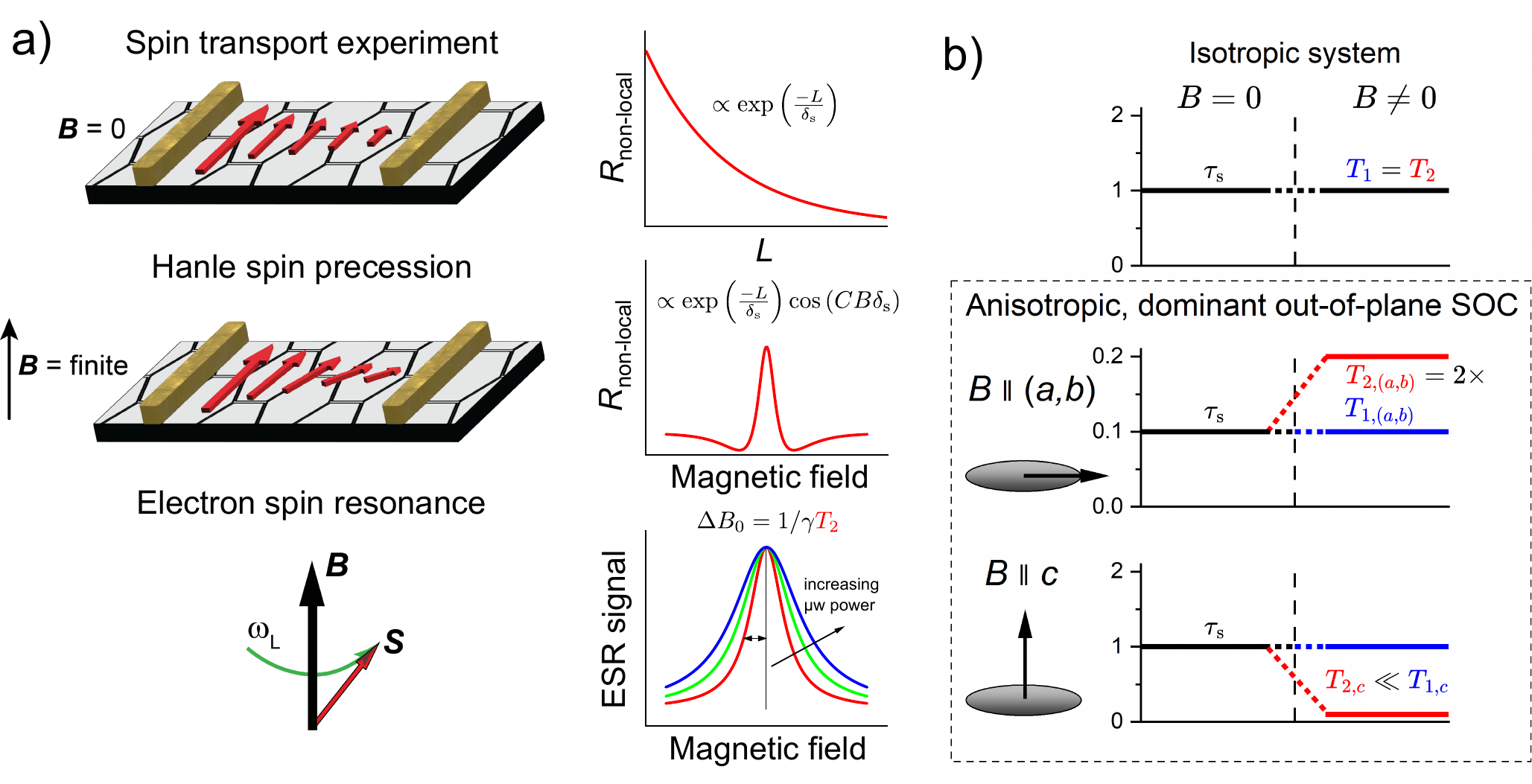}
	\caption{\textbf{Spin-relaxation measurement schemes, measurement of $T_1$, and the effect of anisotropy.} \textbf{a}, The spin-diffusion time and the related spin-relaxation time can be determined from the measurement of the non-local resistance in spin-injection studies, in a Hanle-type spin precession experiment, or using electron spin resonance, where the low microwave power linewidth gives the transversal relaxation time, $T_2$. $T_1$ can be obtained through a controlled saturation of the ESR line. \textbf{b}, In a fully isotropic system, both relaxation times are equal to the zero-field spin-relaxation time. However, in an anisotropic system, this simple relation fails \cite{yafet1963g}; in a two-dimensional system, such as graphite, where the out-of-plane SOC dominates \cite{MarkusNatComm}, the spin-relaxation times will differ depending on the direction of the external magnetic field. When it is along the plane with the weak SOC, both relaxation times are small and show a $1{:}2$ ratio. However, when the magnetic field is perpendicular to the plane, the $T_1$ becomes much longer than $T_2$. The ratio of the two values depends on the strength of the anisotropy \cite{FabianActaPhysSlovaca}.}
	\label{Fig1}
\end{figure*}


The spin-diffusion length, $\delta_{\text{s}}$, and $\tau_{\text{s}}$ are closely related\cite{janossy1976, SilsbeeJanossyMonod}: $\delta_{\text{s}}=\sqrt{D\tau_{\text{s}}}$ (Refs. \onlinecite{FabianRMP, FabianActaPhysSlovaca}). The diffusion constant, $D$, in the mean free-path approximation is $D=\frac{1}{d}\overline{v}\overline{\ell}$, where $d$ is $2$ in one and two dimensions and $3$ in three dimensions. In a metal, the Fermi velocity, $v_{\text{F}}$, replaces the average velocity, $\overline{v}$, and the mean free path is $\overline{\ell}=v_{\text{F}}\tau$. Here, $\tau$ is the momentum relaxation time. Thus, the relation between the material parameters and the spin-diffusion length is $\delta_{\text{s}}=\frac{1}{\sqrt{d}}v_{\text{F}}\sqrt{\tau\tau_{\text{s}}}$ (Refs. \onlinecite{FabianRMP, FabianActaPhysSlovaca}). 

The spin-diffusion length is measured from the non-local resistance in spin-injection experiments \cite{TombrosNAT2007} (see Fig. \ref{Fig1}a) or in a Hanle-type precession \cite{TombrosPRL2008}, while fitting the solutions of a one-dimensional Bloch--Torrey equation to the data (discussed in the Supporting Information). The spin transport method does not require an external magnetic field; however, it necessitates several devices with various electrode lengths. The Hanle-type spin precession experiment is a combined transport-spectroscopy approach, works well on a single device and measures both the spin-diffusion length and the spin lifetime. However, the rotation of the spin magnetization in the magnetic field complicates the measurement of spin lifetime in a single direction. This is unimportant in isotropic metals, but limits its use in two-dimensional materials. Both spin-transport methods require ferromagnetic contacts that influence the spin lifetime.

The present ESR spectroscopic approach yields $\tau_{\text{s}}$ and its anisotropy directly; $\delta_{\text{s}}$ is derived from known values of $D$. For the ESR, a microwave field is applied near the Larmor frequency. It is polarized in the plane perpendicular to a large static field, $B_0 (\theta)$. Here $\theta$ is the polar angle with respect to the out-of-plane direction, $c$ of graphite. By measuring the ESR spectrum as a function of microwave power, two spin relaxation times, $T_1$ and $T_2$ are measured as a function of $\theta$. $T_1$ is the lifetime of spin magnetization parallel, while $T_2$ is the lifetime perpendicular to $B_0$.  

For a system with an anisotropic SOC, $T_1 \neq T_2$. Fig. \ref{Fig1}b. discusses the case when the out-of-plane SOC is dominant\cite{yafet1963g}. In general, the SOC, perpendicular to a given spin direction, leads to relaxation \cite{SlichterBook, AbragamBook}. Therefore, for this type of "easy-axis SOC" anisotropy, the zero-field spin-relaxation time, or $\tau_{\text{s}}$, also has a significant anisotropy such that $\tau_{\text{s},(a,b)} \ll \tau_{\text{s},c}$ and $T_1$ and $T_2$ differ. Both relaxation times are relatively short and $T_{2,ab}= 2\tau_{\text{s},ab}=2T_{1,ab}$ holds when the magnetic field is in the $(a,b)$ plane. However, for the magnetic field is along $c$, perpendicular to the $(a,b)$ plane, the $T_1$ spin-relaxation time is much longer and equals $\tau_{\text{s},c}$. 

$T_2$ is obtained in ESR spectroscopy from the linewidth, $\Delta B_0$. at low microwave powers. In this case, the lineshape is independent of power intensity (i.e., in the absence of saturation) and $T_2=1/\gamma \Delta B_0$. Here $\gamma~\approx~2\pi \cdot28~\text{GHz}/\text{T}$ is the electron gyromagnetic ratio.  $T_1$ and $T_2$ are determined from the microwave excitation power dependence of the linewidth $\Delta B$ of the continuous wave ESR spectrum:
\begin{equation}
    \Delta B=\Delta B_0 \sqrt{1+{\color{red}\boldsymbol{T_1}}\cdot \gamma B_1^2/\Delta B_0},
    \label{T1_determination}
\end{equation}
where $B_1$ is the strength of the microwave magnetic field, whose square is proportional to the microwave power, $p$. This effect is shown in Fig. \ref{Fig1}a. A hand-waving description of the broadening is that intensive irradiation decreases the ESR signal intensity, which is known as saturation\cite{SlichterBook, AbragamBook}. The magnitude of the saturation is strongest in the middle of the ESR line and progressively diminishes away from the resonance condition, which acts as if the middle of the line were "pushed in". The equation is valid provided there is no inhomogeneous broadening. In the high purity, well oriented graphite the residual linewdths at low powers of different quality samples change little (see Supporting Information). In most localized electron systems, small inhomogeneities broaden the line and the lifetimes cannot be extracted from Eq. \eqref{T1_determination}. 

\begin{figure*}[!ht]
	\centering
    \includegraphics[width=1\textwidth]{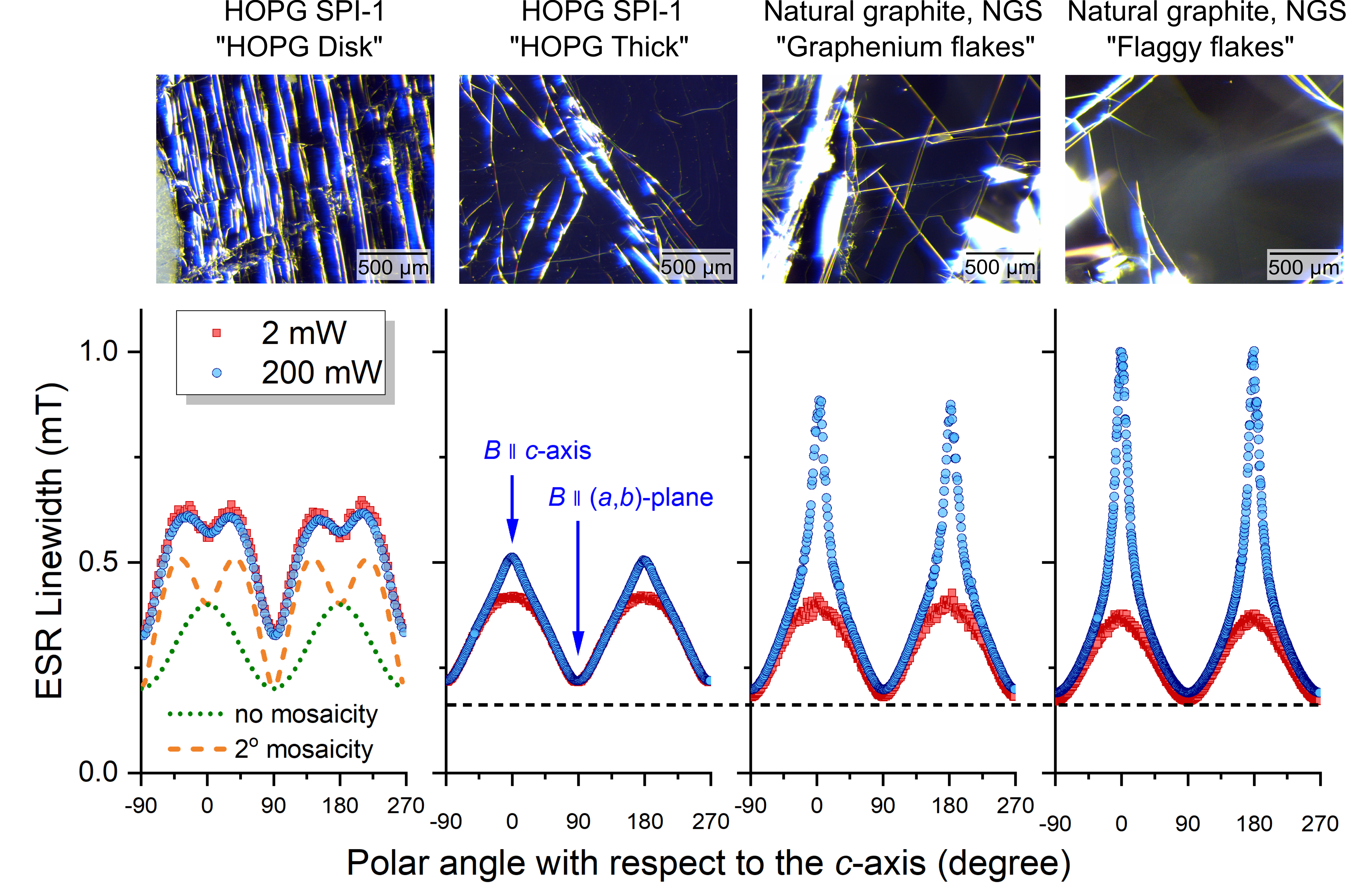}
	\caption{\textbf{Photographs of the studied samples and ESR linewidth data.} We studied two different types of samples: synthetic highly oriented pyrolytic graphite (HOPG) and natural graphite crystals. From both types, two subtypes were studied: disk and a thick piece from the HOPG and "Graphenium flakes" and "Flaggy flakes" from the natural graphite. Top panel: optical images, from left to right, the four types of samples show a progressively increasing individual crystallite size. Bottom panel: the ESR linewidth data also shows a gradual improvement from left to right. A strong mosaicity is evident for the HOPG disk sample, supported by lineshape simulations (dashed and dotted curves) and also therein no microwave power induced saturation of the linewidth is observed as well as a large residual linewidth when $B \parallel (a,b)$. The line broadening under intensive microwave irradiation is largest for the "Flaggy flakes" sample, where it is almost a factor $3$, indicating the presence of an exceptionally long $T_1$. The residual linewidth value for $B \parallel (a,b)$ also improves from left to the right as indicated by the dotted horizontal line. Data on the HOPG disk sample is equivalent to that reported in Ref. \onlinecite{MarkusNatComm}.}
	\label{Fig2}
\end{figure*}

The high-quality synthetic highly-oriented graphite (HOPG) and natural graphite crystals (Methods) studied here verify the predictions of a very long intrinsic perpendicular lifetime and large anisotropy.  The angular-dependent ESR linewidth measured at high and low microwave powers (Fig. \ref{Fig2}) reveal a significant microwave field-induced broadening along $c$ corresponding to a long $T_1$. In accordance with previous observations, the extra broadening appears in the higher quality "HOPG Thick" samples but is absent in the saw-cut "HOPG Disk" samples. Surprisingly, the natural graphite crystals show an unparalleled microwave field-induced broadening with about $2.5$ times linewidth increase when saturated by high microwave power. The ESR spectrum retains a perfect Lorentzian lineshape at all powers, indicating that the linewidths correspond to a long lifetime and are not due to inhomogeneities. Optical microscopy shows in natural graphite crystals graphite flakes with lateral dimensions of a few millimeters, whereas the HOPG samples show a broken surface with lateral dimensions ranging between $10-100$ micrometers. A few degrees disorder in the orientation of mosaic crystals has a well observable effect on the ESR spectra. Mosaicity is evident in the inhomogeneous ESR linewidth of the lower-quality "HOPG Disk" sample (Fig. \ref{Fig2}). (See details in the Supporting Information).

\begin{figure}[!htb]
	\centering
    \includegraphics[width=\linewidth]{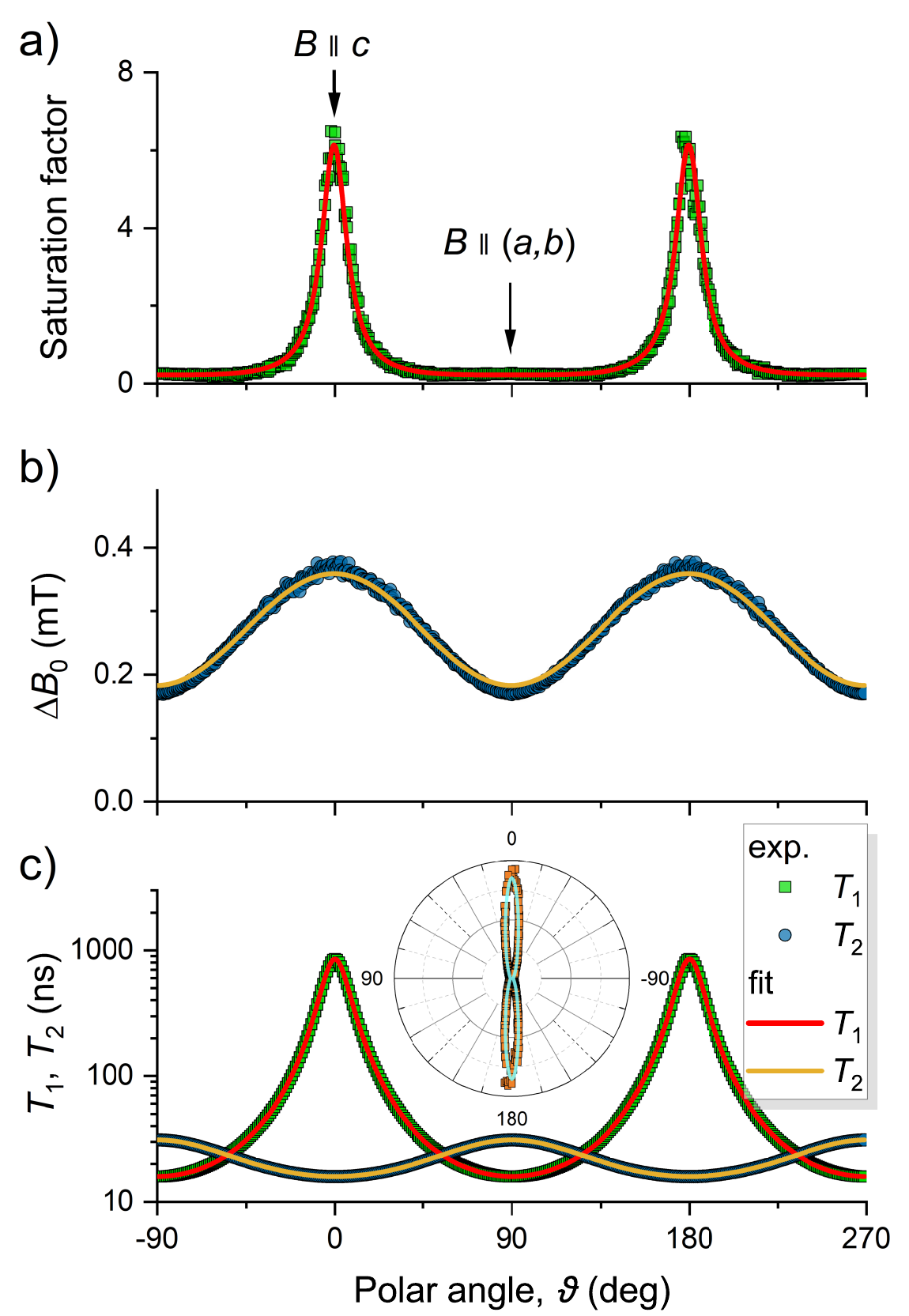}
	\caption{\textbf{Angular-dependent relaxation time in graphite.} \textbf{a,} Angular dependence of the saturation factor as determined from the $200$~mW and $2$~mW ESR linewidth for the "Flaggy flake" natural graphite crystal. \textbf{b,} Angular dependence of the low-power ESR linewidth, $\Delta B_0$ as obtained from the $2$ mW measurement. Solid lines in \textbf{a} and \textbf{b} are fits with the model explained in the text. \textbf{c,} The respective relaxation times obtained from the data and the model as shown on a logarithmic scale. Note the $1{:}2$ variation of $T_2$ between the two directions of the magnetic field and also that $T_1$ becomes ultralong when $B \parallel c$. Inset shows the experimental and modeled $T_1$ data on a polar plot (on a linear scale).}
	\label{Fig3}
\end{figure}

The significant microwave power-induced broadening allows an accurate measurement of the angular dependence of $T_1$ using Eq. \eqref{T1_determination}. To this end, we calculate the angular-dependent saturation factor from the microwave power-dependent linewidth, $\Delta B(p)$, as: $s\left(p,\vartheta \right)=\left[\left(\frac{\Delta B(p)}{\Delta B_0}\right)^2-1\right]$. Thus $T_1(\vartheta)=s(p,\vartheta) \frac{\Delta B_0}{\gamma {B_1}^2}$ and the result is shown in Fig. \ref{Fig3} for the "Flaggy flakes" graphite crystal sample. The low-power ESR linewidth (taken at $p=2$~mW), $\Delta B_0$ is also shown, as these data are used to obtain $T_2\left(\vartheta \right)=1/\gamma \Delta B_0$. 

In the next step, we follow the seminal paper of Yafet \cite{yafet1963g} to obtain the relation between the measured angular dependence of the ESR relaxation rates and the zero magnetic field spin relaxation times due to SOC in graphite, $\tau_{\text{s},c}$ and $\tau_{\text{s},ab}$. Yafet \cite{yafet1963g} suggested a model in which effective anisotropic fluctuating magnetic fields determine the angular-dependent relaxation. In graphite, the fluctuations arise from the fluctuating SOC acting on diffusing electrons. After some algebra (details are given in the Supporting Information), we obtain:
\begin{align}
        T_1^{-1}\left(\vartheta \right)&= \tau_{\text{s},c}^{-1}\frac{1+\cos 2\vartheta}{2}+\tau_{\text{s},(a,b)}^{-1}\sin^2\vartheta, \label{AngDepRelaxationTimesRewritten1}\\
        T_2^{-1}\left(\vartheta \right)&=\tau_{\text{s},c}^{-1}\frac{1-\cos 2\vartheta}{4}+\tau_{\text{s},(a,b)}^{-1}\frac{1+\cos^2 \vartheta}{2}.
        \label{AngDepRelaxationTimesRewritten}
\end{align}
Here, $\tau_{\text{s},c}$ and $\tau_{\text{s},(a,b)}$ are the spin-relaxation times for spins polarized along the graphite $c$ axis or in the $(a,b)$ plane, respectively.

The fit using Eqs. \eqref{AngDepRelaxationTimesRewritten1} and \eqref{AngDepRelaxationTimesRewritten} describes well the full angular-dependence of the ESR relaxation rates for several samples the with three free parameters (see Fig. \ref{Fig3}). For the high quality "Flaggy flakes" graphite crystal sample the fit gives $\tau_{\text{s},c}=850(10)~\text{ns}$, $\tau_{\text{s},(a,b)}=15.83(2)~\text{ns}$, and $B_1=0.271(1)~\text{mT}\cdot\sqrt{p[\text{W}]}$. The value of $T_{2,c}$ is robust in this determination; however, $\tau_{\text{s},c}$ depends on the value of the microwave field $B_1$. The manufacturer specified that $B_1[\text{mT}]=0.2\sqrt{\frac{Q}{7{,}500}~p[\text{W}]}$, where $Q$ is the quality factor of the microwave cavity and $p$ is measured in watts. We measured $Q=11{,}500$ using a frequency sweep method \cite{Klein1993, Donovan1993}, which corresponds to $B_1=0.248~\text{mT}\cdot\sqrt{p[\text{W}]}$) and gives $\tau_{\text{s},c}=1{,}050(10)~\text{ns}$. However, deviations from manufacturer values are possible, and letting $B_1$ be a free parameter improves the fit quality: the adjusted $R^2$ value increases from $0.992$ to $0.994$, while the change in $B_1$ is $10\%$.

Two further observations support the ultralong $\tau_{\text{s},c}$. (See details in Supporting Information). The saturation factor was also determined from the ESR line intensity, albeit with an accuracy inferior to that of the ESR linewidth. The linewidth is a purely spectroscopic measurable and less prone to instrumental errors. Furthermore, we observed an anomalously large out-of-phase component of the magnetic field modulated ESR signal, which is compatible with a long $T_1$ when compared to numerical solutions of the Bloch equations.
 
In theory, long relaxation times of $T_1 \sim 1{,}000~\text{ns}$ could be detected using various approaches such as spin-echo ESR \cite{PooleBook, SlichterBook}, longitudinally detected ESR \cite{Martinelly_LOD, Atsarkin, Schweiger1, Schweiger2}, or saturation recovery ESR \cite{EatonsRSI1992, EatonsAMR2017, EatonsRSI2019}. However, the very small spin-susceptibility from the semimetallic low density of states and the short $T_2\sim 15~\text{ns}$ accompanying the long $T_1$, renders these techniques impractical and near impossible to apply for graphite. 

The observation of an ultralong $T_1$ for $B\parallel c$ predicts a macroscopic spin-diffusion length, $\delta_{\text{s}}=\sqrt{D T_1}$. The anisotropy of the diffusion constant follows that of the mobility, $\mu$ (and also of the conductivity), according to the Einstein relation: $D=\mu \frac{k_{\text{B}}T}{e}$ ($k_{\text{B}}$ and $e$ are the Boltzmann constant and the elementary charge, respectively). In a Fermi gas \cite{AschcroftMermin}, $k_{\text{B}}T$ is replaced at low temperatures by the Fermi energy, $E_{\text{F}}$. In graphite $E_{\text{F}}\approx 15\dots 25~\text{meV}$ (Refs. \onlinecite{DresselhausAP2002, BrandtBook, thoutam2022temperature}) is nearly equal to $k_{\text{B}}T\approx 26~\text{meV}$ at room temperature, and the two formulae give the same result. More precisely, at finite-temperatures the chemical potential\cite{Solyom2} gives a better description.


With $\sigma_{(a,b)}/\sigma_c\approx 10^4$ (Refs. \onlinecite{DresselhausAP2002, BrandtBook}) and $(a,b)$ plane mobility ranging between $\mu_{(a,b)}=10{,}000 \dots 100{,}000~\frac{\text{cm}^2}{\text{Vs}}$ (Refs. \onlinecite{DresselhausAP2002, KopelevichPRL2003}), we obtain for the anisotropic diffusion constants: $D_{(a,b)}=200\dots 2{,}000~\text{cm}^2/\text{s}$ and $D_{c}=0.02\dots 0.2~\text{cm}^2/\text{s}$. Using the geometric means of these diffusion constant ranges, the spin-diffusion length in the plane is $\delta_{\text{s},(a,b)} \approx 250~\mu\text{m}$ and the spin-diffusion length along the direction $c$ is $\delta_{\text{s},c}\approx 2.5~\mu\text{m}$. The $\delta_{\text{s},(a,b)}$ has a giant, macroscopic value but $\delta_{\text{s},c}$ is also considerable, and samples with similar thickness are within reach for spintronic applications. The in-plane diffusion length is, in fact, comparable to the minority charge-carrier diffusion length in doped Si single crystals (typical lifetimes of $100~\mu\text{s}$ and $\mu_{\text{e}}\approx 1{,}000 ~\frac{\text{cm}^2}{\text{Vs}}$ for electrons and $\mu_{\text{h}}\approx 400~\frac{\text{cm}^2}{\text{Vs}}$ for holes \cite{KittelBook}).


\begin{figure}[!htb]
	\centering
    \includegraphics[width=\linewidth]{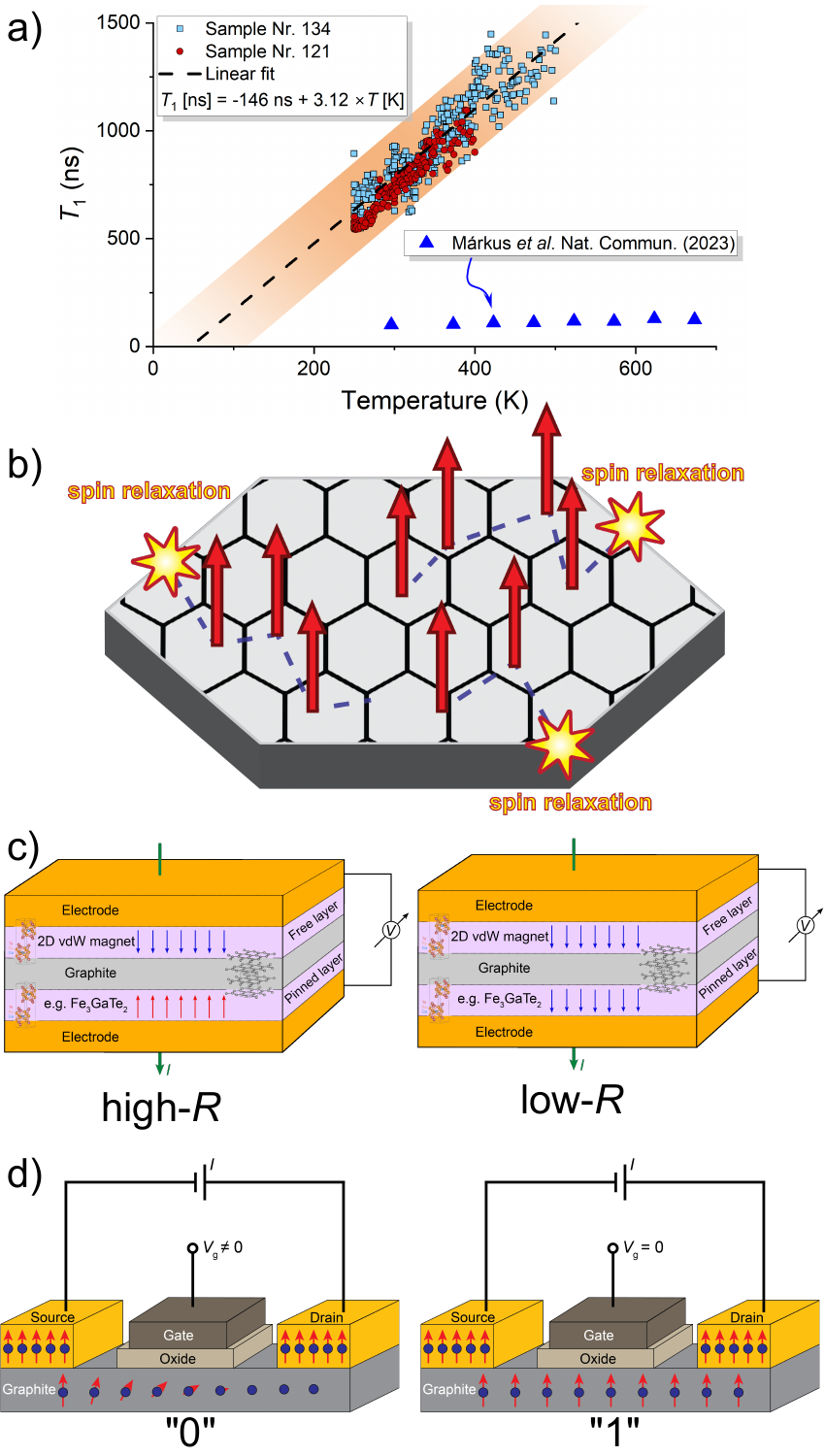}
	\caption{\textbf{Temperature dependent relaxation time.} \textbf{a,} Temperature dependent $T_1$ as obtained from the power-dependent linewidth for two different samples. The solid line is a fit to sample Nr. 134. We show earlier $T_1$ data from Ref. \onlinecite{MarkusNatComm}. \textbf{b,} Schematics of the proposed mechanism for the $T_1$: electrons with spins polarized perpendicular to the $(a,b)$ plane has a long relaxation time, however, they can relax on graphite edges when they reach these through diffusion. \textbf{c,} Proposed \emph{current-perpendicular-to-plane} (CPP) type spin valve design exploiting the ultralong out-of-plane lifetime in graphite combined with a van der Waals ferromagnet with strong out-of-plane magnetization, e.g., Fe$_3$GaTe$_2$ \cite{ChenNanoLett2024}. \textbf{d,} Hypothesized Datta--Das SFET using graphite as transport layer. Electron spins are injected parallel to the graphite $c$ axis. The electric fields from the gate electrode generate an in-plane magnetic field, rotating the spins away from the $c$-direction, where spin relaxation is short, thereby nullifying the signal.}
	\label{Fig4}
\end{figure}

The temperature dependence of the spin-relaxation time for electrons polarized perpendicular to the graphite $(a,b)$ plane is shown in Fig. \ref{Fig4}a for the "Flaggy flakes" sample. The $T_1$ data were obtained in the $250-500$ K range from the temperature-dependent saturation factor, $s=\left[\left(\frac{\Delta B(200~\text{mW})}{\Delta B(2~\text{mW})}\right)^2-1\right]$ using the linewidths at various powers (the raw data are given in the Supporting Information). As the linewidth (or $T_2$) is also temperature dependent, the $T_1$ vs. temperature is obtained by scaling $s\times\Delta B(2~\text{mW})$ to the room-temperature $T_1$ value. For comparison, previous data from Ref. \onlinecite{MarkusNatComm} is also presented in Fig. \ref{Fig4}a.

$T_1$ is strongly temperature dependent; it increases by a factor $2$ from $250$ K to $500$ K. It is tempting to fit the approximately linearly increase of the data with a straight line. In principle, such a behavior could be explained by the D'yakonov--Perel'-like spin-relaxation mechanism \cite{DyakonovPerelSPSS1972, FabianRMP}. This would predict $\left(T_1\right)^{-1}=\overline{\delta \omega^2} \tau$, that is, the momentum relaxation time $\tau$ is inversely proportional to $T_1$. (Here $\overline{\delta \omega^2}$ is the mean value of the squared SOC-related fluctuating fields). 

Nevertheless, the strong dependence on the sample preparation method, sample morphology, and the fact that the spin-diffusion length is macroscopic, leads us to suggest a diffusion length limited spin-relaxation time. As discussed in the Supporting Information, deliberate breaking of the samples shortens $T_1$. It is well known that in semiconductors, minority charge carrier recombination can take place on surfaces, heterojunctions, or any other parts that break the global crystal symmetry \cite{pierret2002advanced, schroder2006semiconductor}. The \emph{surface recombination-limited charge carrier lifetime} in semiconductors is $\tau_{\text{c}}=L^2/D$, where $L$ is the characteristic size of the sample and $D$ is the diffusion constant. On the other hand, for this effect the quality of surfaces plays an important role; properly terminated edges in passivated silicon\cite{pierret2002advanced, schroder2006semiconductor} do not limit the charge carrier lifetime.

We suggest that a similar effect dominates the spin lifetime in our graphite samples. The increase of $T_1$ with increasing temperature could be explained by a \emph{sample length limit} of the spin diffusion constant. For this scenario, we assume that in high quality graphite, disorder at sample edges is the dominant factor, while diffusion to the top or bottom terminating graphene planes does not contribute significantly to the spin relaxation.

The Einstein relation for an itinerant electron system, $D=\mu E_{\text{F}}/e$, implies a roughly $T^{-1}$ temperature dependence\cite{SouleMcClure, BrandtBook} for the charge-carrier mobility, $\mu$. Furthermore semimetal graphite has a small Fermi surface where $E_{\text{F}}/e$ is temperature dependent. Magnetotransport studies suggest \cite{thoutam2022temperature} that $E_{\text{F}}$ \emph{decreases} with increasing temperature by an uncertain amount. This reinforces the idea that $D$ decreases with increasing temperature. To explain the observed lengthening of $T_1$ despite the reduction of the bulk diffusion constant, we suggest that the spin relaxation is diffusion length limited. In this case surfaces perpendicular to the $(a,b)$ planes limit the spin relaxation time, $\tau_s,c$ in samples smaller than the bulk spin diffusion length. The experimentally observed $T_1$ lifetime in this work is not necessarily the ultimate limit. For graphite crystals with even larger lateral sizes or with well-terminated edges, the spin lifetime may be even longer. 

We recall that the ultralong $T_1$ for electron spins polarized perpendicular to the graphite $(a,b)$ plane predicts a long zero-field limit of the spin-relaxation time, $\tau_{\text{s},c}$. Consequently, the spin diffusion lengths are long, in the range of micrometers perpendicular to the planes and up to millimeters parallel to the planes. Thus graphite is a promising candidate for the non-magnetic intermediate layer in spintronic devices at ambient temperature. We present in Fig. \ref{Fig4}c the schematics of a proposed spin-valve device mimicking the existing current-perpendicular-to-plane (CPP) device structure. This geometry is superior to the current-in-plane (CIP) arrangement and is widely used in tunnel magnetoresistance (TMR) junctions \cite{TapariaSTAM2024}, such as hard-drive read heads. It requires a compatible ferromagnet, with an out-of-plane magnetization easy axis, e.g., CrI$_3$ \cite{DillonJAP1965, HuangNatNanot2018}, Cr$_2$Ge$_2$Te$_6$ \cite{CarteauxJPCM1995, VerzhbitskiyNatElec2020}, Cr$_2$Si$_2$Te$_6$ \cite{CarteauxEPL1995}, Fe$_3$GeTe$_2$ \cite{DeiserothEJIC2006, FeiNatMater2018}, Fe$_5$GeTe$_2$ \cite{Chen2DMater2022}, or the room temperature ferromagnet, Fe$_3$GaTe$_2$ \cite{ZhangNatComm2022, YinCEC2023, HuAdvMater2024}. The polarized spins provided by the "free layer" easy-axis ferromagnet are pumped through the graphite layers by a finite electric current, $I$. The spin orientation of electrons is maintained, while they propagate through the graphite layer. The spin diffusion coefficient in the $c$-direction allows a couple of micrometers thick graphite layer. Depending on the matching orientation of the "free" and "pinned layers" the structure has a high-$R$ (antiparallel) or a low-$R$ state (parallel).

In addition to the spin-valve-based magnetic field sensor, we introduce the concept of a graphite-based Datta--Das spin field-effect transistor (SFET) \cite{Datta1990}, as shown in Fig.~\ref{Fig4}d. In the original concept, the electric field from the gate generates a magnetic field in the rest frame of the electrons moving with velocity $\mathbf{v}$: $\mathbf{B}=-(\mathbf{v}\times \mathbf{E})/c^2$. Normally, significant rotations by strong electric fields are required to achieve noticeable spin signals. However for the proposed graphite-based SFET a much weaker electric field is sufficient. Due to the large spin lifetime anisotropy, a slight rotation away from the $c$-direction is enough to increase the spin relaxation substantially and thereby diminish the signal. The device operates as a spin transistor with highly sensitive control over operation under ambient conditions.

\section*{Conclusions}\vspace{-4mm}
Our findings reveal that in natural graphite the electron spin lifetime is exceptionally long for polarization perpendicular to the $(a,b)$ plane. It is about $1~\mu$s at room temperature, a remarkable milestone for spintronics. The strong anisotropy in spin relaxation underscores the crucial influence of electron diffusion and crystallite edges on spin lifetime. The results position graphite, and multilayer AB-stacked graphene, as compelling candidates for spin-based technologies, particularly within 2D van der Waals heterostructures. The long lifetimes enable the development of high-performance spintronic devices that operate under ambient conditions.
 
\section*{Methods}\vspace{-4mm}
We studied natural graphite samples (NGS Trading \& Consulting GmbH). Unfortunately, the exact origin of the latter samples was not revealed. Two different manufacturer grades were examined: "Graphenium flakes" and "Flaggy flakes". Of these, we studied several starting sample sizes from the manufacturer (ranging from $2-30$ mm), and generally found that the larger pieces gave better results. In addition, we observed a less fractured surface for the "Flaggy flakes" and longer spin-relaxation times than for the "Graphenium" samples.

Furthermore, we investigated various grades (SPI-1, SPI-2, SPI-3 from Structure Probe, Inc. and ZYA grade from UniversityWafer) from synthetic, highly oriented pyrolytic graphite (HOPG). While we overall observed that SPI-1 and ZYA has the highest quality, we also observe batch-dependent variations. From the SPI-1, two different sample types are available, "Disk" and square shaped "HOPG Thick" samples. The disks are usually of inferior quality as these are cut from larger HOPG samples, which results in a broken surface. Also for the samples we studied, SPI-1 was of better quality from the point of view of the ESR lifetime than the ZYA sample. 

The most pronounced effect of sample quality is on the residual ESR linewidth, and the effect of mosaicity, as discussed in the manuscript. Our benchmark for a good quality sample is a $B \parallel (a,b)$ ESR linewidth around $0.18$ mT, the absence of the double-peak structure due to mosaicity around the $B \parallel c$ orientation and also that the linewidth changes rapidly for increases microwave powers for this orientation. In addition, the samples must be sufficiently thin (estimated thickness below $10~\mu$m), which is seen as a symmetric ESR lineshape in contrast to thicker samples where a characteristic Dysonian lineshape is observed \cite{Dyson, WalmsleySynthMet1989, WalmsleyPRB1992, WalmsleyJMR1996}. A summary of the sample quality variations is presented in the Supporting Information.

Elemental analysis was conducted using X-ray photoelectron spectroscopy (XPS, PHI VersaProbe II) and X-ray fluorescence (XRF, EDAX Orbis PC Micro-XRF) at multiple points. The XPS survey spectra of the graphite samples detected only oxygen and silicon, in addition to carbon. The silicon and oxygen content varied across the sample surfaces, but their ratio closely resembled $1{:}2$, suggesting the presence of SiO$_2$ impurities. XRF analysis indicated the possible presence of Si, Cr, Fe, Co, Ni, and O in all samples at concentrations near the detection limit ($1-5$ ppm) and revealed no significant compositional differences between samples. We also attempted to clean some of the graphite crystals by soaking them in concentrated $38\%$ HCl for $2$ hours to remove all metallic contaminants. However, no reaction was observed, and the acid treatment did not affect the observed ESR signal by any means.

The actual samples with about $2-3$ mm lateral size were transferred from the starting graphite using the micromechanical transfer technique (also known as the Scotch tape method and using Scotch Magic Tape) and covered on both sides with the tape, which was then cut with a scalpel to fit to the single crystal suprasil mounting rod (ATS Life Sciences Wilmad WG-856-Q). A large number of samples were prepared, of which about $10 \%$ showed the ultralong relaxation time. These were selected by studying the ESR linewidth at $200$ mW microwave power for the $B \parallel c$ orientation and samples where the linewidth is higher than $0.8$ mT were studied further. The linewidth data for all samples is given in the Supporting Information. 

The ESR experiments were performed on a Bruker Elexsys II E500 continuous-wave ESR spectrometer in X-band (around $9.4$ GHz and a corresponding Zeeman field of about $0.33$ T), equipped with a nitrogen flow cryostat (ER4141VTM) and a programmable goniometer (ER218PG1). The reproducibility of rotations is better than $1^{\circ}$. Care was taken to appropriately set the magnetic field resolution, modulation amplitude, and sampling frequency to obtain undistorted ESR lineshapes. Spectral parameters were obtained by fitting derivative Lorentzian curves to the observed data, an inspection of the fitted lineshapes showed a proper match in all cases. The microwave cavity was kept at a constant temperature during the temperature-dependent studies and it was continuously purged with dry nitrogen. This guarantees that the power to $B_1$ conversion factor is unchanged. 

The significant line broadening for $B \parallel c$ at high microwave powers reduces the signal-to-noise ratio (SNR). For the temperature-dependent studies, we therefore offset the sample polar angle on purpose by about $8$ degrees which reduces the linewidth from $1$~mT to $0.7$~mT thus increasing the SNR by a factor $2$ and correspondingly reduces the acquisition time by a factor $4$. We measured the linewidth at each temperature point for three different microwave powers, which provides stability and reproducibility. We studied the $250-500$ K temperature range, where the $T_1$ data is fully reproducible. Our preliminary results indicated that outside this temperature range, the sample quality irreversibly reduces probably due to mechanical stress during the cooling using vacuum grease and due to the proximity of the Scotch tape. We increased the temperature by $2$ K after each measurement and thermalization was about $20$ seconds.

\section*{Acknowledgement}\vspace{-4mm}
We would like to kindly thank P. Makk and Sz. Csonka for providing some of the NGS natural graphite crystals. Work supported by the National Research, Development and Innovation Office of Hungary (NKFIH), and by the Ministry of Culture and Innovation Grants Nr. K137852, 149457, 2022-2.1.1-NL-202200004, TKP2021-EGA-02, and TKP2021-NVA-02.

\bibliography{graphite}

\begin{thebibliography}{110}%
\makeatletter
\providecommand \@ifxundefined [1]{%
 \@ifx{#1\undefined}
}%
\providecommand \@ifnum [1]{%
 \ifnum #1\expandafter \@firstoftwo
 \else \expandafter \@secondoftwo
 \fi
}%
\providecommand \@ifx [1]{%
 \ifx #1\expandafter \@firstoftwo
 \else \expandafter \@secondoftwo
 \fi
}%
\providecommand \natexlab [1]{#1}%
\providecommand \enquote  [1]{``#1''}%
\providecommand \bibnamefont  [1]{#1}%
\providecommand \bibfnamefont [1]{#1}%
\providecommand \citenamefont [1]{#1}%
\providecommand \href@noop [0]{\@secondoftwo}%
\providecommand \href [0]{\begingroup \@sanitize@url \@href}%
\providecommand \@href[1]{\@@startlink{#1}\@@href}%
\providecommand \@@href[1]{\endgroup#1\@@endlink}%
\providecommand \@sanitize@url [0]{\catcode `\\12\catcode `\$12\catcode
  `\&12\catcode `\#12\catcode `\^12\catcode `\_12\catcode `\%12\relax}%
\providecommand \@@startlink[1]{}%
\providecommand \@@endlink[0]{}%
\providecommand \url  [0]{\begingroup\@sanitize@url \@url }%
\providecommand \@url [1]{\endgroup\@href {#1}{\urlprefix }}%
\providecommand \urlprefix  [0]{URL }%
\providecommand \Eprint [0]{\href }%
\providecommand \doibase [0]{https://doi.org/}%
\providecommand \selectlanguage [0]{\@gobble}%
\providecommand \bibinfo  [0]{\@secondoftwo}%
\providecommand \bibfield  [0]{\@secondoftwo}%
\providecommand \translation [1]{[#1]}%
\providecommand \BibitemOpen [0]{}%
\providecommand \bibitemStop [0]{}%
\providecommand \bibitemNoStop [0]{.\EOS\space}%
\providecommand \EOS [0]{\spacefactor3000\relax}%
\providecommand \BibitemShut  [1]{\csname bibitem#1\endcsname}%
\let\auto@bib@innerbib\@empty
\bibitem [{\citenamefont {\v{Z}uti\'c}, \citenamefont {Fabian},\ and\
  \citenamefont {Sarma}(2004)}]{FabianRMP}%
  \BibitemOpen
  \bibfield  {author} {\bibinfo {author} {\bibfnamefont {I.}~\bibnamefont
  {\v{Z}uti\'c}}, \bibinfo {author} {\bibfnamefont {J.}~\bibnamefont
  {Fabian}},\ and\ \bibinfo {author} {\bibfnamefont {S.~D.}\ \bibnamefont
  {Sarma}},\ }\bibfield  {title} {\enquote {\bibinfo {title} {{Spintronics:
  Fundamentals and applications}},}\ }\href@noop {} {\bibfield  {journal}
  {\bibinfo  {journal} {Rev. Mod. Phys.}\ }\textbf {\bibinfo {volume} {76}},\
  \bibinfo {pages} {323--410} (\bibinfo {year} {2004})}\BibitemShut {NoStop}%
\bibitem [{\citenamefont {Wolf}\ \emph {et~al.}(2001)\citenamefont {Wolf},
  \citenamefont {Awschalom}, \citenamefont {Buhrman}, \citenamefont {Daughton},
  \citenamefont {von Moln{\'a}r}, \citenamefont {Roukes}, \citenamefont
  {Chtchelkanova},\ and\ \citenamefont {Treger}}]{WolfSCI}%
  \BibitemOpen
  \bibfield  {author} {\bibinfo {author} {\bibfnamefont {S.~A.}\ \bibnamefont
  {Wolf}}, \bibinfo {author} {\bibfnamefont {D.~D.}\ \bibnamefont {Awschalom}},
  \bibinfo {author} {\bibfnamefont {R.~A.}\ \bibnamefont {Buhrman}}, \bibinfo
  {author} {\bibfnamefont {J.~M.}\ \bibnamefont {Daughton}}, \bibinfo {author}
  {\bibfnamefont {S.}~\bibnamefont {von Moln{\'a}r}}, \bibinfo {author}
  {\bibfnamefont {M.~L.}\ \bibnamefont {Roukes}}, \bibinfo {author}
  {\bibfnamefont {A.~Y.}\ \bibnamefont {Chtchelkanova}},\ and\ \bibinfo
  {author} {\bibfnamefont {D.~M.}\ \bibnamefont {Treger}},\ }\bibfield  {title}
  {\enquote {\bibinfo {title} {{Spintronics: A Spin-Based Electronics Vision
  for the Future}},}\ }\href@noop {} {\bibfield  {journal} {\bibinfo  {journal}
  {Science}\ }\textbf {\bibinfo {volume} {294}},\ \bibinfo {pages} {1488--1495}
  (\bibinfo {year} {2001})}\BibitemShut {NoStop}%
\bibitem [{\citenamefont {Wu}, \citenamefont {Jiang},\ and\ \citenamefont
  {Weng}(2010)}]{WuReview}%
  \BibitemOpen
  \bibfield  {author} {\bibinfo {author} {\bibfnamefont {M.~W.}\ \bibnamefont
  {Wu}}, \bibinfo {author} {\bibfnamefont {J.~H.}\ \bibnamefont {Jiang}},\ and\
  \bibinfo {author} {\bibfnamefont {M.~Q.}\ \bibnamefont {Weng}},\ }\bibfield
  {title} {\enquote {\bibinfo {title} {{Spin dynamics in semiconductors}},}\
  }\href@noop {} {\bibfield  {journal} {\bibinfo  {journal} {Phys. Rep.}\
  }\textbf {\bibinfo {volume} {493}},\ \bibinfo {pages} {61--236} (\bibinfo
  {year} {2010})}\BibitemShut {NoStop}%
\bibitem [{\citenamefont {Castro~Neto}\ \emph {et~al.}(2009)\citenamefont
  {Castro~Neto}, \citenamefont {Guinea}, \citenamefont {Peres}, \citenamefont
  {Novoselov},\ and\ \citenamefont {Geim}}]{CastroNetoRMP2009}%
  \BibitemOpen
  \bibfield  {author} {\bibinfo {author} {\bibfnamefont {A.~H.}\ \bibnamefont
  {Castro~Neto}}, \bibinfo {author} {\bibfnamefont {F.}~\bibnamefont {Guinea}},
  \bibinfo {author} {\bibfnamefont {N.~M.~R.}\ \bibnamefont {Peres}}, \bibinfo
  {author} {\bibfnamefont {K.~S.}\ \bibnamefont {Novoselov}},\ and\ \bibinfo
  {author} {\bibfnamefont {A.~K.}\ \bibnamefont {Geim}},\ }\bibfield  {title}
  {\enquote {\bibinfo {title} {The electronic properties of graphene},}\
  }\href@noop {} {\bibfield  {journal} {\bibinfo  {journal} {Reviews of Modern
  Physics}\ }\textbf {\bibinfo {volume} {81}},\ \bibinfo {pages} {109--162}
  (\bibinfo {year} {2009})}\BibitemShut {NoStop}%
\bibitem [{\citenamefont {D\'ora}, \citenamefont {Mur\'anyi},\ and\
  \citenamefont {Simon}(2010)}]{DoraEPL2010}%
  \BibitemOpen
  \bibfield  {author} {\bibinfo {author} {\bibfnamefont {B.}~\bibnamefont
  {D\'ora}}, \bibinfo {author} {\bibfnamefont {F.}~\bibnamefont {Mur\'anyi}},\
  and\ \bibinfo {author} {\bibfnamefont {F.}~\bibnamefont {Simon}},\ }\bibfield
   {title} {\enquote {\bibinfo {title} {{Electron spin dynamics and electron
  spin resonance in graphene}},}\ }\href@noop {} {\bibfield  {journal}
  {\bibinfo  {journal} {Europhysics Letters}\ }\textbf {\bibinfo {volume}
  {92}},\ \bibinfo {pages} {17002} (\bibinfo {year} {2010})}\BibitemShut
  {NoStop}%
\bibitem [{\citenamefont {Han}, \citenamefont {Gmitra},\ and\ \citenamefont
  {Fabian}(2014)}]{KawakamiFabianNatNanotechn2014}%
  \BibitemOpen
  \bibfield  {author} {\bibinfo {author} {\bibfnamefont {R.~K.}\ \bibnamefont
  {Han}, \bibfnamefont {W.~Kawakami}}, \bibinfo {author} {\bibfnamefont
  {M.}~\bibnamefont {Gmitra}},\ and\ \bibinfo {author} {\bibfnamefont
  {J.}~\bibnamefont {Fabian}},\ }\bibfield  {title} {\enquote {\bibinfo {title}
  {{Graphene spintronics}},}\ }\href@noop {} {\bibfield  {journal} {\bibinfo
  {journal} {Nature Nanotechnology}\ }\textbf {\bibinfo {volume} {9}},\
  \bibinfo {pages} {794–807} (\bibinfo {year} {2014})}\BibitemShut {NoStop}%
\bibitem [{\citenamefont {Ghising}, \citenamefont {Biswas},\ and\ \citenamefont
  {Lee}(2023)}]{Graphene_Spintronics_Review}%
  \BibitemOpen
  \bibfield  {author} {\bibinfo {author} {\bibfnamefont {P.}~\bibnamefont
  {Ghising}}, \bibinfo {author} {\bibfnamefont {C.}~\bibnamefont {Biswas}},\
  and\ \bibinfo {author} {\bibfnamefont {Y.~H.}\ \bibnamefont {Lee}},\
  }\bibfield  {title} {\enquote {\bibinfo {title} {Graphene spin valves for
  spin logic devices},}\ }\href@noop {} {\bibfield  {journal} {\bibinfo
  {journal} {Advanced Materials}\ }\textbf {\bibinfo {volume} {35}},\ \bibinfo
  {pages} {2209137} (\bibinfo {year} {2023})}\BibitemShut {NoStop}%
\bibitem [{\citenamefont {Geim}\ and\ \citenamefont
  {Grigorieva}(2013)}]{GeimNat2013}%
  \BibitemOpen
  \bibfield  {author} {\bibinfo {author} {\bibfnamefont {A.~K.}\ \bibnamefont
  {Geim}}\ and\ \bibinfo {author} {\bibfnamefont {I.~V.}\ \bibnamefont
  {Grigorieva}},\ }\bibfield  {title} {\enquote {\bibinfo {title} {{Van der
  Waals heterostructures}},}\ }\href@noop {} {\bibfield  {journal} {\bibinfo
  {journal} {Nature}\ }\textbf {\bibinfo {volume} {499}},\ \bibinfo {pages}
  {419--425} (\bibinfo {year} {2013})}\BibitemShut {NoStop}%
\bibitem [{\citenamefont {Avsar}\ \emph {et~al.}(2014)\citenamefont {Avsar},
  \citenamefont {Tan}, \citenamefont {Taychatanapat}, \citenamefont
  {Balakrishnan}, \citenamefont {Koon}, \citenamefont {Yeo}, \citenamefont
  {Lahiri}, \citenamefont {Carvalho}, \citenamefont {Rodin}, \citenamefont
  {O’Farrell}, \citenamefont {Eda}, \citenamefont {Castro~Neto},\ and\
  \citenamefont {\"{O}zyilmaz}}]{OzyilmazNatComm2011}%
  \BibitemOpen
  \bibfield  {author} {\bibinfo {author} {\bibfnamefont {A.}~\bibnamefont
  {Avsar}}, \bibinfo {author} {\bibfnamefont {J.~Y.}\ \bibnamefont {Tan}},
  \bibinfo {author} {\bibfnamefont {T.}~\bibnamefont {Taychatanapat}}, \bibinfo
  {author} {\bibfnamefont {J.}~\bibnamefont {Balakrishnan}}, \bibinfo {author}
  {\bibfnamefont {G.}~\bibnamefont {Koon}}, \bibinfo {author} {\bibfnamefont
  {Y.}~\bibnamefont {Yeo}}, \bibinfo {author} {\bibfnamefont {J.}~\bibnamefont
  {Lahiri}}, \bibinfo {author} {\bibfnamefont {A.}~\bibnamefont {Carvalho}},
  \bibinfo {author} {\bibfnamefont {A.~S.}\ \bibnamefont {Rodin}}, \bibinfo
  {author} {\bibfnamefont {E.}~\bibnamefont {O’Farrell}}, \bibinfo {author}
  {\bibfnamefont {G.}~\bibnamefont {Eda}}, \bibinfo {author} {\bibfnamefont
  {A.~H.}\ \bibnamefont {Castro~Neto}},\ and\ \bibinfo {author} {\bibfnamefont
  {B.}~\bibnamefont {\"{O}zyilmaz}},\ }\bibfield  {title} {\enquote {\bibinfo
  {title} {{Spin–orbit proximity effect in graphene}},}\ }\href@noop {}
  {\bibfield  {journal} {\bibinfo  {journal} {Nature Communications}\ }\textbf
  {\bibinfo {volume} {5}},\ \bibinfo {pages} {4875} (\bibinfo {year}
  {2014})}\BibitemShut {NoStop}%
\bibitem [{\citenamefont {Wang}\ \emph {et~al.}(2015)\citenamefont {Wang},
  \citenamefont {Ki}, \citenamefont {Chen}, \citenamefont {Berger},
  \citenamefont {MacDonald},\ and\ \citenamefont
  {Morpurgo}}]{MorpurgoNatComm2015}%
  \BibitemOpen
  \bibfield  {author} {\bibinfo {author} {\bibfnamefont {Z.}~\bibnamefont
  {Wang}}, \bibinfo {author} {\bibfnamefont {D.-K.}\ \bibnamefont {Ki}},
  \bibinfo {author} {\bibfnamefont {H.}~\bibnamefont {Chen}}, \bibinfo {author}
  {\bibfnamefont {H.}~\bibnamefont {Berger}}, \bibinfo {author} {\bibfnamefont
  {A.~H.}\ \bibnamefont {MacDonald}},\ and\ \bibinfo {author} {\bibfnamefont
  {A.~F.}\ \bibnamefont {Morpurgo}},\ }\bibfield  {title} {\enquote {\bibinfo
  {title} {{Strong interface-induced spin–orbit interaction in graphene on
  WS$_2$}},}\ }\href@noop {} {\bibfield  {journal} {\bibinfo  {journal} {Nature
  Communications}\ }\textbf {\bibinfo {volume} {6}},\ \bibinfo {pages} {8339}
  (\bibinfo {year} {2015})}\BibitemShut {NoStop}%
\bibitem [{\citenamefont {Wang}\ \emph {et~al.}(2016)\citenamefont {Wang},
  \citenamefont {Ki}, \citenamefont {Khoo}, \citenamefont {Mauro},
  \citenamefont {Berger}, \citenamefont {Levitov},\ and\ \citenamefont
  {Morpurgo}}]{MorpurgoPRX2016}%
  \BibitemOpen
  \bibfield  {author} {\bibinfo {author} {\bibfnamefont {Z.}~\bibnamefont
  {Wang}}, \bibinfo {author} {\bibfnamefont {D.-K.}\ \bibnamefont {Ki}},
  \bibinfo {author} {\bibfnamefont {J.~Y.}\ \bibnamefont {Khoo}}, \bibinfo
  {author} {\bibfnamefont {D.}~\bibnamefont {Mauro}}, \bibinfo {author}
  {\bibfnamefont {H.}~\bibnamefont {Berger}}, \bibinfo {author} {\bibfnamefont
  {L.~S.}\ \bibnamefont {Levitov}},\ and\ \bibinfo {author} {\bibfnamefont
  {A.~F.}\ \bibnamefont {Morpurgo}},\ }\bibfield  {title} {\enquote {\bibinfo
  {title} {{Origin and Magnitude of 'Designer' Spin-Orbit Interaction in
  Graphene on Semiconducting Transition Metal Dichalcogenides}},}\ }\href@noop
  {} {\bibfield  {journal} {\bibinfo  {journal} {Phys. Rev. X}\ }\textbf
  {\bibinfo {volume} {6}},\ \bibinfo {pages} {041020} (\bibinfo {year}
  {2016})}\BibitemShut {NoStop}%
\bibitem [{\citenamefont {Yang}\ \emph {et~al.}(2016)\citenamefont {Yang},
  \citenamefont {Tu}, \citenamefont {Kim}, \citenamefont {Wu}, \citenamefont
  {Wang}, \citenamefont {Alicea}, \citenamefont {Wu}, \citenamefont
  {Bockrath},\ and\ \citenamefont {Shi}}]{Graphene_WS2_hetero_1}%
  \BibitemOpen
  \bibfield  {author} {\bibinfo {author} {\bibfnamefont {B.}~\bibnamefont
  {Yang}}, \bibinfo {author} {\bibfnamefont {M.-F.}\ \bibnamefont {Tu}},
  \bibinfo {author} {\bibfnamefont {J.}~\bibnamefont {Kim}}, \bibinfo {author}
  {\bibfnamefont {Y.}~\bibnamefont {Wu}}, \bibinfo {author} {\bibfnamefont
  {H.}~\bibnamefont {Wang}}, \bibinfo {author} {\bibfnamefont {J.}~\bibnamefont
  {Alicea}}, \bibinfo {author} {\bibfnamefont {R.}~\bibnamefont {Wu}}, \bibinfo
  {author} {\bibfnamefont {M.}~\bibnamefont {Bockrath}},\ and\ \bibinfo
  {author} {\bibfnamefont {J.}~\bibnamefont {Shi}},\ }\bibfield  {title}
  {\enquote {\bibinfo {title} {{Tunable spin{\textendash}orbit coupling and
  symmetry-protected edge states in graphene/WS$_2$}},}\ }\href@noop {}
  {\bibfield  {journal} {\bibinfo  {journal} {2D Materials}\ }\textbf {\bibinfo
  {volume} {3}},\ \bibinfo {pages} {031012} (\bibinfo {year}
  {2016})}\BibitemShut {NoStop}%
\bibitem [{\citenamefont {Yang}\ \emph {et~al.}(2017)\citenamefont {Yang},
  \citenamefont {Lohmann}, \citenamefont {Barroso}, \citenamefont {Liao},
  \citenamefont {Lin}, \citenamefont {Liu}, \citenamefont {Bartels},
  \citenamefont {Watanabe}, \citenamefont {Taniguchi},\ and\ \citenamefont
  {Shi}}]{Graphene_WS2_hetero_2}%
  \BibitemOpen
  \bibfield  {author} {\bibinfo {author} {\bibfnamefont {B.}~\bibnamefont
  {Yang}}, \bibinfo {author} {\bibfnamefont {M.}~\bibnamefont {Lohmann}},
  \bibinfo {author} {\bibfnamefont {D.}~\bibnamefont {Barroso}}, \bibinfo
  {author} {\bibfnamefont {I.}~\bibnamefont {Liao}}, \bibinfo {author}
  {\bibfnamefont {Z.}~\bibnamefont {Lin}}, \bibinfo {author} {\bibfnamefont
  {Y.}~\bibnamefont {Liu}}, \bibinfo {author} {\bibfnamefont {L.}~\bibnamefont
  {Bartels}}, \bibinfo {author} {\bibfnamefont {K.}~\bibnamefont {Watanabe}},
  \bibinfo {author} {\bibfnamefont {T.}~\bibnamefont {Taniguchi}},\ and\
  \bibinfo {author} {\bibfnamefont {J.}~\bibnamefont {Shi}},\ }\bibfield
  {title} {\enquote {\bibinfo {title} {{Strong electron-hole symmetric Rashba
  spin-orbit coupling in graphene/monolayer transition metal dichalcogenide
  heterostructures}},}\ }\href@noop {} {\bibfield  {journal} {\bibinfo
  {journal} {Phys. Rev. B}\ }\textbf {\bibinfo {volume} {96}},\ \bibinfo
  {pages} {041409} (\bibinfo {year} {2017})}\BibitemShut {NoStop}%
\bibitem [{\citenamefont {Torres}\ \emph {et~al.}(2017)\citenamefont {Torres},
  \citenamefont {Sierra}, \citenamefont {Ben{\'{\i}}tez}, \citenamefont
  {Bonell}, \citenamefont {Costache},\ and\ \citenamefont
  {Valenzuela}}]{Valenzuela2DMat2017}%
  \BibitemOpen
  \bibfield  {author} {\bibinfo {author} {\bibfnamefont {W.~S.}\ \bibnamefont
  {Torres}}, \bibinfo {author} {\bibfnamefont {J.~F.}\ \bibnamefont {Sierra}},
  \bibinfo {author} {\bibfnamefont {L.~A.}\ \bibnamefont {Ben{\'{\i}}tez}},
  \bibinfo {author} {\bibfnamefont {F.}~\bibnamefont {Bonell}}, \bibinfo
  {author} {\bibfnamefont {M.~V.}\ \bibnamefont {Costache}},\ and\ \bibinfo
  {author} {\bibfnamefont {S.~O.}\ \bibnamefont {Valenzuela}},\ }\bibfield
  {title} {\enquote {\bibinfo {title} {{Spin precession and spin Hall effect in
  monolayer graphene/Pt nanostructures}},}\ }\href@noop {} {\bibfield
  {journal} {\bibinfo  {journal} {2D Materials}\ }\textbf {\bibinfo {volume}
  {4}},\ \bibinfo {pages} {041008} (\bibinfo {year} {2017})}\BibitemShut
  {NoStop}%
\bibitem [{\citenamefont {Offidani}\ \emph {et~al.}(2017)\citenamefont
  {Offidani}, \citenamefont {Milletar\`{\i}}, \citenamefont {Raimondi},\ and\
  \citenamefont {Ferreira}}]{FerreiraPRL2017}%
  \BibitemOpen
  \bibfield  {author} {\bibinfo {author} {\bibfnamefont {M.}~\bibnamefont
  {Offidani}}, \bibinfo {author} {\bibfnamefont {M.}~\bibnamefont
  {Milletar\`{\i}}}, \bibinfo {author} {\bibfnamefont {R.}~\bibnamefont
  {Raimondi}},\ and\ \bibinfo {author} {\bibfnamefont {A.}~\bibnamefont
  {Ferreira}},\ }\bibfield  {title} {\enquote {\bibinfo {title} {{Optimal
  Charge-to-Spin Conversion in Graphene on Transition-Metal
  Dichalcogenides}},}\ }\href@noop {} {\bibfield  {journal} {\bibinfo
  {journal} {Phys. Rev. Lett.}\ }\textbf {\bibinfo {volume} {119}},\ \bibinfo
  {pages} {196801} (\bibinfo {year} {2017})}\BibitemShut {NoStop}%
\bibitem [{\citenamefont {Dankert}\ and\ \citenamefont
  {Dash}(2017)}]{DashNatComm2017}%
  \BibitemOpen
  \bibfield  {author} {\bibinfo {author} {\bibfnamefont {A.}~\bibnamefont
  {Dankert}}\ and\ \bibinfo {author} {\bibfnamefont {S.~P.}\ \bibnamefont
  {Dash}},\ }\bibfield  {title} {\enquote {\bibinfo {title} {{Electrical gate
  control of spin current in van der Waals heterostructures at room
  temperature}},}\ }\href@noop {} {\bibfield  {journal} {\bibinfo  {journal}
  {Nature Communications}\ }\textbf {\bibinfo {volume} {8}},\ \bibinfo {pages}
  {16093} (\bibinfo {year} {2017})}\BibitemShut {NoStop}%
\bibitem [{\citenamefont {Ben\'{i}tez}\ \emph {et~al.}(2020)\citenamefont
  {Ben\'{i}tez}, \citenamefont {Savero~Torres}, \citenamefont {Sierra},
  \citenamefont {Timmermans}, \citenamefont {Garcia}, \citenamefont {Roche},
  \citenamefont {Costache},\ and\ \citenamefont
  {Valenzuela}}]{ValenzuelaNatMat2020}%
  \BibitemOpen
  \bibfield  {author} {\bibinfo {author} {\bibfnamefont {L.~A.}\ \bibnamefont
  {Ben\'{i}tez}}, \bibinfo {author} {\bibfnamefont {W.}~\bibnamefont
  {Savero~Torres}}, \bibinfo {author} {\bibfnamefont {J.~F.}\ \bibnamefont
  {Sierra}}, \bibinfo {author} {\bibfnamefont {M.}~\bibnamefont {Timmermans}},
  \bibinfo {author} {\bibfnamefont {J.~H.}\ \bibnamefont {Garcia}}, \bibinfo
  {author} {\bibfnamefont {S.}~\bibnamefont {Roche}}, \bibinfo {author}
  {\bibfnamefont {M.~V.}\ \bibnamefont {Costache}},\ and\ \bibinfo {author}
  {\bibfnamefont {S.~O.}\ \bibnamefont {Valenzuela}},\ }\bibfield  {title}
  {\enquote {\bibinfo {title} {{Tunable room-temperature spin galvanic and spin
  Hall effects in van der Waals heterostructures}},}\ }\href@noop {} {\bibfield
   {journal} {\bibinfo  {journal} {Nat. Mat.}\ }\textbf {\bibinfo {volume}
  {19}},\ \bibinfo {pages} {170–175} (\bibinfo {year} {2020})}\BibitemShut
  {NoStop}%
\bibitem [{\citenamefont {Gmitra}\ and\ \citenamefont
  {Fabian}(2015)}]{FabianGmitraPRB2015}%
  \BibitemOpen
  \bibfield  {author} {\bibinfo {author} {\bibfnamefont {M.}~\bibnamefont
  {Gmitra}}\ and\ \bibinfo {author} {\bibfnamefont {J.}~\bibnamefont
  {Fabian}},\ }\bibfield  {title} {\enquote {\bibinfo {title} {{Graphene on
  transition-metal dichalcogenides: A platform for proximity spin-orbit physics
  and optospintronics}},}\ }\href@noop {} {\bibfield  {journal} {\bibinfo
  {journal} {Phys. Rev. B}\ }\textbf {\bibinfo {volume} {92}},\ \bibinfo
  {pages} {155403} (\bibinfo {year} {2015})}\BibitemShut {NoStop}%
\bibitem [{\citenamefont {Gmitra}\ and\ \citenamefont
  {Fabian}(2017)}]{FabianGmitraPRL2017}%
  \BibitemOpen
  \bibfield  {author} {\bibinfo {author} {\bibfnamefont {M.}~\bibnamefont
  {Gmitra}}\ and\ \bibinfo {author} {\bibfnamefont {J.}~\bibnamefont
  {Fabian}},\ }\bibfield  {title} {\enquote {\bibinfo {title} {{Proximity
  Effects in Bilayer Graphene on Monolayer ${\mathrm{WSe}}_{2}$: Field-Effect
  Spin Valley Locking, Spin-Orbit Valve, and Spin Transistor}},}\ }\href@noop
  {} {\bibfield  {journal} {\bibinfo  {journal} {Phys. Rev. Lett.}\ }\textbf
  {\bibinfo {volume} {119}},\ \bibinfo {pages} {146401} (\bibinfo {year}
  {2017})}\BibitemShut {NoStop}%
\bibitem [{\citenamefont {\v{Z}uti\'{c}}\ \emph {et~al.}(2019)\citenamefont
  {\v{Z}uti\'{c}}, \citenamefont {A.}, \citenamefont {B.}, \citenamefont
  {Dery},\ and\ \citenamefont {Belashchenko}}]{ZuticMatToday2019}%
  \BibitemOpen
  \bibfield  {author} {\bibinfo {author} {\bibfnamefont {I.}~\bibnamefont
  {\v{Z}uti\'{c}}}, \bibinfo {author} {\bibfnamefont {M.-A.}\ \bibnamefont
  {A.}}, \bibinfo {author} {\bibfnamefont {S.}~\bibnamefont {B.}}, \bibinfo
  {author} {\bibfnamefont {H.}~\bibnamefont {Dery}},\ and\ \bibinfo {author}
  {\bibfnamefont {K.}~\bibnamefont {Belashchenko}},\ }\bibfield  {title}
  {\enquote {\bibinfo {title} {{Proximitized materials"}},}\ }\href@noop {}
  {\bibfield  {journal} {\bibinfo  {journal} {Materials Today}\ }\textbf
  {\bibinfo {volume} {22}},\ \bibinfo {pages} {85 -- 107} (\bibinfo {year}
  {2019})}\BibitemShut {NoStop}%
\bibitem [{\citenamefont {H\"ogl}\ \emph {et~al.}(2020)\citenamefont {H\"ogl},
  \citenamefont {Frank}, \citenamefont {Zollner}, \citenamefont {Kochan},
  \citenamefont {Gmitra},\ and\ \citenamefont {Fabian}}]{FabianPRL2020}%
  \BibitemOpen
  \bibfield  {author} {\bibinfo {author} {\bibfnamefont {P.}~\bibnamefont
  {H\"ogl}}, \bibinfo {author} {\bibfnamefont {T.}~\bibnamefont {Frank}},
  \bibinfo {author} {\bibfnamefont {K.}~\bibnamefont {Zollner}}, \bibinfo
  {author} {\bibfnamefont {D.}~\bibnamefont {Kochan}}, \bibinfo {author}
  {\bibfnamefont {M.}~\bibnamefont {Gmitra}},\ and\ \bibinfo {author}
  {\bibfnamefont {J.}~\bibnamefont {Fabian}},\ }\bibfield  {title} {\enquote
  {\bibinfo {title} {{Quantum Anomalous Hall Effects in Graphene from
  Proximity-Induced Uniform and Staggered Spin-Orbit and Exchange Coupling}},}\
  }\href@noop {} {\bibfield  {journal} {\bibinfo  {journal} {Phys. Rev. Lett.}\
  }\textbf {\bibinfo {volume} {124}},\ \bibinfo {pages} {136403} (\bibinfo
  {year} {2020})}\BibitemShut {NoStop}%
\bibitem [{\citenamefont {Sierra}\ \emph {et~al.}(2025)\citenamefont {Sierra},
  \citenamefont {Sv\v{e}tl\'ik}, \citenamefont {Torres}, \citenamefont
  {Camosi}, \citenamefont {Herling}, \citenamefont {Guillet}, \citenamefont
  {Xu}, \citenamefont {Reparaz}, \citenamefont {Marinova}, \citenamefont
  {Dimitrov},\ and\ \citenamefont {Valenzuela}}]{ValenzuelaNatMater2025}%
  \BibitemOpen
  \bibfield  {author} {\bibinfo {author} {\bibfnamefont {J.~F.}\ \bibnamefont
  {Sierra}}, \bibinfo {author} {\bibfnamefont {J.}~\bibnamefont
  {Sv\v{e}tl\'ik}}, \bibinfo {author} {\bibfnamefont {W.~S.}\ \bibnamefont
  {Torres}}, \bibinfo {author} {\bibfnamefont {L.}~\bibnamefont {Camosi}},
  \bibinfo {author} {\bibfnamefont {F.}~\bibnamefont {Herling}}, \bibinfo
  {author} {\bibfnamefont {T.}~\bibnamefont {Guillet}}, \bibinfo {author}
  {\bibfnamefont {K.}~\bibnamefont {Xu}}, \bibinfo {author} {\bibfnamefont
  {J.~S.}\ \bibnamefont {Reparaz}}, \bibinfo {author} {\bibfnamefont
  {V.}~\bibnamefont {Marinova}}, \bibinfo {author} {\bibfnamefont
  {D.}~\bibnamefont {Dimitrov}},\ and\ \bibinfo {author} {\bibfnamefont
  {S.~O.}\ \bibnamefont {Valenzuela}},\ }\bibfield  {title} {\enquote {\bibinfo
  {title} {{Room-temperature anisotropic in-plane spin dynamics in graphene
  induced by PdSe$_2$ proximity}},}\ }\href@noop {} {\bibfield  {journal}
  {\bibinfo  {journal} {Nature Materials}\ } (\bibinfo {year}
  {2025})}\BibitemShut {NoStop}%
\bibitem [{\citenamefont {Cummings}\ \emph {et~al.}(2017)\citenamefont
  {Cummings}, \citenamefont {Garcia}, \citenamefont {Fabian},\ and\
  \citenamefont {Roche}}]{FabianRochePRL2017}%
  \BibitemOpen
  \bibfield  {author} {\bibinfo {author} {\bibfnamefont {A.~W.}\ \bibnamefont
  {Cummings}}, \bibinfo {author} {\bibfnamefont {J.~H.}\ \bibnamefont
  {Garcia}}, \bibinfo {author} {\bibfnamefont {J.}~\bibnamefont {Fabian}},\
  and\ \bibinfo {author} {\bibfnamefont {S.}~\bibnamefont {Roche}},\ }\bibfield
   {title} {\enquote {\bibinfo {title} {{Giant Spin Lifetime Anisotropy in
  Graphene Induced by Proximity Effects}},}\ }\href@noop {} {\bibfield
  {journal} {\bibinfo  {journal} {Phys. Rev. Lett.}\ }\textbf {\bibinfo
  {volume} {119}},\ \bibinfo {pages} {206601} (\bibinfo {year}
  {2017})}\BibitemShut {NoStop}%
\bibitem [{\citenamefont {Benítez}\ \emph {et~al.}(2018)\citenamefont
  {Benítez}, \citenamefont {Sierra}, \citenamefont {Savero~Torres},
  \citenamefont {Arrighi}, \citenamefont {Bonell}, \citenamefont {Costache},\
  and\ \citenamefont {Valenzuela}}]{ValenzuelaNatPhys2018}%
  \BibitemOpen
  \bibfield  {author} {\bibinfo {author} {\bibfnamefont {L.~A.}\ \bibnamefont
  {Benítez}}, \bibinfo {author} {\bibfnamefont {J.~F.}\ \bibnamefont
  {Sierra}}, \bibinfo {author} {\bibfnamefont {W.}~\bibnamefont
  {Savero~Torres}}, \bibinfo {author} {\bibfnamefont {A.}~\bibnamefont
  {Arrighi}}, \bibinfo {author} {\bibfnamefont {F.}~\bibnamefont {Bonell}},
  \bibinfo {author} {\bibfnamefont {M.~V.}\ \bibnamefont {Costache}},\ and\
  \bibinfo {author} {\bibfnamefont {S.~O.}\ \bibnamefont {Valenzuela}},\
  }\bibfield  {title} {\enquote {\bibinfo {title} {{Strongly anisotropic spin
  relaxation in graphene–transition metal dichalcogenide heterostructures at
  room temperature}},}\ }\href@noop {} {\bibfield  {journal} {\bibinfo
  {journal} {Nat. Phys.}\ }\textbf {\bibinfo {volume} {14}},\ \bibinfo {pages}
  {303–308} (\bibinfo {year} {2018})}\BibitemShut {NoStop}%
\bibitem [{\citenamefont {Leutenantsmeyer}\ \emph {et~al.}(2018)\citenamefont
  {Leutenantsmeyer}, \citenamefont {Ingla-Ayn\'es}, \citenamefont {Fabian},\
  and\ \citenamefont {van Wees}}]{Fabian_vanWeesPRL2018}%
  \BibitemOpen
  \bibfield  {author} {\bibinfo {author} {\bibfnamefont {J.~C.}\ \bibnamefont
  {Leutenantsmeyer}}, \bibinfo {author} {\bibfnamefont {J.}~\bibnamefont
  {Ingla-Ayn\'es}}, \bibinfo {author} {\bibfnamefont {J.}~\bibnamefont
  {Fabian}},\ and\ \bibinfo {author} {\bibfnamefont {B.~J.}\ \bibnamefont {van
  Wees}},\ }\bibfield  {title} {\enquote {\bibinfo {title} {{Observation of
  Spin-Valley-Coupling-Induced Large Spin-Lifetime Anisotropy in Bilayer
  Graphene}},}\ }\href@noop {} {\bibfield  {journal} {\bibinfo  {journal}
  {Phys. Rev. Lett.}\ }\textbf {\bibinfo {volume} {121}},\ \bibinfo {pages}
  {127702} (\bibinfo {year} {2018})}\BibitemShut {NoStop}%
\bibitem [{\citenamefont {Zihlmann}\ \emph {et~al.}(2018)\citenamefont
  {Zihlmann}, \citenamefont {Cummings}, \citenamefont {Garcia}, \citenamefont
  {Kedves}, \citenamefont {Watanabe}, \citenamefont {Taniguchi}, \citenamefont
  {Sch\"onenberger},\ and\ \citenamefont {Makk}}]{MakkPRB2018}%
  \BibitemOpen
  \bibfield  {author} {\bibinfo {author} {\bibfnamefont {S.}~\bibnamefont
  {Zihlmann}}, \bibinfo {author} {\bibfnamefont {A.~W.}\ \bibnamefont
  {Cummings}}, \bibinfo {author} {\bibfnamefont {J.~H.}\ \bibnamefont
  {Garcia}}, \bibinfo {author} {\bibfnamefont {M.}~\bibnamefont {Kedves}},
  \bibinfo {author} {\bibfnamefont {K.}~\bibnamefont {Watanabe}}, \bibinfo
  {author} {\bibfnamefont {T.}~\bibnamefont {Taniguchi}}, \bibinfo {author}
  {\bibfnamefont {C.}~\bibnamefont {Sch\"onenberger}},\ and\ \bibinfo {author}
  {\bibfnamefont {P.}~\bibnamefont {Makk}},\ }\bibfield  {title} {\enquote
  {\bibinfo {title} {{Large spin relaxation anisotropy and valley-Zeeman
  spin-orbit coupling in ${\mathrm{WSe}}_{2}$/graphene/$h$-BN
  heterostructures}},}\ }\href@noop {} {\bibfield  {journal} {\bibinfo
  {journal} {Phys. Rev. B}\ }\textbf {\bibinfo {volume} {97}},\ \bibinfo
  {pages} {075434} (\bibinfo {year} {2018})}\BibitemShut {NoStop}%
\bibitem [{\citenamefont {Xu}\ \emph {et~al.}(2018)\citenamefont {Xu},
  \citenamefont {Zhu}, \citenamefont {Luo}, \citenamefont {Lu},\ and\
  \citenamefont {Kawakami}}]{KawakamiPRL2018}%
  \BibitemOpen
  \bibfield  {author} {\bibinfo {author} {\bibfnamefont {J.}~\bibnamefont
  {Xu}}, \bibinfo {author} {\bibfnamefont {T.}~\bibnamefont {Zhu}}, \bibinfo
  {author} {\bibfnamefont {Y.~K.}\ \bibnamefont {Luo}}, \bibinfo {author}
  {\bibfnamefont {Y.-M.}\ \bibnamefont {Lu}},\ and\ \bibinfo {author}
  {\bibfnamefont {R.~K.}\ \bibnamefont {Kawakami}},\ }\bibfield  {title}
  {\enquote {\bibinfo {title} {{Strong and Tunable Spin-Lifetime Anisotropy in
  Dual-Gated Bilayer Graphene}},}\ }\href@noop {} {\bibfield  {journal}
  {\bibinfo  {journal} {Phys. Rev. Lett.}\ }\textbf {\bibinfo {volume} {121}},\
  \bibinfo {pages} {127703} (\bibinfo {year} {2018})}\BibitemShut {NoStop}%
\bibitem [{\citenamefont {Wakamura}\ \emph {et~al.}(2018)\citenamefont
  {Wakamura}, \citenamefont {Reale}, \citenamefont {Palczynski}, \citenamefont
  {Gu\'eron}, \citenamefont {Mattevi},\ and\ \citenamefont
  {Bouchiat}}]{BouchiatAnis_PRL2018}%
  \BibitemOpen
  \bibfield  {author} {\bibinfo {author} {\bibfnamefont {T.}~\bibnamefont
  {Wakamura}}, \bibinfo {author} {\bibfnamefont {F.}~\bibnamefont {Reale}},
  \bibinfo {author} {\bibfnamefont {P.}~\bibnamefont {Palczynski}}, \bibinfo
  {author} {\bibfnamefont {S.}~\bibnamefont {Gu\'eron}}, \bibinfo {author}
  {\bibfnamefont {C.}~\bibnamefont {Mattevi}},\ and\ \bibinfo {author}
  {\bibfnamefont {H.}~\bibnamefont {Bouchiat}},\ }\bibfield  {title} {\enquote
  {\bibinfo {title} {{Strong Anisotropic Spin-Orbit Interaction Induced in
  Graphene by Monolayer ${\mathrm{WS}}_{2}$}},}\ }\href@noop {} {\bibfield
  {journal} {\bibinfo  {journal} {Phys. Rev. Lett.}\ }\textbf {\bibinfo
  {volume} {120}},\ \bibinfo {pages} {106802} (\bibinfo {year}
  {2018})}\BibitemShut {NoStop}%
\bibitem [{\citenamefont {Omar}, \citenamefont {Madhushankar},\ and\
  \citenamefont {van Wees}(2019)}]{vanWeesPRB2019}%
  \BibitemOpen
  \bibfield  {author} {\bibinfo {author} {\bibfnamefont {S.}~\bibnamefont
  {Omar}}, \bibinfo {author} {\bibfnamefont {B.~N.}\ \bibnamefont
  {Madhushankar}},\ and\ \bibinfo {author} {\bibfnamefont {B.~J.}\ \bibnamefont
  {van Wees}},\ }\bibfield  {title} {\enquote {\bibinfo {title} {{Large
  spin-relaxation anisotropy in bilayer-graphene/${\mathrm{WS}}_{2}$
  heterostructures}},}\ }\href@noop {} {\bibfield  {journal} {\bibinfo
  {journal} {Phys. Rev. B}\ }\textbf {\bibinfo {volume} {100}},\ \bibinfo
  {pages} {155415} (\bibinfo {year} {2019})}\BibitemShut {NoStop}%
\bibitem [{\citenamefont {Tombros}\ \emph {et~al.}(2008)\citenamefont
  {Tombros}, \citenamefont {Tanabe}, \citenamefont {Veligura}, \citenamefont
  {J\'ozsa}, \citenamefont {Popinciuc}, \citenamefont {Jonkman},\ and\
  \citenamefont {van Wees}}]{TombrosPRL2008}%
  \BibitemOpen
  \bibfield  {author} {\bibinfo {author} {\bibfnamefont {N.}~\bibnamefont
  {Tombros}}, \bibinfo {author} {\bibfnamefont {S.}~\bibnamefont {Tanabe}},
  \bibinfo {author} {\bibfnamefont {A.}~\bibnamefont {Veligura}}, \bibinfo
  {author} {\bibfnamefont {C.}~\bibnamefont {J\'ozsa}}, \bibinfo {author}
  {\bibfnamefont {M.}~\bibnamefont {Popinciuc}}, \bibinfo {author}
  {\bibfnamefont {H.~T.}\ \bibnamefont {Jonkman}},\ and\ \bibinfo {author}
  {\bibfnamefont {B.~J.}\ \bibnamefont {van Wees}},\ }\bibfield  {title}
  {\enquote {\bibinfo {title} {{Anisotropic Spin Relaxation in Graphene}},}\
  }\href@noop {} {\bibfield  {journal} {\bibinfo  {journal} {Phys. Rev. Lett.}\
  }\textbf {\bibinfo {volume} {101}},\ \bibinfo {pages} {046601} (\bibinfo
  {year} {2008})}\BibitemShut {NoStop}%
\bibitem [{\citenamefont {Raes}\ \emph {et~al.}(2016)\citenamefont {Raes},
  \citenamefont {Scheerder}, \citenamefont {Costache}, \citenamefont {Bonell},
  \citenamefont {Sierra}, \citenamefont {Cuppens}, \citenamefont {Van~de
  Vondel},\ and\ \citenamefont {Valenzuela}}]{ValenzuelaNatComm2016}%
  \BibitemOpen
  \bibfield  {author} {\bibinfo {author} {\bibfnamefont {B.}~\bibnamefont
  {Raes}}, \bibinfo {author} {\bibfnamefont {J.}~\bibnamefont {Scheerder}},
  \bibinfo {author} {\bibfnamefont {M.~V.}\ \bibnamefont {Costache}}, \bibinfo
  {author} {\bibfnamefont {F.}~\bibnamefont {Bonell}}, \bibinfo {author}
  {\bibfnamefont {J.~F.}\ \bibnamefont {Sierra}}, \bibinfo {author}
  {\bibfnamefont {J.}~\bibnamefont {Cuppens}}, \bibinfo {author} {\bibfnamefont
  {J.}~\bibnamefont {Van~de Vondel}},\ and\ \bibinfo {author} {\bibfnamefont
  {S.~O.}\ \bibnamefont {Valenzuela}},\ }\bibfield  {title} {\enquote {\bibinfo
  {title} {{Determination of the spin-lifetime anisotropy in graphene using
  oblique spin precession}},}\ }\href@noop {} {\bibfield  {journal} {\bibinfo
  {journal} {Nature Communications}\ }\textbf {\bibinfo {volume} {7}},\
  \bibinfo {pages} {11444} (\bibinfo {year} {2016})}\BibitemShut {NoStop}%
\bibitem [{\citenamefont {Zhu}\ and\ \citenamefont
  {Kawakami}(2018)}]{KawakamiAnisPRB2018}%
  \BibitemOpen
  \bibfield  {author} {\bibinfo {author} {\bibfnamefont {T.}~\bibnamefont
  {Zhu}}\ and\ \bibinfo {author} {\bibfnamefont {R.~K.}\ \bibnamefont
  {Kawakami}},\ }\bibfield  {title} {\enquote {\bibinfo {title} {{Modeling the
  oblique spin precession in lateral spin valves for accurate determination of
  the spin lifetime anisotropy: Effect of finite contact resistance and channel
  length}},}\ }\href@noop {} {\bibfield  {journal} {\bibinfo  {journal} {Phys.
  Rev. B}\ }\textbf {\bibinfo {volume} {97}},\ \bibinfo {pages} {144413}
  (\bibinfo {year} {2018})}\BibitemShut {NoStop}%
\bibitem [{\citenamefont {Ringer}\ \emph {et~al.}(2018)\citenamefont {Ringer},
  \citenamefont {Hartl}, \citenamefont {Rosenauer}, \citenamefont {V\"olkl},
  \citenamefont {Kadur}, \citenamefont {Hopperdietzel}, \citenamefont {Weiss},\
  and\ \citenamefont {Eroms}}]{GraphAnisPRB2018}%
  \BibitemOpen
  \bibfield  {author} {\bibinfo {author} {\bibfnamefont {S.}~\bibnamefont
  {Ringer}}, \bibinfo {author} {\bibfnamefont {S.}~\bibnamefont {Hartl}},
  \bibinfo {author} {\bibfnamefont {M.}~\bibnamefont {Rosenauer}}, \bibinfo
  {author} {\bibfnamefont {T.}~\bibnamefont {V\"olkl}}, \bibinfo {author}
  {\bibfnamefont {M.}~\bibnamefont {Kadur}}, \bibinfo {author} {\bibfnamefont
  {F.}~\bibnamefont {Hopperdietzel}}, \bibinfo {author} {\bibfnamefont
  {D.}~\bibnamefont {Weiss}},\ and\ \bibinfo {author} {\bibfnamefont
  {J.}~\bibnamefont {Eroms}},\ }\bibfield  {title} {\enquote {\bibinfo {title}
  {{Measuring anisotropic spin relaxation in graphene}},}\ }\href@noop {}
  {\bibfield  {journal} {\bibinfo  {journal} {Phys. Rev. B}\ }\textbf {\bibinfo
  {volume} {97}},\ \bibinfo {pages} {205439} (\bibinfo {year}
  {2018})}\BibitemShut {NoStop}%
\bibitem [{\citenamefont {Tombros}\ \emph {et~al.}(2007)\citenamefont
  {Tombros}, \citenamefont {J\'ozsa}, \citenamefont {Popinciuc}, \citenamefont
  {Jonkman},\ and\ \citenamefont {van Wees}}]{TombrosNAT2007}%
  \BibitemOpen
  \bibfield  {author} {\bibinfo {author} {\bibfnamefont {N.}~\bibnamefont
  {Tombros}}, \bibinfo {author} {\bibfnamefont {C.}~\bibnamefont {J\'ozsa}},
  \bibinfo {author} {\bibfnamefont {M.}~\bibnamefont {Popinciuc}}, \bibinfo
  {author} {\bibfnamefont {H.~T.}\ \bibnamefont {Jonkman}},\ and\ \bibinfo
  {author} {\bibfnamefont {B.~J.}\ \bibnamefont {van Wees}},\ }\bibfield
  {title} {\enquote {\bibinfo {title} {{Electronic spin transport and spin
  precession in single graphene layers at room temperature}},}\ }\href@noop {}
  {\bibfield  {journal} {\bibinfo  {journal} {Nature}\ }\textbf {\bibinfo
  {volume} {448}},\ \bibinfo {pages} {571--574} (\bibinfo {year}
  {2007})}\BibitemShut {NoStop}%
\bibitem [{\citenamefont {Han}\ and\ \citenamefont
  {Kawakami}(2011)}]{KawakamiPRL2011}%
  \BibitemOpen
  \bibfield  {author} {\bibinfo {author} {\bibfnamefont {W.}~\bibnamefont
  {Han}}\ and\ \bibinfo {author} {\bibfnamefont {R.~K.}\ \bibnamefont
  {Kawakami}},\ }\bibfield  {title} {\enquote {\bibinfo {title} {{Spin
  Relaxation in Single-Layer and Bilayer Graphene}},}\ }\href@noop {}
  {\bibfield  {journal} {\bibinfo  {journal} {Phys. Rev. Lett.}\ }\textbf
  {\bibinfo {volume} {107}},\ \bibinfo {pages} {047207} (\bibinfo {year}
  {2011})}\BibitemShut {NoStop}%
\bibitem [{\citenamefont {Yang}\ \emph {et~al.}(2011)\citenamefont {Yang},
  \citenamefont {Balakrishnan}, \citenamefont {Volmer}, \citenamefont {Avsar},
  \citenamefont {Jaiswal}, \citenamefont {Samm}, \citenamefont {Ali},
  \citenamefont {Pachoud}, \citenamefont {Zeng}, \citenamefont {Popinciuc},
  \citenamefont {G\"untherodt}, \citenamefont {Beschoten},\ and\ \citenamefont
  {\"Ozyilmaz}}]{OzyilmazPRL2011}%
  \BibitemOpen
  \bibfield  {author} {\bibinfo {author} {\bibfnamefont {T.-Y.}\ \bibnamefont
  {Yang}}, \bibinfo {author} {\bibfnamefont {J.}~\bibnamefont {Balakrishnan}},
  \bibinfo {author} {\bibfnamefont {F.}~\bibnamefont {Volmer}}, \bibinfo
  {author} {\bibfnamefont {A.}~\bibnamefont {Avsar}}, \bibinfo {author}
  {\bibfnamefont {M.}~\bibnamefont {Jaiswal}}, \bibinfo {author} {\bibfnamefont
  {J.}~\bibnamefont {Samm}}, \bibinfo {author} {\bibfnamefont {S.~R.}\
  \bibnamefont {Ali}}, \bibinfo {author} {\bibfnamefont {A.}~\bibnamefont
  {Pachoud}}, \bibinfo {author} {\bibfnamefont {M.}~\bibnamefont {Zeng}},
  \bibinfo {author} {\bibfnamefont {M.}~\bibnamefont {Popinciuc}}, \bibinfo
  {author} {\bibfnamefont {G.}~\bibnamefont {G\"untherodt}}, \bibinfo {author}
  {\bibfnamefont {B.}~\bibnamefont {Beschoten}},\ and\ \bibinfo {author}
  {\bibfnamefont {B.}~\bibnamefont {\"Ozyilmaz}},\ }\bibfield  {title}
  {\enquote {\bibinfo {title} {{Observation of Long Spin-Relaxation Times in
  Bilayer Graphene at Room Temperature}},}\ }\href@noop {} {\bibfield
  {journal} {\bibinfo  {journal} {Phys. Rev. Lett.}\ }\textbf {\bibinfo
  {volume} {107}},\ \bibinfo {pages} {047206} (\bibinfo {year}
  {2011})}\BibitemShut {NoStop}%
\bibitem [{\citenamefont {Zomer}\ \emph {et~al.}(2012)\citenamefont {Zomer},
  \citenamefont {{Guimar\~{a}es}}, \citenamefont {Tombros},\ and\ \citenamefont
  {van Wees}}]{ZomerPRB2012}%
  \BibitemOpen
  \bibfield  {author} {\bibinfo {author} {\bibfnamefont {P.~J.}\ \bibnamefont
  {Zomer}}, \bibinfo {author} {\bibfnamefont {M.~H.~D.}\ \bibnamefont
  {{Guimar\~{a}es}}}, \bibinfo {author} {\bibfnamefont {N.}~\bibnamefont
  {Tombros}},\ and\ \bibinfo {author} {\bibfnamefont {B.~J.}\ \bibnamefont {van
  Wees}},\ }\bibfield  {title} {\enquote {\bibinfo {title} {{Long-distance spin
  transport in high-mobility graphene on hexagonal boron nitride}},}\
  }\href@noop {} {\bibfield  {journal} {\bibinfo  {journal} {Physical Review
  B}\ }\textbf {\bibinfo {volume} {86}},\ \bibinfo {pages} {161416} (\bibinfo
  {year} {2012})}\BibitemShut {NoStop}%
\bibitem [{\citenamefont {Roche}\ and\ \citenamefont
  {Valenzuela}(2014)}]{RocheValenzuela2014}%
  \BibitemOpen
  \bibfield  {author} {\bibinfo {author} {\bibfnamefont {S.}~\bibnamefont
  {Roche}}\ and\ \bibinfo {author} {\bibfnamefont {S.~O.}\ \bibnamefont
  {Valenzuela}},\ }\bibfield  {title} {\enquote {\bibinfo {title} {{Graphene
  spintronics: puzzling controversies and challenges for spin manipulation}},}\
  }\href@noop {} {\bibfield  {journal} {\bibinfo  {journal} {Journal of Physics
  D: Applied Physics}\ }\textbf {\bibinfo {volume} {47}},\ \bibinfo {pages}
  {094011} (\bibinfo {year} {2014})}\BibitemShut {NoStop}%
\bibitem [{\citenamefont {Venkata~Kamalakar}\ \emph {et~al.}(2015)\citenamefont
  {Venkata~Kamalakar}, \citenamefont {Groenveld}, \citenamefont {A.},\ and\
  \citenamefont {Dash}}]{KamalakarNatComm2015}%
  \BibitemOpen
  \bibfield  {author} {\bibinfo {author} {\bibfnamefont {M.}~\bibnamefont
  {Venkata~Kamalakar}}, \bibinfo {author} {\bibfnamefont {C.}~\bibnamefont
  {Groenveld}}, \bibinfo {author} {\bibfnamefont {D.}~\bibnamefont {A.}},\ and\
  \bibinfo {author} {\bibfnamefont {S.~P.}\ \bibnamefont {Dash}},\ }\bibfield
  {title} {\enquote {\bibinfo {title} {{Long distance spin communication in
  chemical vapour deposited graphene}},}\ }\href@noop {} {\bibfield  {journal}
  {\bibinfo  {journal} {Nature Communications}\ }\textbf {\bibinfo {volume}
  {6}},\ \bibinfo {pages} {6766} (\bibinfo {year} {2015})}\BibitemShut
  {NoStop}%
\bibitem [{\citenamefont {Dr\"ogeler}\ \emph {et~al.}(2016)\citenamefont
  {Dr\"ogeler}, \citenamefont {Franzen}, \citenamefont {Volmer}, \citenamefont
  {Pohlmann}, \citenamefont {Banszerus}, \citenamefont {Wolter}, \citenamefont
  {Watanabe}, \citenamefont {Taniguchi}, \citenamefont {Stampfer},\ and\
  \citenamefont {Beschoten}}]{BeschotenNL2016}%
  \BibitemOpen
  \bibfield  {author} {\bibinfo {author} {\bibfnamefont {M.}~\bibnamefont
  {Dr\"ogeler}}, \bibinfo {author} {\bibfnamefont {C.}~\bibnamefont {Franzen}},
  \bibinfo {author} {\bibfnamefont {F.}~\bibnamefont {Volmer}}, \bibinfo
  {author} {\bibfnamefont {T.}~\bibnamefont {Pohlmann}}, \bibinfo {author}
  {\bibfnamefont {L.}~\bibnamefont {Banszerus}}, \bibinfo {author}
  {\bibfnamefont {M.}~\bibnamefont {Wolter}}, \bibinfo {author} {\bibfnamefont
  {K.}~\bibnamefont {Watanabe}}, \bibinfo {author} {\bibfnamefont
  {T.}~\bibnamefont {Taniguchi}}, \bibinfo {author} {\bibfnamefont
  {C.}~\bibnamefont {Stampfer}},\ and\ \bibinfo {author} {\bibfnamefont
  {B.}~\bibnamefont {Beschoten}},\ }\bibfield  {title} {\enquote {\bibinfo
  {title} {{Spin Lifetimes Exceeding 12 ns in Graphene Nonlocal Spin Valve
  Devices}},}\ }\href@noop {} {\bibfield  {journal} {\bibinfo  {journal} {Nano
  Letters}\ }\textbf {\bibinfo {volume} {16}},\ \bibinfo {pages} {3533--3539}
  (\bibinfo {year} {2016})}\BibitemShut {NoStop}%
\bibitem [{\citenamefont {Zhang}\ \emph {et~al.}(2024)\citenamefont {Zhang},
  \citenamefont {Wu}, \citenamefont {Yang}, \citenamefont {Jin}, \citenamefont
  {Zhang},\ and\ \citenamefont {Chang}}]{GrapheneSpintronicsReview2024}%
  \BibitemOpen
  \bibfield  {author} {\bibinfo {author} {\bibfnamefont {G.}~\bibnamefont
  {Zhang}}, \bibinfo {author} {\bibfnamefont {H.}~\bibnamefont {Wu}}, \bibinfo
  {author} {\bibfnamefont {L.}~\bibnamefont {Yang}}, \bibinfo {author}
  {\bibfnamefont {W.}~\bibnamefont {Jin}}, \bibinfo {author} {\bibfnamefont
  {W.}~\bibnamefont {Zhang}},\ and\ \bibinfo {author} {\bibfnamefont
  {H.}~\bibnamefont {Chang}},\ }\bibfield  {title} {\enquote {\bibinfo {title}
  {Graphene-based spintronics},}\ }\href@noop {} {\bibfield  {journal}
  {\bibinfo  {journal} {Applied Physics Reviews}\ }\textbf {\bibinfo {volume}
  {11}},\ \bibinfo {pages} {021308} (\bibinfo {year} {2024})}\BibitemShut
  {NoStop}%
\bibitem [{\citenamefont {Chen}\ \emph {et~al.}(2025)\citenamefont {Chen},
  \citenamefont {Xie}, \citenamefont {Wu}, \citenamefont {Lu},\ and\
  \citenamefont {Zhang}}]{ChenACSAEM2025}%
  \BibitemOpen
  \bibfield  {author} {\bibinfo {author} {\bibfnamefont {P.}~\bibnamefont
  {Chen}}, \bibinfo {author} {\bibfnamefont {G.}~\bibnamefont {Xie}}, \bibinfo
  {author} {\bibfnamefont {S.}~\bibnamefont {Wu}}, \bibinfo {author}
  {\bibfnamefont {X.}~\bibnamefont {Lu}},\ and\ \bibinfo {author}
  {\bibfnamefont {G.}~\bibnamefont {Zhang}},\ }\bibfield  {title} {\enquote
  {\bibinfo {title} {{Room Temperature Spin Transport in Sub-50 nm Bilayer
  Graphene Nanoribbons}},}\ }\href@noop {} {\bibfield  {journal} {\bibinfo
  {journal} {ACS Applied Electronic Materials}\ }\textbf {\bibinfo {volume}
  {7}},\ \bibinfo {pages} {1392--1397} (\bibinfo {year} {2025})}\BibitemShut
  {NoStop}%
\bibitem [{\citenamefont {Kochan}, \citenamefont {Gmitra},\ and\ \citenamefont
  {Fabian}(2014)}]{FabianPRL2015}%
  \BibitemOpen
  \bibfield  {author} {\bibinfo {author} {\bibfnamefont {D.}~\bibnamefont
  {Kochan}}, \bibinfo {author} {\bibfnamefont {M.}~\bibnamefont {Gmitra}},\
  and\ \bibinfo {author} {\bibfnamefont {J.}~\bibnamefont {Fabian}},\
  }\bibfield  {title} {\enquote {\bibinfo {title} {{Spin Relaxation Mechanism
  in Graphene: Resonant Scattering by Magnetic Impurities}},}\ }\href@noop {}
  {\bibfield  {journal} {\bibinfo  {journal} {Phys. Rev. Lett.}\ }\textbf
  {\bibinfo {volume} {112}},\ \bibinfo {pages} {116602} (\bibinfo {year}
  {2014})}\BibitemShut {NoStop}%
\bibitem [{\citenamefont {Cummings}\ \emph {et~al.}(2025)\citenamefont
  {Cummings}, \citenamefont {Dubois}, \citenamefont {Guerrero}, \citenamefont
  {Charlier},\ and\ \citenamefont {Roche}}]{UpperLimit_T1_Cummings2025}%
  \BibitemOpen
  \bibfield  {author} {\bibinfo {author} {\bibfnamefont {A.~W.}\ \bibnamefont
  {Cummings}}, \bibinfo {author} {\bibfnamefont {S.~M.-M.}\ \bibnamefont
  {Dubois}}, \bibinfo {author} {\bibfnamefont {P.~A.}\ \bibnamefont
  {Guerrero}}, \bibinfo {author} {\bibfnamefont {J.-C.}\ \bibnamefont
  {Charlier}},\ and\ \bibinfo {author} {\bibfnamefont {S.}~\bibnamefont
  {Roche}},\ }\bibfield  {title} {\enquote {\bibinfo {title} {Upper limit of
  spin relaxation in suspended graphene},}\ }\href@noop {} {\bibfield
  {journal} {\bibinfo  {journal} {Carbon}\ }\textbf {\bibinfo {volume} {234}},\
  \bibinfo {pages} {119920} (\bibinfo {year} {2025})}\BibitemShut {NoStop}%
\bibitem [{\citenamefont {Lee}\ \emph {et~al.}(2008)\citenamefont {Lee},
  \citenamefont {Maeng}, \citenamefont {Son}, \citenamefont {Park},
  \citenamefont {Ahn}, \citenamefont {Lee}, \citenamefont {Lee}, \citenamefont
  {Park},\ and\ \citenamefont {Park}}]{Kawakami_electrode1}%
  \BibitemOpen
  \bibfield  {author} {\bibinfo {author} {\bibfnamefont {J.~H.}\ \bibnamefont
  {Lee}}, \bibinfo {author} {\bibfnamefont {S.~L.}\ \bibnamefont {Maeng}},
  \bibinfo {author} {\bibfnamefont {J.~Y.}\ \bibnamefont {Son}}, \bibinfo
  {author} {\bibfnamefont {C.~Y.}\ \bibnamefont {Park}}, \bibinfo {author}
  {\bibfnamefont {J.~H.}\ \bibnamefont {Ahn}}, \bibinfo {author} {\bibfnamefont
  {S.~W.}\ \bibnamefont {Lee}}, \bibinfo {author} {\bibfnamefont
  {T.}~\bibnamefont {Lee}}, \bibinfo {author} {\bibfnamefont {J.~H.}\
  \bibnamefont {Park}},\ and\ \bibinfo {author} {\bibfnamefont {J.~H.}\
  \bibnamefont {Park}},\ }\bibfield  {title} {\enquote {\bibinfo {title}
  {Growth of atomically smooth mgo films on graphene by molecular beam
  epitaxy},}\ }\href@noop {} {\bibfield  {journal} {\bibinfo  {journal}
  {Applied Physics Letters}\ }\textbf {\bibinfo {volume} {93}},\ \bibinfo
  {pages} {183107} (\bibinfo {year} {2008})}\BibitemShut {NoStop}%
\bibitem [{\citenamefont {Han}\ \emph {et~al.}(2009)\citenamefont {Han},
  \citenamefont {Pi}, \citenamefont {Bao}, \citenamefont {McCreary},
  \citenamefont {Li}, \citenamefont {Wang}, \citenamefont {Lau},\ and\
  \citenamefont {Kawakami}}]{Kawakami_electrode2}%
  \BibitemOpen
  \bibfield  {author} {\bibinfo {author} {\bibfnamefont {W.}~\bibnamefont
  {Han}}, \bibinfo {author} {\bibfnamefont {K.}~\bibnamefont {Pi}}, \bibinfo
  {author} {\bibfnamefont {W.}~\bibnamefont {Bao}}, \bibinfo {author}
  {\bibfnamefont {K.~M.}\ \bibnamefont {McCreary}}, \bibinfo {author}
  {\bibfnamefont {Y.}~\bibnamefont {Li}}, \bibinfo {author} {\bibfnamefont
  {W.~H.}\ \bibnamefont {Wang}}, \bibinfo {author} {\bibfnamefont {C.~N.}\
  \bibnamefont {Lau}},\ and\ \bibinfo {author} {\bibfnamefont {R.~K.}\
  \bibnamefont {Kawakami}},\ }\bibfield  {title} {\enquote {\bibinfo {title}
  {Electrical detection of spin precession in single layer graphene spin valves
  with transparent contacts},}\ }\href {https://doi.org/10.1063/1.3147203}
  {\bibfield  {journal} {\bibinfo  {journal} {Applied Physics Letters}\
  }\textbf {\bibinfo {volume} {94}},\ \bibinfo {pages} {222109} (\bibinfo
  {year} {2009})}\BibitemShut {NoStop}%
\bibitem [{\citenamefont {Gmitra}\ \emph {et~al.}(2009)\citenamefont {Gmitra},
  \citenamefont {Konschuh}, \citenamefont {Ertler}, \citenamefont
  {Ambrosch-Draxl},\ and\ \citenamefont {Fabian}}]{GmitraPRB2009}%
  \BibitemOpen
  \bibfield  {author} {\bibinfo {author} {\bibfnamefont {M.}~\bibnamefont
  {Gmitra}}, \bibinfo {author} {\bibfnamefont {S.}~\bibnamefont {Konschuh}},
  \bibinfo {author} {\bibfnamefont {C.}~\bibnamefont {Ertler}}, \bibinfo
  {author} {\bibfnamefont {C.}~\bibnamefont {Ambrosch-Draxl}},\ and\ \bibinfo
  {author} {\bibfnamefont {J.}~\bibnamefont {Fabian}},\ }\bibfield  {title}
  {\enquote {\bibinfo {title} {{Band-structure topologies of graphene:
  Spin-orbit coupling effects from first principles}},}\ }\href@noop {}
  {\bibfield  {journal} {\bibinfo  {journal} {Phys. Rev. B}\ }\textbf {\bibinfo
  {volume} {80}},\ \bibinfo {pages} {235431} (\bibinfo {year}
  {2009})}\BibitemShut {NoStop}%
\bibitem [{\citenamefont {Kane}\ and\ \citenamefont
  {Mele}(2005)}]{KaneMelePRL2005}%
  \BibitemOpen
  \bibfield  {author} {\bibinfo {author} {\bibfnamefont {C.~L.}\ \bibnamefont
  {Kane}}\ and\ \bibinfo {author} {\bibfnamefont {E.~J.}\ \bibnamefont
  {Mele}},\ }\bibfield  {title} {\enquote {\bibinfo {title} {{Quantum Spin Hall
  Effect in Graphene}},}\ }\href@noop {} {\bibfield  {journal} {\bibinfo
  {journal} {Phys. Rev. Lett.}\ }\textbf {\bibinfo {volume} {95}},\ \bibinfo
  {pages} {226801} (\bibinfo {year} {2005})}\BibitemShut {NoStop}%
\bibitem [{\citenamefont {M\'arkus}\ \emph {et~al.}(2023)\citenamefont
  {M\'arkus}, \citenamefont {Gmitra}, \citenamefont {D\'ora}, \citenamefont
  {Cs\H{o}sz}, \citenamefont {Feh\'er}, \citenamefont {Szirmai}, \citenamefont
  {N\'afr\'adi}, \citenamefont {Z\'olyomi}, \citenamefont {Forr\'o},
  \citenamefont {Fabian},\ and\ \citenamefont {Simon}}]{MarkusNatComm}%
  \BibitemOpen
  \bibfield  {author} {\bibinfo {author} {\bibfnamefont {B.~G.}\ \bibnamefont
  {M\'arkus}}, \bibinfo {author} {\bibfnamefont {M.}~\bibnamefont {Gmitra}},
  \bibinfo {author} {\bibfnamefont {B.}~\bibnamefont {D\'ora}}, \bibinfo
  {author} {\bibfnamefont {G.}~\bibnamefont {Cs\H{o}sz}}, \bibinfo {author}
  {\bibfnamefont {T.}~\bibnamefont {Feh\'er}}, \bibinfo {author} {\bibfnamefont
  {P.}~\bibnamefont {Szirmai}}, \bibinfo {author} {\bibfnamefont
  {B.}~\bibnamefont {N\'afr\'adi}}, \bibinfo {author} {\bibfnamefont
  {V.}~\bibnamefont {Z\'olyomi}}, \bibinfo {author} {\bibfnamefont
  {L.}~\bibnamefont {Forr\'o}}, \bibinfo {author} {\bibfnamefont
  {J.}~\bibnamefont {Fabian}},\ and\ \bibinfo {author} {\bibfnamefont
  {F.}~\bibnamefont {Simon}},\ }\bibfield  {title} {\enquote {\bibinfo {title}
  {{Ultralong 100 ns spin relaxation time in graphite at room temperature}},}\
  }\href@noop {} {\bibfield  {journal} {\bibinfo  {journal} {Nature
  Communications}\ }\textbf {\bibinfo {volume} {14}} (\bibinfo {year}
  {2023})}\BibitemShut {NoStop}%
\bibitem [{\citenamefont {Xu}(2024)}]{XuTheory}%
  \BibitemOpen
  \bibfield  {author} {\bibinfo {author} {\bibfnamefont {J.}~\bibnamefont
  {Xu}},\ }\bibfield  {title} {\enquote {\bibinfo {title} {Spin relaxation in
  graphite due to spin-orbit--phonon interaction from a first-principles
  density matrix approach},}\ }\href@noop {} {\bibfield  {journal} {\bibinfo
  {journal} {Phys. Rev. B}\ }\textbf {\bibinfo {volume} {110}},\ \bibinfo
  {pages} {144307} (\bibinfo {year} {2024})}\BibitemShut {NoStop}%
\bibitem [{\citenamefont {Gächter}\ \emph {et~al.}(2022)\citenamefont
  {Gächter}, \citenamefont {Kurzmann}, \citenamefont {Gargiulo}, \citenamefont
  {Kuhlmann}, \citenamefont {Lee}, \citenamefont {Overweg}, \citenamefont
  {Mohrmann}, \citenamefont {Sestoft}, \citenamefont {Watanabe}, \citenamefont
  {Taniguchi}, \citenamefont {Ihn}, \citenamefont {Ensslin},\ and\
  \citenamefont {Desjardins}}]{Gachter2022}%
  \BibitemOpen
  \bibfield  {author} {\bibinfo {author} {\bibfnamefont {L.~M.}\ \bibnamefont
  {Gächter}}, \bibinfo {author} {\bibfnamefont {A.}~\bibnamefont {Kurzmann}},
  \bibinfo {author} {\bibfnamefont {F.}~\bibnamefont {Gargiulo}}, \bibinfo
  {author} {\bibfnamefont {M.}~\bibnamefont {Kuhlmann}}, \bibinfo {author}
  {\bibfnamefont {D.~K.~K.}\ \bibnamefont {Lee}}, \bibinfo {author}
  {\bibfnamefont {K.}~\bibnamefont {Overweg}}, \bibinfo {author} {\bibfnamefont
  {J.}~\bibnamefont {Mohrmann}}, \bibinfo {author} {\bibfnamefont {J.~E.}\
  \bibnamefont {Sestoft}}, \bibinfo {author} {\bibfnamefont {K.}~\bibnamefont
  {Watanabe}}, \bibinfo {author} {\bibfnamefont {T.}~\bibnamefont {Taniguchi}},
  \bibinfo {author} {\bibfnamefont {T.}~\bibnamefont {Ihn}}, \bibinfo {author}
  {\bibfnamefont {K.}~\bibnamefont {Ensslin}},\ and\ \bibinfo {author}
  {\bibfnamefont {M.~M.}\ \bibnamefont {Desjardins}},\ }\bibfield  {title}
  {\enquote {\bibinfo {title} {Single-shot spin readout in graphene quantum
  dots},}\ }\href@noop {} {\bibfield  {journal} {\bibinfo  {journal} {PRX
  Quantum}\ }\textbf {\bibinfo {volume} {3}},\ \bibinfo {pages} {020343}
  (\bibinfo {year} {2022})}\BibitemShut {NoStop}%
\bibitem [{\citenamefont {Garreis}\ \emph {et~al.}(2024)\citenamefont
  {Garreis}, \citenamefont {Tong}, \citenamefont {Huang}, \citenamefont
  {Terle}, \citenamefont {Ruckriegel}, \citenamefont {Gerber}, \citenamefont
  {Gächter}, \citenamefont {Watanabe}, \citenamefont {Taniguchi},
  \citenamefont {Ihn},\ and\ \citenamefont {Ensslin}}]{Garreis2024}%
  \BibitemOpen
  \bibfield  {author} {\bibinfo {author} {\bibfnamefont {R.}~\bibnamefont
  {Garreis}}, \bibinfo {author} {\bibfnamefont {C.}~\bibnamefont {Tong}},
  \bibinfo {author} {\bibfnamefont {W.~W.}\ \bibnamefont {Huang}}, \bibinfo
  {author} {\bibfnamefont {J.}~\bibnamefont {Terle}}, \bibinfo {author}
  {\bibfnamefont {M.~J.}\ \bibnamefont {Ruckriegel}}, \bibinfo {author}
  {\bibfnamefont {J.~D.}\ \bibnamefont {Gerber}}, \bibinfo {author}
  {\bibfnamefont {L.~M.}\ \bibnamefont {Gächter}}, \bibinfo {author}
  {\bibfnamefont {K.}~\bibnamefont {Watanabe}}, \bibinfo {author}
  {\bibfnamefont {T.}~\bibnamefont {Taniguchi}}, \bibinfo {author}
  {\bibfnamefont {T.}~\bibnamefont {Ihn}},\ and\ \bibinfo {author}
  {\bibfnamefont {K.}~\bibnamefont {Ensslin}},\ }\bibfield  {title} {\enquote
  {\bibinfo {title} {Long-lived valley states in bilayer graphene quantum
  dots},}\ }\href@noop {} {\bibfield  {journal} {\bibinfo  {journal} {Nature
  Physics}\ }\textbf {\bibinfo {volume} {20}},\ \bibinfo {pages} {428--434}
  (\bibinfo {year} {2024})}\BibitemShut {NoStop}%
\bibitem [{\citenamefont {Denisov}\ \emph {et~al.}(2024)\citenamefont {Denisov}
  \emph {et~al.}}]{Denisov2024}%
  \BibitemOpen
  \bibfield  {author} {\bibinfo {author} {\bibfnamefont {A.~O.}\ \bibnamefont
  {Denisov}} \emph {et~al.},\ }\bibfield  {title} {\enquote {\bibinfo {title}
  {Ultra-long relaxation of a kramers qubit formed in a bilayer graphene
  quantum dot},}\ }\href {http://arxiv.org/abs/2403.08143} {\bibfield
  {journal} {\bibinfo  {journal} {arXiv preprint}\ } (\bibinfo {year}
  {2024})},\ \Eprint {https://arxiv.org/abs/2403.08143} {arXiv:2403.08143
  [cond-mat.mes-hall]} \BibitemShut {NoStop}%
\bibitem [{\citenamefont {N\'afr\'adi}\ \emph {et~al.}(2016)\citenamefont
  {N\'afr\'adi}, \citenamefont {Choucair}, \citenamefont {Dinse},\ and\
  \citenamefont {Forr\'o}}]{NafradiNatComm}%
  \BibitemOpen
  \bibfield  {author} {\bibinfo {author} {\bibfnamefont {B.}~\bibnamefont
  {N\'afr\'adi}}, \bibinfo {author} {\bibfnamefont {M.}~\bibnamefont
  {Choucair}}, \bibinfo {author} {\bibfnamefont {K.-P.}\ \bibnamefont
  {Dinse}},\ and\ \bibinfo {author} {\bibfnamefont {L.}~\bibnamefont
  {Forr\'o}},\ }\bibfield  {title} {\enquote {\bibinfo {title} {{Room
  temperature manipulation of long lifetime spins in metallic-like carbon
  nanospheres}},}\ }\href@noop {} {\bibfield  {journal} {\bibinfo  {journal}
  {Nature Communications}\ }\textbf {\bibinfo {volume} {7}},\ \bibinfo {pages}
  {12232} (\bibinfo {year} {2016})}\BibitemShut {NoStop}%
\bibitem [{\citenamefont {Rao}\ \emph {et~al.}(2012)\citenamefont {Rao},
  \citenamefont {Stesmans}, \citenamefont {Van~Tol}, \citenamefont {Agarwal},
  \citenamefont {Chowdhury}, \citenamefont {Hong}, \citenamefont {Dutta},\ and\
  \citenamefont {Srivastava}}]{GrapheneNanoRibbon}%
  \BibitemOpen
  \bibfield  {author} {\bibinfo {author} {\bibfnamefont {S.~S.}\ \bibnamefont
  {Rao}}, \bibinfo {author} {\bibfnamefont {A.}~\bibnamefont {Stesmans}},
  \bibinfo {author} {\bibfnamefont {J.}~\bibnamefont {Van~Tol}}, \bibinfo
  {author} {\bibfnamefont {K.}~\bibnamefont {Agarwal}}, \bibinfo {author}
  {\bibfnamefont {S.}~\bibnamefont {Chowdhury}}, \bibinfo {author}
  {\bibfnamefont {A.}~\bibnamefont {Hong}}, \bibinfo {author} {\bibfnamefont
  {M.}~\bibnamefont {Dutta}},\ and\ \bibinfo {author} {\bibfnamefont
  {A.}~\bibnamefont {Srivastava}},\ }\bibfield  {title} {\enquote {\bibinfo
  {title} {Spin dynamics and relaxation in graphene nanoribbons: Electron spin
  resonance probing},}\ }\href@noop {} {\bibfield  {journal} {\bibinfo
  {journal} {ACS Nano}\ }\textbf {\bibinfo {volume} {6}},\ \bibinfo {pages}
  {7615--7623} (\bibinfo {year} {2012})}\BibitemShut {NoStop}%
\bibitem [{\citenamefont {Yafet}(1963)}]{yafet1963g}%
  \BibitemOpen
  \bibfield  {author} {\bibinfo {author} {\bibfnamefont {Y.}~\bibnamefont
  {Yafet}},\ }\bibfield  {title} {\enquote {\bibinfo {title} {{$g$-Factors and
  Spin-Lattice Relaxation of Conduction Electrons}},}\ }\href@noop {}
  {\bibfield  {journal} {\bibinfo  {journal} {Solid State Physics}\ }\textbf
  {\bibinfo {volume} {14}},\ \bibinfo {pages} {1--98} (\bibinfo {year}
  {1963})}\BibitemShut {NoStop}%
\bibitem [{\citenamefont {Fabian}\ \emph {et~al.}(2007)\citenamefont {Fabian},
  \citenamefont {Matos-Abiaguea}, \citenamefont {Ertlera}, \citenamefont
  {Stano},\ and\ \citenamefont {Zutic}}]{FabianActaPhysSlovaca}%
  \BibitemOpen
  \bibfield  {author} {\bibinfo {author} {\bibfnamefont {J.}~\bibnamefont
  {Fabian}}, \bibinfo {author} {\bibfnamefont {A.}~\bibnamefont
  {Matos-Abiaguea}}, \bibinfo {author} {\bibfnamefont {C.}~\bibnamefont
  {Ertlera}}, \bibinfo {author} {\bibfnamefont {P.}~\bibnamefont {Stano}},\
  and\ \bibinfo {author} {\bibfnamefont {I.}~\bibnamefont {Zutic}},\ }\bibfield
   {title} {\enquote {\bibinfo {title} {{Semiconductor Spintronics}},}\
  }\href@noop {} {\bibfield  {journal} {\bibinfo  {journal} {Acta Physica
  Slovaca}\ }\textbf {\bibinfo {volume} {57}} (\bibinfo {year}
  {2007})}\BibitemShut {NoStop}%
\bibitem [{\citenamefont {J\'anossy}\ and\ \citenamefont
  {Monod}(1976)}]{janossy1976}%
  \BibitemOpen
  \bibfield  {author} {\bibinfo {author} {\bibfnamefont {A.}~\bibnamefont
  {J\'anossy}}\ and\ \bibinfo {author} {\bibfnamefont {P.}~\bibnamefont
  {Monod}},\ }\bibfield  {title} {\enquote {\bibinfo {title} {{Spin Diffusion
  in Magnetic Materials}},}\ }\href@noop {} {\bibfield  {journal} {\bibinfo
  {journal} {Physical Review Letters}\ }\textbf {\bibinfo {volume} {37}},\
  \bibinfo {pages} {612--614} (\bibinfo {year} {1976})}\BibitemShut {NoStop}%
\bibitem [{\citenamefont {Silsbee}, \citenamefont {Janossy},\ and\
  \citenamefont {Monod}(1979)}]{SilsbeeJanossyMonod}%
  \BibitemOpen
  \bibfield  {author} {\bibinfo {author} {\bibfnamefont {R.~H.}\ \bibnamefont
  {Silsbee}}, \bibinfo {author} {\bibfnamefont {A.}~\bibnamefont {Janossy}},\
  and\ \bibinfo {author} {\bibfnamefont {P.}~\bibnamefont {Monod}},\ }\bibfield
   {title} {\enquote {\bibinfo {title} {Coupling between ferromagnetic and
  conduction-spin-resonance modes at a ferromagnetic---normal-metal
  interface},}\ }\href@noop {} {\bibfield  {journal} {\bibinfo  {journal}
  {Phys. Rev. B}\ }\textbf {\bibinfo {volume} {19}},\ \bibinfo {pages}
  {4382--4399} (\bibinfo {year} {1979})}\BibitemShut {NoStop}%
\bibitem [{\citenamefont {Slichter}(1989)}]{SlichterBook}%
  \BibitemOpen
  \bibfield  {author} {\bibinfo {author} {\bibfnamefont {C.~P.}\ \bibnamefont
  {Slichter}},\ }\href@noop {} {\emph {\bibinfo {title} {{Principles of
  Magnetic Resonance}}}},\ \bibinfo {edition} {3rd}\ ed.\ (\bibinfo
  {publisher} {Spinger-Verlag},\ \bibinfo {address} {New York},\ \bibinfo
  {year} {1989})\BibitemShut {NoStop}%
\bibitem [{\citenamefont {Abragam}(1961)}]{AbragamBook}%
  \BibitemOpen
  \bibfield  {author} {\bibinfo {author} {\bibfnamefont {A.}~\bibnamefont
  {Abragam}},\ }\href@noop {} {\emph {\bibinfo {title} {{Principles of Nuclear
  Magnetism}}}}\ (\bibinfo  {publisher} {Oxford University Press},\ \bibinfo
  {address} {Oxford, England},\ \bibinfo {year} {1961})\BibitemShut {NoStop}%
\bibitem [{\citenamefont {Klein}\ \emph {et~al.}(1993)\citenamefont {Klein},
  \citenamefont {Donovan}, \citenamefont {Dressel},\ and\ \citenamefont
  {Gr\"uner}}]{Klein1993}%
  \BibitemOpen
  \bibfield  {author} {\bibinfo {author} {\bibfnamefont {O.}~\bibnamefont
  {Klein}}, \bibinfo {author} {\bibfnamefont {S.}~\bibnamefont {Donovan}},
  \bibinfo {author} {\bibfnamefont {M.}~\bibnamefont {Dressel}},\ and\ \bibinfo
  {author} {\bibfnamefont {G.}~\bibnamefont {Gr\"uner}},\ }\bibfield  {title}
  {\enquote {\bibinfo {title} {{Microwave cavity perturbation technique: Part
  I: Principles}},}\ }\href@noop {} {\bibfield  {journal} {\bibinfo  {journal}
  {International Journal of Infrared and Millimeter Waves}\ }\textbf {\bibinfo
  {volume} {14}},\ \bibinfo {pages} {2423--2457} (\bibinfo {year}
  {1993})}\BibitemShut {NoStop}%
\bibitem [{\citenamefont {Donovan}\ \emph {et~al.}(1993)\citenamefont
  {Donovan}, \citenamefont {Klein}, \citenamefont {Dressel}, \citenamefont
  {Holczer},\ and\ \citenamefont {Gr\"uner}}]{Donovan1993}%
  \BibitemOpen
  \bibfield  {author} {\bibinfo {author} {\bibfnamefont {S.}~\bibnamefont
  {Donovan}}, \bibinfo {author} {\bibfnamefont {O.}~\bibnamefont {Klein}},
  \bibinfo {author} {\bibfnamefont {M.}~\bibnamefont {Dressel}}, \bibinfo
  {author} {\bibfnamefont {K.}~\bibnamefont {Holczer}},\ and\ \bibinfo {author}
  {\bibfnamefont {G.}~\bibnamefont {Gr\"uner}},\ }\bibfield  {title} {\enquote
  {\bibinfo {title} {{Microwave cavity perturbation technique: Part II:
  Experimental scheme}},}\ }\href@noop {} {\bibfield  {journal} {\bibinfo
  {journal} {International Journal of Infrared and Millimeter Waves}\ }\textbf
  {\bibinfo {volume} {14}},\ \bibinfo {pages} {2459--2487} (\bibinfo {year}
  {1993})}\BibitemShut {NoStop}%
\bibitem [{\citenamefont {Poole}(1983)}]{PooleBook}%
  \BibitemOpen
  \bibfield  {author} {\bibinfo {author} {\bibfnamefont {C.~P.}\ \bibnamefont
  {Poole}},\ }\href@noop {} {\emph {\bibinfo {title} {{Electron Spin
  Resonance}}}},\ \bibinfo {edition} {1983rd}\ ed.\ (\bibinfo  {publisher}
  {John Wiley \& Sons},\ \bibinfo {address} {New York},\ \bibinfo {year}
  {1983})\BibitemShut {NoStop}%
\bibitem [{\citenamefont {Colligiani}\ \emph {et~al.}(1992)\citenamefont
  {Colligiani}, \citenamefont {Giordano}, \citenamefont {Leporini},
  \citenamefont {Lucchesi}, \citenamefont {Martinelli}, \citenamefont {Pardi},\
  and\ \citenamefont {Santucci}}]{Martinelly_LOD}%
  \BibitemOpen
  \bibfield  {author} {\bibinfo {author} {\bibfnamefont {A.}~\bibnamefont
  {Colligiani}}, \bibinfo {author} {\bibfnamefont {M.}~\bibnamefont
  {Giordano}}, \bibinfo {author} {\bibfnamefont {D.}~\bibnamefont {Leporini}},
  \bibinfo {author} {\bibfnamefont {M.}~\bibnamefont {Lucchesi}}, \bibinfo
  {author} {\bibfnamefont {M.}~\bibnamefont {Martinelli}}, \bibinfo {author}
  {\bibfnamefont {L.}~\bibnamefont {Pardi}},\ and\ \bibinfo {author}
  {\bibfnamefont {S.}~\bibnamefont {Santucci}},\ }\bibfield  {title} {\enquote
  {\bibinfo {title} {{Longitudinally Detected Electron Spin Resonance: Recent
  Developments}},}\ }\href@noop {} {\bibfield  {journal} {\bibinfo  {journal}
  {APPLIED MAGNETIC RESONANCE}\ }\textbf {\bibinfo {volume} {3}},\ \bibinfo
  {pages} {107--129} (\bibinfo {year} {1992})}\BibitemShut {NoStop}%
\bibitem [{\citenamefont {Atsarkin}, \citenamefont {Demidov},\ and\
  \citenamefont {Vasneva}(1995)}]{Atsarkin}%
  \BibitemOpen
  \bibfield  {author} {\bibinfo {author} {\bibfnamefont {V.}~\bibnamefont
  {Atsarkin}}, \bibinfo {author} {\bibfnamefont {V.}~\bibnamefont {Demidov}},\
  and\ \bibinfo {author} {\bibfnamefont {G.}~\bibnamefont {Vasneva}},\
  }\bibfield  {title} {\enquote {\bibinfo {title} {{Electron-Spin-Lattice
  Relaxation in GdBa$_2$Cu$_3$O$_{6+\text{x}}$}},}\ }\href@noop {} {\bibfield
  {journal} {\bibinfo  {journal} {Physical Review B}\ }\textbf {\bibinfo
  {volume} {52}},\ \bibinfo {pages} {1290--1296} (\bibinfo {year}
  {1995})}\BibitemShut {NoStop}%
\bibitem [{\citenamefont {Granwehr}, \citenamefont {Forrer},\ and\
  \citenamefont {Schweiger}(2001)}]{Schweiger1}%
  \BibitemOpen
  \bibfield  {author} {\bibinfo {author} {\bibfnamefont {J.}~\bibnamefont
  {Granwehr}}, \bibinfo {author} {\bibfnamefont {J.}~\bibnamefont {Forrer}},\
  and\ \bibinfo {author} {\bibfnamefont {A.}~\bibnamefont {Schweiger}},\
  }\bibfield  {title} {\enquote {\bibinfo {title} {{Longitudinally detected
  EPR: Improved instrumentation and new pulse schemes}},}\ }\href@noop {}
  {\bibfield  {journal} {\bibinfo  {journal} {Journal of Magnetic Resonance}\
  }\textbf {\bibinfo {volume} {151}},\ \bibinfo {pages} {78--84} (\bibinfo
  {year} {2001})}\BibitemShut {NoStop}%
\bibitem [{\citenamefont {Granwehr}\ and\ \citenamefont
  {Schweiger}(2001)}]{Schweiger2}%
  \BibitemOpen
  \bibfield  {author} {\bibinfo {author} {\bibfnamefont {J.}~\bibnamefont
  {Granwehr}}\ and\ \bibinfo {author} {\bibfnamefont {A.}~\bibnamefont
  {Schweiger}},\ }\bibfield  {title} {\enquote {\bibinfo {title} {{Measurement
  of spin-lattice relaxation times with longitudinal detection}},}\ }\href@noop
  {} {\bibfield  {journal} {\bibinfo  {journal} {Applied Magnetic Resonance}\
  }\textbf {\bibinfo {volume} {20}},\ \bibinfo {pages} {137--150} (\bibinfo
  {year} {2001})}\BibitemShut {NoStop}%
\bibitem [{\citenamefont {Quine}, \citenamefont {Eaton},\ and\ \citenamefont
  {Eaton}(1992)}]{EatonsRSI1992}%
  \BibitemOpen
  \bibfield  {author} {\bibinfo {author} {\bibfnamefont {R.}~\bibnamefont
  {Quine}}, \bibinfo {author} {\bibfnamefont {S.}~\bibnamefont {Eaton}},\ and\
  \bibinfo {author} {\bibfnamefont {G.}~\bibnamefont {Eaton}},\ }\bibfield
  {title} {\enquote {\bibinfo {title} {{Saturation Recovery
  Electron-Paramagnetic Resonance Spectrometer}},}\ }\href@noop {} {\bibfield
  {journal} {\bibinfo  {journal} {Review of Scientific Instruments}\ }\textbf
  {\bibinfo {volume} {63}},\ \bibinfo {pages} {4251--4262} (\bibinfo {year}
  {1992})}\BibitemShut {NoStop}%
\bibitem [{\citenamefont {Rinard}\ \emph {et~al.}(2017)\citenamefont {Rinard},
  \citenamefont {Quine}, \citenamefont {McPeak}, \citenamefont {Buchanan},
  \citenamefont {Eaton},\ and\ \citenamefont {Eaton}}]{EatonsAMR2017}%
  \BibitemOpen
  \bibfield  {author} {\bibinfo {author} {\bibfnamefont {G.~A.}\ \bibnamefont
  {Rinard}}, \bibinfo {author} {\bibfnamefont {R.~W.}\ \bibnamefont {Quine}},
  \bibinfo {author} {\bibfnamefont {J.}~\bibnamefont {McPeak}}, \bibinfo
  {author} {\bibfnamefont {L.}~\bibnamefont {Buchanan}}, \bibinfo {author}
  {\bibfnamefont {S.~S.}\ \bibnamefont {Eaton}},\ and\ \bibinfo {author}
  {\bibfnamefont {G.~R.}\ \bibnamefont {Eaton}},\ }\bibfield  {title} {\enquote
  {\bibinfo {title} {{An X-Band Crossed-Loop EPR Resonator}},}\ }\href@noop {}
  {\bibfield  {journal} {\bibinfo  {journal} {Applied MAgnetic Resonance}\
  }\textbf {\bibinfo {volume} {48}},\ \bibinfo {pages} {1219--1226} (\bibinfo
  {year} {2017})}\BibitemShut {NoStop}%
\bibitem [{\citenamefont {McPeak}\ \emph {et~al.}(2019)\citenamefont {McPeak},
  \citenamefont {Quine}, \citenamefont {Eaton},\ and\ \citenamefont
  {Eaton}}]{EatonsRSI2019}%
  \BibitemOpen
  \bibfield  {author} {\bibinfo {author} {\bibfnamefont {J.~E.}\ \bibnamefont
  {McPeak}}, \bibinfo {author} {\bibfnamefont {R.~W.}\ \bibnamefont {Quine}},
  \bibinfo {author} {\bibfnamefont {S.~S.}\ \bibnamefont {Eaton}},\ and\
  \bibinfo {author} {\bibfnamefont {G.~R.}\ \bibnamefont {Eaton}},\ }\bibfield
  {title} {\enquote {\bibinfo {title} {{An X-band continuous wave saturation
  recovery electron paramagnetic resonance spectrometer based on an arbitrary
  waveform generator}},}\ }\href@noop {} {\bibfield  {journal} {\bibinfo
  {journal} {Review of Scientific Instruments}\ }\textbf {\bibinfo {volume}
  {90}},\ \bibinfo {pages} {024102} (\bibinfo {year} {2019})}\BibitemShut
  {NoStop}%
\bibitem [{\citenamefont {Ashcroft}\ and\ \citenamefont
  {Mermin}(1976)}]{AschcroftMermin}%
  \BibitemOpen
  \bibfield  {author} {\bibinfo {author} {\bibfnamefont {N.~W.}\ \bibnamefont
  {Ashcroft}}\ and\ \bibinfo {author} {\bibfnamefont {N.~D.}\ \bibnamefont
  {Mermin}},\ }\href@noop {} {\emph {\bibinfo {title} {Solid State Physics}}}\
  (\bibinfo  {publisher} {Harcourt College Publishers},\ \bibinfo {year}
  {1976})\BibitemShut {NoStop}%
\bibitem [{\citenamefont {Dresselhaus}\ and\ \citenamefont
  {Dresselhaus}(2002)}]{DresselhausAP2002}%
  \BibitemOpen
  \bibfield  {author} {\bibinfo {author} {\bibfnamefont {M.~S.}\ \bibnamefont
  {Dresselhaus}}\ and\ \bibinfo {author} {\bibfnamefont {G.}~\bibnamefont
  {Dresselhaus}},\ }\bibfield  {title} {\enquote {\bibinfo {title}
  {{Intercalation compounds of graphite}},}\ }\href@noop {} {\bibfield
  {journal} {\bibinfo  {journal} {Adv. Phys.}\ }\textbf {\bibinfo {volume}
  {51}},\ \bibinfo {pages} {1--186} (\bibinfo {year} {2002})}\BibitemShut
  {NoStop}%
\bibitem [{\citenamefont {Brandt}, \citenamefont {Chudinov},\ and\
  \citenamefont {Ponomarev}(1988)}]{BrandtBook}%
  \BibitemOpen
  \bibfield  {author} {\bibinfo {author} {\bibfnamefont {N.}~\bibnamefont
  {Brandt}}, \bibinfo {author} {\bibfnamefont {S.}~\bibnamefont {Chudinov}},\
  and\ \bibinfo {author} {\bibfnamefont {Y.}~\bibnamefont {Ponomarev}},\
  }\href@noop {} {\emph {\bibinfo {title} {{Semimetals - 1. Graphite and its
  Compounds}}}}\ (\bibinfo  {publisher} {Elsevier Science Publishers},\
  \bibinfo {address} {Amsterdam},\ \bibinfo {year} {1988})\BibitemShut
  {NoStop}%
\bibitem [{\citenamefont {Thoutam}\ \emph {et~al.}(2022)\citenamefont
  {Thoutam}, \citenamefont {Pate}, \citenamefont {Wang}, \citenamefont {Wang},
  \citenamefont {Divan}, \citenamefont {Martin}, \citenamefont {Luican-Mayer},
  \citenamefont {Welp}, \citenamefont {Kwok},\ and\ \citenamefont
  {Xiao}}]{thoutam2022temperature}%
  \BibitemOpen
  \bibfield  {author} {\bibinfo {author} {\bibfnamefont {L.~R.}\ \bibnamefont
  {Thoutam}}, \bibinfo {author} {\bibfnamefont {S.~E.}\ \bibnamefont {Pate}},
  \bibinfo {author} {\bibfnamefont {T.}~\bibnamefont {Wang}}, \bibinfo {author}
  {\bibfnamefont {Y.-L.}\ \bibnamefont {Wang}}, \bibinfo {author}
  {\bibfnamefont {R.}~\bibnamefont {Divan}}, \bibinfo {author} {\bibfnamefont
  {I.}~\bibnamefont {Martin}}, \bibinfo {author} {\bibfnamefont
  {A.}~\bibnamefont {Luican-Mayer}}, \bibinfo {author} {\bibfnamefont
  {U.}~\bibnamefont {Welp}}, \bibinfo {author} {\bibfnamefont {W.-K.}\
  \bibnamefont {Kwok}},\ and\ \bibinfo {author} {\bibfnamefont {Z.-L.}\
  \bibnamefont {Xiao}},\ }\bibfield  {title} {\enquote {\bibinfo {title}
  {Temperature-driven changes in the fermi surface of graphite},}\ }\href@noop
  {} {\bibfield  {journal} {\bibinfo  {journal} {Physical Review B}\ }\textbf
  {\bibinfo {volume} {106}},\ \bibinfo {pages} {155117} (\bibinfo {year}
  {2022})}\BibitemShut {NoStop}%
\bibitem [{\citenamefont {S\'olyom}(2009)}]{Solyom2}%
  \BibitemOpen
  \bibfield  {author} {\bibinfo {author} {\bibfnamefont {J.}~\bibnamefont
  {S\'olyom}},\ }\href@noop {} {\emph {\bibinfo {title} {{Fundamentals of the
  Physics of Solids -- Volume II: Electronic Properties}}}}\ (\bibinfo
  {publisher} {Spinger-Verlag},\ \bibinfo {address} {Berlin},\ \bibinfo {year}
  {2009})\BibitemShut {NoStop}%
\bibitem [{\citenamefont {Kopelevich}\ \emph {et~al.}(2003)\citenamefont
  {Kopelevich}, \citenamefont {Torres}, \citenamefont {da~Silva}, \citenamefont
  {Mrowka}, \citenamefont {Kempa},\ and\ \citenamefont
  {Esquinazi}}]{KopelevichPRL2003}%
  \BibitemOpen
  \bibfield  {author} {\bibinfo {author} {\bibfnamefont {Y.}~\bibnamefont
  {Kopelevich}}, \bibinfo {author} {\bibfnamefont {J.~H.~S.}\ \bibnamefont
  {Torres}}, \bibinfo {author} {\bibfnamefont {R.~R.}\ \bibnamefont
  {da~Silva}}, \bibinfo {author} {\bibfnamefont {F.}~\bibnamefont {Mrowka}},
  \bibinfo {author} {\bibfnamefont {H.}~\bibnamefont {Kempa}},\ and\ \bibinfo
  {author} {\bibfnamefont {P.}~\bibnamefont {Esquinazi}},\ }\bibfield  {title}
  {\enquote {\bibinfo {title} {Reentrant metallic behavior of graphite in the
  quantum limit},}\ }\href@noop {} {\bibfield  {journal} {\bibinfo  {journal}
  {Phys. Rev. Lett.}\ }\textbf {\bibinfo {volume} {90}},\ \bibinfo {pages}
  {156402} (\bibinfo {year} {2003})}\BibitemShut {NoStop}%
\bibitem [{\citenamefont {Kittel}(2004)}]{KittelBook}%
  \BibitemOpen
  \bibfield  {author} {\bibinfo {author} {\bibfnamefont {C.}~\bibnamefont
  {Kittel}},\ }\href@noop {} {\emph {\bibinfo {title} {Introduction to Solid
  State Physics}}},\ \bibinfo {edition} {8th}\ ed.\ (\bibinfo  {publisher}
  {Wiley},\ \bibinfo {address} {Hoboken, NJ},\ \bibinfo {year}
  {2004})\BibitemShut {NoStop}%
\bibitem [{\citenamefont {Chen}\ \emph {et~al.}(2024)\citenamefont {Chen},
  \citenamefont {Yang}, \citenamefont {Ying},\ and\ \citenamefont {gang
  Guo}}]{ChenNanoLett2024}%
  \BibitemOpen
  \bibfield  {author} {\bibinfo {author} {\bibfnamefont {Z.}~\bibnamefont
  {Chen}}, \bibinfo {author} {\bibfnamefont {Y.}~\bibnamefont {Yang}}, \bibinfo
  {author} {\bibfnamefont {T.}~\bibnamefont {Ying}},\ and\ \bibinfo {author}
  {\bibfnamefont {J.}~\bibnamefont {gang Guo}},\ }\bibfield  {title} {\enquote
  {\bibinfo {title} {{High‑$T_c$ Ferromagnetic Semiconductor in Thinned 3D
  Ising Ferromagnetic Metal Fe$_3$GaTe$_2$}},}\ }\href@noop {} {\bibfield
  {journal} {\bibinfo  {journal} {Nano Letters}\ }\textbf {\bibinfo {volume}
  {24}},\ \bibinfo {pages} {993--1000} (\bibinfo {year} {2024})}\BibitemShut
  {NoStop}%
\bibitem [{\citenamefont {Dyakonov}\ and\ \citenamefont
  {Perel}(1972)}]{DyakonovPerelSPSS1972}%
  \BibitemOpen
  \bibfield  {author} {\bibinfo {author} {\bibfnamefont {M.}~\bibnamefont
  {Dyakonov}}\ and\ \bibinfo {author} {\bibfnamefont {V.}~\bibnamefont
  {Perel}},\ }\bibfield  {title} {\enquote {\bibinfo {title} {Spin relaxation
  of conduction electrons in noncentrosymmetric semiconductors},}\ }\href@noop
  {} {\bibfield  {journal} {\bibinfo  {journal} {Soviet Physics Solid State,
  USSR}\ }\textbf {\bibinfo {volume} {13}},\ \bibinfo {pages} {3023--3026}
  (\bibinfo {year} {1972})}\BibitemShut {NoStop}%
\bibitem [{\citenamefont {Pierret}(2002)}]{pierret2002advanced}%
  \BibitemOpen
  \bibfield  {author} {\bibinfo {author} {\bibfnamefont {R.~F.}\ \bibnamefont
  {Pierret}},\ }\href@noop {} {\emph {\bibinfo {title} {Advanced Semiconductor
  Fundamentals}}},\ \bibinfo {edition} {2nd}\ ed.\ (\bibinfo  {publisher}
  {Pearson Education},\ \bibinfo {address} {Upper Saddle River, NJ},\ \bibinfo
  {year} {2002})\BibitemShut {NoStop}%
\bibitem [{\citenamefont {Schroder}(2006)}]{schroder2006semiconductor}%
  \BibitemOpen
  \bibfield  {author} {\bibinfo {author} {\bibfnamefont {D.~K.}\ \bibnamefont
  {Schroder}},\ }\href@noop {} {\emph {\bibinfo {title} {Semiconductor Material
  and Device Characterization}}},\ \bibinfo {edition} {3rd}\ ed.\ (\bibinfo
  {publisher} {Wiley},\ \bibinfo {address} {Hoboken, NJ},\ \bibinfo {year}
  {2006})\BibitemShut {NoStop}%
\bibitem [{\citenamefont {Soule}\ and\ \citenamefont
  {McClure}(1959)}]{SouleMcClure}%
  \BibitemOpen
  \bibfield  {author} {\bibinfo {author} {\bibfnamefont {D.~E.}\ \bibnamefont
  {Soule}}\ and\ \bibinfo {author} {\bibfnamefont {J.~W.}\ \bibnamefont
  {McClure}},\ }\bibfield  {title} {\enquote {\bibinfo {title} {{Band structure
  and transport properties of single-crystal graphite }},}\ }\href@noop {}
  {\bibfield  {journal} {\bibinfo  {journal} {Journal of Physics and Chemistry
  of Solids}\ }\textbf {\bibinfo {volume} {8}},\ \bibinfo {pages} {29 -- 35}
  (\bibinfo {year} {1959})}\BibitemShut {NoStop}%
\bibitem [{\citenamefont {Taparia}\ \emph {et~al.}(2024)\citenamefont
  {Taparia}, \citenamefont {Sasaki}, \citenamefont {Nakatani}, \citenamefont
  {Suto}, \citenamefont {Miura}, \citenamefont {Li}, \citenamefont {Kushwaha},
  \citenamefont {Inubushi}, \citenamefont {Ichikawa}, \citenamefont {Nakada},
  \citenamefont {Sasaki}, \citenamefont {Mitani},\ and\ \citenamefont
  {Sakuraba}}]{TapariaSTAM2024}%
  \BibitemOpen
  \bibfield  {author} {\bibinfo {author} {\bibfnamefont {D.}~\bibnamefont
  {Taparia}}, \bibinfo {author} {\bibfnamefont {T.}~\bibnamefont {Sasaki}},
  \bibinfo {author} {\bibfnamefont {T.}~\bibnamefont {Nakatani}}, \bibinfo
  {author} {\bibfnamefont {H.}~\bibnamefont {Suto}}, \bibinfo {author}
  {\bibfnamefont {Y.}~\bibnamefont {Miura}}, \bibinfo {author} {\bibfnamefont
  {Z.}~\bibnamefont {Li}}, \bibinfo {author} {\bibfnamefont {V.~K.}\
  \bibnamefont {Kushwaha}}, \bibinfo {author} {\bibfnamefont {K.}~\bibnamefont
  {Inubushi}}, \bibinfo {author} {\bibfnamefont {S.}~\bibnamefont {Ichikawa}},
  \bibinfo {author} {\bibfnamefont {K.}~\bibnamefont {Nakada}}, \bibinfo
  {author} {\bibfnamefont {T.}~\bibnamefont {Sasaki}}, \bibinfo {author}
  {\bibfnamefont {S.}~\bibnamefont {Mitani}},\ and\ \bibinfo {author}
  {\bibfnamefont {Y.}~\bibnamefont {Sakuraba}},\ }\bibfield  {title} {\enquote
  {\bibinfo {title} {{Improvement in CPP-GMR read head sensor performance using
  [001]-oriented polycrystalline halfmetallic Heusler alloy
  Co$_2$FeGa$_{0.5}$Ge$_{0.5}$ and CoFe bilayer electrode}},}\ }\href@noop {}
  {\bibfield  {journal} {\bibinfo  {journal} {Science and Technology of
  Advanced Materials}\ }\textbf {\bibinfo {volume} {25}},\ \bibinfo {pages}
  {2388503} (\bibinfo {year} {2024})}\BibitemShut {NoStop}%
\bibitem [{\citenamefont {Dillon}\ and\ \citenamefont
  {Olson}(1965)}]{DillonJAP1965}%
  \BibitemOpen
  \bibfield  {author} {\bibinfo {author} {\bibfnamefont {J.~F.}\ \bibnamefont
  {Dillon}}\ and\ \bibinfo {author} {\bibfnamefont {C.~E.}\ \bibnamefont
  {Olson}},\ }\bibfield  {title} {\enquote {\bibinfo {title} {{Magnetization,
  Resonance, and Optical Properties of the Ferromagnet CrI$_3$}},}\ }\href@noop
  {} {\bibfield  {journal} {\bibinfo  {journal} {Journal of Applied Physics}\
  }\textbf {\bibinfo {volume} {36}},\ \bibinfo {pages} {1259--1260} (\bibinfo
  {year} {1965})}\BibitemShut {NoStop}%
\bibitem [{\citenamefont {Huang}\ \emph {et~al.}(2018)\citenamefont {Huang},
  \citenamefont {Clark}, \citenamefont {Klein}, \citenamefont {MacNeill},
  \citenamefont {Navarro-Moratalla}, \citenamefont {Seyler}, \citenamefont
  {Wilson}, \citenamefont {McGuire}, \citenamefont {Cobden}, \citenamefont
  {Xiao}, \citenamefont {Yao}, \citenamefont {Jarillo-Herrero},\ and\
  \citenamefont {Xu}}]{HuangNatNanot2018}%
  \BibitemOpen
  \bibfield  {author} {\bibinfo {author} {\bibfnamefont {B.}~\bibnamefont
  {Huang}}, \bibinfo {author} {\bibfnamefont {G.}~\bibnamefont {Clark}},
  \bibinfo {author} {\bibfnamefont {D.~R.}\ \bibnamefont {Klein}}, \bibinfo
  {author} {\bibfnamefont {D.}~\bibnamefont {MacNeill}}, \bibinfo {author}
  {\bibfnamefont {E.}~\bibnamefont {Navarro-Moratalla}}, \bibinfo {author}
  {\bibfnamefont {K.~L.}\ \bibnamefont {Seyler}}, \bibinfo {author}
  {\bibfnamefont {N.}~\bibnamefont {Wilson}}, \bibinfo {author} {\bibfnamefont
  {M.~A.}\ \bibnamefont {McGuire}}, \bibinfo {author} {\bibfnamefont {D.~H.}\
  \bibnamefont {Cobden}}, \bibinfo {author} {\bibfnamefont {D.}~\bibnamefont
  {Xiao}}, \bibinfo {author} {\bibfnamefont {W.}~\bibnamefont {Yao}}, \bibinfo
  {author} {\bibfnamefont {P.}~\bibnamefont {Jarillo-Herrero}},\ and\ \bibinfo
  {author} {\bibfnamefont {X.}~\bibnamefont {Xu}},\ }\bibfield  {title}
  {\enquote {\bibinfo {title} {{Electrical control of 2D magnetism in bilayer
  CrI$_3$}},}\ }\href@noop {} {\bibfield  {journal} {\bibinfo  {journal}
  {Nature Nanotechnology}\ }\textbf {\bibinfo {volume} {13}},\ \bibinfo {pages}
  {544--548} (\bibinfo {year} {2018})}\BibitemShut {NoStop}%
\bibitem [{\citenamefont {Carteaux}\ \emph {et~al.}(1995)\citenamefont
  {Carteaux}, \citenamefont {Brunet}, \citenamefont {Ouvrard},\ and\
  \citenamefont {Andr\'e}}]{CarteauxJPCM1995}%
  \BibitemOpen
  \bibfield  {author} {\bibinfo {author} {\bibfnamefont {V.}~\bibnamefont
  {Carteaux}}, \bibinfo {author} {\bibfnamefont {D.}~\bibnamefont {Brunet}},
  \bibinfo {author} {\bibfnamefont {G.}~\bibnamefont {Ouvrard}},\ and\ \bibinfo
  {author} {\bibfnamefont {G.}~\bibnamefont {Andr\'e}},\ }\bibfield  {title}
  {\enquote {\bibinfo {title} {{Crystallographic,magnetic and
  electronicstructuresof a new layered ferromagneticcompound
  Cr$_2$Ge$_2$Te$_6$}},}\ }\href@noop {} {\bibfield  {journal} {\bibinfo
  {journal} {J. Phys.: Condens. Matter}\ }\textbf {\bibinfo {volume} {7}},\
  \bibinfo {pages} {69--87} (\bibinfo {year} {1995})}\BibitemShut {NoStop}%
\bibitem [{\citenamefont {Verzhbitskiy}\ \emph {et~al.}(2020)\citenamefont
  {Verzhbitskiy}, \citenamefont {Kurebayashi}, \citenamefont {Cheng},
  \citenamefont {Zhou}, \citenamefont {Khan}, \citenamefont {Feng},\ and\
  \citenamefont {Eda}}]{VerzhbitskiyNatElec2020}%
  \BibitemOpen
  \bibfield  {author} {\bibinfo {author} {\bibfnamefont {I.~A.}\ \bibnamefont
  {Verzhbitskiy}}, \bibinfo {author} {\bibfnamefont {H.}~\bibnamefont
  {Kurebayashi}}, \bibinfo {author} {\bibfnamefont {H.}~\bibnamefont {Cheng}},
  \bibinfo {author} {\bibfnamefont {J.}~\bibnamefont {Zhou}}, \bibinfo {author}
  {\bibfnamefont {S.}~\bibnamefont {Khan}}, \bibinfo {author} {\bibfnamefont
  {Y.~P.}\ \bibnamefont {Feng}},\ and\ \bibinfo {author} {\bibfnamefont
  {G.}~\bibnamefont {Eda}},\ }\bibfield  {title} {\enquote {\bibinfo {title}
  {{Controlling the magnetic anisotropy in Cr$_2$Ge$_2$Te$_6$ by electrostatic
  gating}},}\ }\href@noop {} {\bibfield  {journal} {\bibinfo  {journal} {Nature
  Electronics}\ }\textbf {\bibinfo {volume} {3}},\ \bibinfo {pages} {460--465}
  (\bibinfo {year} {2020})}\BibitemShut {NoStop}%
\bibitem [{\citenamefont {Carteaux}, \citenamefont {Moussa},\ and\
  \citenamefont {Spiesser}(1995)}]{CarteauxEPL1995}%
  \BibitemOpen
  \bibfield  {author} {\bibinfo {author} {\bibfnamefont {V.}~\bibnamefont
  {Carteaux}}, \bibinfo {author} {\bibfnamefont {F.}~\bibnamefont {Moussa}},\
  and\ \bibinfo {author} {\bibfnamefont {M.}~\bibnamefont {Spiesser}},\
  }\bibfield  {title} {\enquote {\bibinfo {title} {{2D Ising-like Ferromagnetic
  Behaviour for the Lamellar Cr$_2$Si$_2$Te$_6$ Compound: a Neutron Scattering
  Investigation}},}\ }\href@noop {} {\bibfield  {journal} {\bibinfo  {journal}
  {Europhys. Lett.}\ }\textbf {\bibinfo {volume} {29}},\ \bibinfo {pages}
  {251--256} (\bibinfo {year} {1995})}\BibitemShut {NoStop}%
\bibitem [{\citenamefont {Deiseroth}\ \emph {et~al.}(2006)\citenamefont
  {Deiseroth}, \citenamefont {Aleksandrov}, \citenamefont {Reiner},
  \citenamefont {Kienle},\ and\ \citenamefont {Kremer}}]{DeiserothEJIC2006}%
  \BibitemOpen
  \bibfield  {author} {\bibinfo {author} {\bibfnamefont {H.-J.}\ \bibnamefont
  {Deiseroth}}, \bibinfo {author} {\bibfnamefont {K.}~\bibnamefont
  {Aleksandrov}}, \bibinfo {author} {\bibfnamefont {C.}~\bibnamefont {Reiner}},
  \bibinfo {author} {\bibfnamefont {L.}~\bibnamefont {Kienle}},\ and\ \bibinfo
  {author} {\bibfnamefont {R.~K.}\ \bibnamefont {Kremer}},\ }\bibfield  {title}
  {\enquote {\bibinfo {title} {{Fe$_3$GeTe$_2$ and Ni$_3$GeTe$_2$ - Two New
  Layered Transition-Metal Compounds: Crystal Structures, HRTEM Investigations,
  and Magnetic and Electrical Properties}},}\ }\href@noop {} {\bibfield
  {journal} {\bibinfo  {journal} {European Journal of Inorganic Chemistry}\
  }\textbf {\bibinfo {volume} {8}},\ \bibinfo {pages} {1561--1567} (\bibinfo
  {year} {2006})}\BibitemShut {NoStop}%
\bibitem [{\citenamefont {Fei}\ \emph {et~al.}(2018)\citenamefont {Fei},
  \citenamefont {Huang}, \citenamefont {Malinowski}, \citenamefont {Wang},
  \citenamefont {Song}, \citenamefont {Sanchez}, \citenamefont {Yao},
  \citenamefont {Xiao}, \citenamefont {Zhu}, \citenamefont {May}, \citenamefont
  {Wu}, \citenamefont {Cobden}, \citenamefont {Chu},\ and\ \citenamefont
  {Xu}}]{FeiNatMater2018}%
  \BibitemOpen
  \bibfield  {author} {\bibinfo {author} {\bibfnamefont {Z.}~\bibnamefont
  {Fei}}, \bibinfo {author} {\bibfnamefont {B.}~\bibnamefont {Huang}}, \bibinfo
  {author} {\bibfnamefont {P.}~\bibnamefont {Malinowski}}, \bibinfo {author}
  {\bibfnamefont {W.}~\bibnamefont {Wang}}, \bibinfo {author} {\bibfnamefont
  {T.}~\bibnamefont {Song}}, \bibinfo {author} {\bibfnamefont {J.}~\bibnamefont
  {Sanchez}}, \bibinfo {author} {\bibfnamefont {W.}~\bibnamefont {Yao}},
  \bibinfo {author} {\bibfnamefont {D.}~\bibnamefont {Xiao}}, \bibinfo {author}
  {\bibfnamefont {X.}~\bibnamefont {Zhu}}, \bibinfo {author} {\bibfnamefont
  {A.~F.}\ \bibnamefont {May}}, \bibinfo {author} {\bibfnamefont
  {W.}~\bibnamefont {Wu}}, \bibinfo {author} {\bibfnamefont {D.~H.}\
  \bibnamefont {Cobden}}, \bibinfo {author} {\bibfnamefont {J.-H.}\
  \bibnamefont {Chu}},\ and\ \bibinfo {author} {\bibfnamefont {X.}~\bibnamefont
  {Xu}},\ }\bibfield  {title} {\enquote {\bibinfo {title} {{Two-dimensional
  itinerant ferromagnetism in atomically thin Fe$_3$GeTe$_2$}},}\ }\href@noop
  {} {\bibfield  {journal} {\bibinfo  {journal} {Nature Materials}\ }\textbf
  {\bibinfo {volume} {17}},\ \bibinfo {pages} {778--782} (\bibinfo {year}
  {2018})}\BibitemShut {NoStop}%
\bibitem [{\citenamefont {Chen}\ \emph {et~al.}(2022)\citenamefont {Chen},
  \citenamefont {Asif}, \citenamefont {Whalen}, \citenamefont {T\'amara-Isaza},
  \citenamefont {Luetke}, \citenamefont {Wang}, \citenamefont {Wang},
  \citenamefont {Ayako}, \citenamefont {Lamsal}, \citenamefont {May},
  \citenamefont {McGuire}, \citenamefont {Chakraborty}, \citenamefont {Xiao},\
  and\ \citenamefont {Ku}}]{Chen2DMater2022}%
  \BibitemOpen
  \bibfield  {author} {\bibinfo {author} {\bibfnamefont {H.}~\bibnamefont
  {Chen}}, \bibinfo {author} {\bibfnamefont {S.}~\bibnamefont {Asif}}, \bibinfo
  {author} {\bibfnamefont {M.}~\bibnamefont {Whalen}}, \bibinfo {author}
  {\bibfnamefont {J.}~\bibnamefont {T\'amara-Isaza}}, \bibinfo {author}
  {\bibfnamefont {B.}~\bibnamefont {Luetke}}, \bibinfo {author} {\bibfnamefont
  {Y.}~\bibnamefont {Wang}}, \bibinfo {author} {\bibfnamefont {X.}~\bibnamefont
  {Wang}}, \bibinfo {author} {\bibfnamefont {M.}~\bibnamefont {Ayako}},
  \bibinfo {author} {\bibfnamefont {S.}~\bibnamefont {Lamsal}}, \bibinfo
  {author} {\bibfnamefont {A.~F.}\ \bibnamefont {May}}, \bibinfo {author}
  {\bibfnamefont {M.~A.}\ \bibnamefont {McGuire}}, \bibinfo {author}
  {\bibfnamefont {C.}~\bibnamefont {Chakraborty}}, \bibinfo {author}
  {\bibfnamefont {J.~Q.}\ \bibnamefont {Xiao}},\ and\ \bibinfo {author}
  {\bibfnamefont {M.~J.~H.}\ \bibnamefont {Ku}},\ }\bibfield  {title} {\enquote
  {\bibinfo {title} {{Revealing room temperature ferromagnetism in exfoliated
  Fe$_5$GeTe$_2$ flakes with quantum magnetic imaging}},}\ }\href@noop {}
  {\bibfield  {journal} {\bibinfo  {journal} {2D Materials}\ }\textbf {\bibinfo
  {volume} {9}},\ \bibinfo {pages} {025017} (\bibinfo {year}
  {2022})}\BibitemShut {NoStop}%
\bibitem [{\citenamefont {Zhang}\ \emph {et~al.}(2022)\citenamefont {Zhang},
  \citenamefont {Guo}, \citenamefont {Wu}, \citenamefont {Wen}, \citenamefont
  {Yang}, \citenamefont {Jin}, \citenamefont {Zhang},\ and\ \citenamefont
  {Chang}}]{ZhangNatComm2022}%
  \BibitemOpen
  \bibfield  {author} {\bibinfo {author} {\bibfnamefont {G.}~\bibnamefont
  {Zhang}}, \bibinfo {author} {\bibfnamefont {F.}~\bibnamefont {Guo}}, \bibinfo
  {author} {\bibfnamefont {H.}~\bibnamefont {Wu}}, \bibinfo {author}
  {\bibfnamefont {X.}~\bibnamefont {Wen}}, \bibinfo {author} {\bibfnamefont
  {L.}~\bibnamefont {Yang}}, \bibinfo {author} {\bibfnamefont {W.}~\bibnamefont
  {Jin}}, \bibinfo {author} {\bibfnamefont {W.}~\bibnamefont {Zhang}},\ and\
  \bibinfo {author} {\bibfnamefont {H.}~\bibnamefont {Chang}},\ }\bibfield
  {title} {\enquote {\bibinfo {title} {{Above-room-temperature strong intrinsic
  ferromagnetism in 2D van der Waals Fe$_3$GaTe$_2$ with large perpendicular
  magnetic anisotropy}},}\ }\href@noop {} {\bibfield  {journal} {\bibinfo
  {journal} {Nature Communications}\ }\textbf {\bibinfo {volume} {13}},\
  \bibinfo {pages} {5067} (\bibinfo {year} {2022})}\BibitemShut {NoStop}%
\bibitem [{\citenamefont {Yin}\ \emph {et~al.}(2023)\citenamefont {Yin},
  \citenamefont {Zhang}, \citenamefont {Jin}, \citenamefont {Di}, \citenamefont
  {Wu}, \citenamefont {Zhang}, \citenamefont {Zhang},\ and\ \citenamefont
  {Chang}}]{YinCEC2023}%
  \BibitemOpen
  \bibfield  {author} {\bibinfo {author} {\bibfnamefont {H.}~\bibnamefont
  {Yin}}, \bibinfo {author} {\bibfnamefont {P.}~\bibnamefont {Zhang}}, \bibinfo
  {author} {\bibfnamefont {W.}~\bibnamefont {Jin}}, \bibinfo {author}
  {\bibfnamefont {B.}~\bibnamefont {Di}}, \bibinfo {author} {\bibfnamefont
  {H.}~\bibnamefont {Wu}}, \bibinfo {author} {\bibfnamefont {G.}~\bibnamefont
  {Zhang}}, \bibinfo {author} {\bibfnamefont {W.}~\bibnamefont {Zhang}},\ and\
  \bibinfo {author} {\bibfnamefont {H.}~\bibnamefont {Chang}},\ }\bibfield
  {title} {\enquote {\bibinfo {title} {{Fe$_3$GaTe$_2$/MoSe$_2$
  ferromagnet/semiconductor 2D van der Waals heterojunction for roomtemperature
  spin-valve devices}},}\ }\href@noop {} {\bibfield  {journal} {\bibinfo
  {journal} {CrystEngComm}\ }\textbf {\bibinfo {volume} {25}},\ \bibinfo
  {pages} {1339--1346} (\bibinfo {year} {2023})}\BibitemShut {NoStop}%
\bibitem [{\citenamefont {Hu}\ \emph {et~al.}(2024)\citenamefont {Hu},
  \citenamefont {Guo}, \citenamefont {Lv}, \citenamefont {Li}, \citenamefont
  {Wang}, \citenamefont {Han}, \citenamefont {Pan}, \citenamefont {Xie},
  \citenamefont {Yu}, \citenamefont {Zhu}, \citenamefont {Qi}, \citenamefont
  {Xian}, \citenamefont {Zhu}, \citenamefont {Shi}, \citenamefont {Bao},
  \citenamefont {Lin}, \citenamefont {Zhou}, \citenamefont {Yang},\ and\
  \citenamefont {jun Gao}}]{HuAdvMater2024}%
  \BibitemOpen
  \bibfield  {author} {\bibinfo {author} {\bibfnamefont {G.}~\bibnamefont
  {Hu}}, \bibinfo {author} {\bibfnamefont {H.}~\bibnamefont {Guo}}, \bibinfo
  {author} {\bibfnamefont {S.}~\bibnamefont {Lv}}, \bibinfo {author}
  {\bibfnamefont {L.}~\bibnamefont {Li}}, \bibinfo {author} {\bibfnamefont
  {Y.}~\bibnamefont {Wang}}, \bibinfo {author} {\bibfnamefont {Y.}~\bibnamefont
  {Han}}, \bibinfo {author} {\bibfnamefont {L.}~\bibnamefont {Pan}}, \bibinfo
  {author} {\bibfnamefont {Y.}~\bibnamefont {Xie}}, \bibinfo {author}
  {\bibfnamefont {W.}~\bibnamefont {Yu}}, \bibinfo {author} {\bibfnamefont
  {K.}~\bibnamefont {Zhu}}, \bibinfo {author} {\bibfnamefont {Q.}~\bibnamefont
  {Qi}}, \bibinfo {author} {\bibfnamefont {G.}~\bibnamefont {Xian}}, \bibinfo
  {author} {\bibfnamefont {S.}~\bibnamefont {Zhu}}, \bibinfo {author}
  {\bibfnamefont {J.}~\bibnamefont {Shi}}, \bibinfo {author} {\bibfnamefont
  {L.}~\bibnamefont {Bao}}, \bibinfo {author} {\bibfnamefont {X.}~\bibnamefont
  {Lin}}, \bibinfo {author} {\bibfnamefont {W.}~\bibnamefont {Zhou}}, \bibinfo
  {author} {\bibfnamefont {H.}~\bibnamefont {Yang}},\ and\ \bibinfo {author}
  {\bibfnamefont {H.}~\bibnamefont {jun Gao}},\ }\bibfield  {title} {\enquote
  {\bibinfo {title} {{Room‐Temperature Antisymmetric Magnetoresistance in van
  der Waals Ferromagnet Fe$_3$GaTe$_2$ Nanosheets}},}\ }\href@noop {}
  {\bibfield  {journal} {\bibinfo  {journal} {Advanced Materials}\ }\textbf
  {\bibinfo {volume} {36}},\ \bibinfo {pages} {2403154} (\bibinfo {year}
  {2024})}\BibitemShut {NoStop}%
\bibitem [{\citenamefont {Datta}\ and\ \citenamefont {Das}(1990)}]{Datta1990}%
  \BibitemOpen
  \bibfield  {author} {\bibinfo {author} {\bibfnamefont {S.}~\bibnamefont
  {Datta}}\ and\ \bibinfo {author} {\bibfnamefont {B.}~\bibnamefont {Das}},\
  }\bibfield  {title} {\enquote {\bibinfo {title} {Electronic analog of the
  electro‐optic modulator},}\ }\href@noop {} {\bibfield  {journal} {\bibinfo
  {journal} {Applied Physics Letters}\ }\textbf {\bibinfo {volume} {56}},\
  \bibinfo {pages} {665--667} (\bibinfo {year} {1990})}\BibitemShut {NoStop}%
\bibitem [{\citenamefont {Dyson}(1955)}]{Dyson}%
  \BibitemOpen
  \bibfield  {author} {\bibinfo {author} {\bibfnamefont {F.~J.}\ \bibnamefont
  {Dyson}},\ }\bibfield  {title} {\enquote {\bibinfo {title} {{Electron spin
  resonance absorption in metals II. Theory of electron diffusion and the skin
  effect}},}\ }\href@noop {} {\bibfield  {journal} {\bibinfo  {journal} {Phys.
  Rev.}\ }\textbf {\bibinfo {volume} {98}},\ \bibinfo {pages} {349--359}
  (\bibinfo {year} {1955})}\BibitemShut {NoStop}%
\bibitem [{\citenamefont {Walmsley}\ \emph {et~al.}(1989)\citenamefont
  {Walmsley}, \citenamefont {Ceotto}, \citenamefont {Castilho},\ and\
  \citenamefont {Rettori}}]{WalmsleySynthMet1989}%
  \BibitemOpen
  \bibfield  {author} {\bibinfo {author} {\bibfnamefont {L.}~\bibnamefont
  {Walmsley}}, \bibinfo {author} {\bibfnamefont {G.}~\bibnamefont {Ceotto}},
  \bibinfo {author} {\bibfnamefont {J.~H.}\ \bibnamefont {Castilho}},\ and\
  \bibinfo {author} {\bibfnamefont {C.}~\bibnamefont {Rettori}},\ }\bibfield
  {title} {\enquote {\bibinfo {title} {{Magnetic field modulation frequency,
  sample size and electromagnetic configuration effects on the spin resonance
  spectra of graphite intercalation compounds}},}\ }\href@noop {} {\bibfield
  {journal} {\bibinfo  {journal} {Synthetic Metals}\ }\textbf {\bibinfo
  {volume} {30}},\ \bibinfo {pages} {97--107} (\bibinfo {year}
  {1989})}\BibitemShut {NoStop}%
\bibitem [{\citenamefont {Walmsley}(1992)}]{WalmsleyPRB1992}%
  \BibitemOpen
  \bibfield  {author} {\bibinfo {author} {\bibfnamefont {L.}~\bibnamefont
  {Walmsley}},\ }\bibfield  {title} {\enquote {\bibinfo {title} {{In-plane
  resistivity in graphite intercalation compounds obtained from
  conduction-electron-spin-resonance measurements}},}\ }\href@noop {}
  {\bibfield  {journal} {\bibinfo  {journal} {Phys. Rev. B}\ }\textbf {\bibinfo
  {volume} {46}},\ \bibinfo {pages} {6256--6260} (\bibinfo {year}
  {1992})}\BibitemShut {NoStop}%
\bibitem [{\citenamefont {Walmsley}(1996)}]{WalmsleyJMR1996}%
  \BibitemOpen
  \bibfield  {author} {\bibinfo {author} {\bibfnamefont {L.}~\bibnamefont
  {Walmsley}},\ }\bibfield  {title} {\enquote {\bibinfo {title} {{Translating
  Conduction-Electron Spin-Resonance Lines into Lorentzian Lines}},}\
  }\href@noop {} {\bibfield  {journal} {\bibinfo  {journal} {Journal of
  Magnetic Resonance, Series A}\ }\textbf {\bibinfo {volume} {\textbf{122}}},\
  \bibinfo {pages} {209--213} (\bibinfo {year} {1996})}\BibitemShut {NoStop}%
\bibitem [{\citenamefont {Portis}(1953)}]{PortisPR1953}%
  \BibitemOpen
  \bibfield  {author} {\bibinfo {author} {\bibfnamefont {A.~M.}\ \bibnamefont
  {Portis}},\ }\bibfield  {title} {\enquote {\bibinfo {title} {{Electronic
  Structure of F Centers: Saturation of the Electron Spin Resonance}},}\
  }\href@noop {} {\bibfield  {journal} {\bibinfo  {journal} {Phys. Rev.}\
  }\textbf {\bibinfo {volume} {91}},\ \bibinfo {pages} {1071} (\bibinfo {year}
  {1953})}\BibitemShut {NoStop}%
\bibitem [{\citenamefont {Chen}\ \emph {et~al.}(2004)\citenamefont {Chen},
  \citenamefont {Ong}, \citenamefont {Neo}, \citenamefont {Varadan},\ and\
  \citenamefont {Varadan}}]{chen2004microwave}%
  \BibitemOpen
  \bibfield  {author} {\bibinfo {author} {\bibfnamefont {L.~F.}\ \bibnamefont
  {Chen}}, \bibinfo {author} {\bibfnamefont {C.~K.}\ \bibnamefont {Ong}},
  \bibinfo {author} {\bibfnamefont {C.~P.}\ \bibnamefont {Neo}}, \bibinfo
  {author} {\bibfnamefont {V.~V.}\ \bibnamefont {Varadan}},\ and\ \bibinfo
  {author} {\bibfnamefont {V.~K.}\ \bibnamefont {Varadan}},\ }\href@noop {}
  {\emph {\bibinfo {title} {{Microwave Electronics: Measurement and Materials
  Characterization}}}},\ \bibinfo {edition} {1st}\ ed.\ (\bibinfo  {publisher}
  {John Wiley \& Sons},\ \bibinfo {address} {Chichester, UK},\ \bibinfo {year}
  {2004})\BibitemShut {NoStop}%
\bibitem [{\citenamefont {Boi}\ \emph {et~al.}(2024)\citenamefont {Boi},
  \citenamefont {Lee}, \citenamefont {Wang}, \citenamefont {Wu}, \citenamefont
  {Li}, \citenamefont {Zhang}, \citenamefont {Song}, \citenamefont {Dai},
  \citenamefont {Taallah}, \citenamefont {Odunmbaku}, \citenamefont {Corrias},
  \citenamefont {Baron-Wiechec}, \citenamefont {Zheng},\ and\ \citenamefont
  {Grasso}}]{BoiCT2024}%
  \BibitemOpen
  \bibfield  {author} {\bibinfo {author} {\bibfnamefont {F.}~\bibnamefont
  {Boi}}, \bibinfo {author} {\bibfnamefont {C.-Y.}\ \bibnamefont {Lee}},
  \bibinfo {author} {\bibfnamefont {S.}~\bibnamefont {Wang}}, \bibinfo {author}
  {\bibfnamefont {H.}~\bibnamefont {Wu}}, \bibinfo {author} {\bibfnamefont
  {L.}~\bibnamefont {Li}}, \bibinfo {author} {\bibfnamefont {L.}~\bibnamefont
  {Zhang}}, \bibinfo {author} {\bibfnamefont {J.}~\bibnamefont {Song}},
  \bibinfo {author} {\bibfnamefont {Y.}~\bibnamefont {Dai}}, \bibinfo {author}
  {\bibfnamefont {A.}~\bibnamefont {Taallah}}, \bibinfo {author} {\bibfnamefont
  {O.}~\bibnamefont {Odunmbaku}}, \bibinfo {author} {\bibfnamefont
  {A.}~\bibnamefont {Corrias}}, \bibinfo {author} {\bibfnamefont
  {A.}~\bibnamefont {Baron-Wiechec}}, \bibinfo {author} {\bibfnamefont
  {S.}~\bibnamefont {Zheng}},\ and\ \bibinfo {author} {\bibfnamefont
  {S.}~\bibnamefont {Grasso}},\ }\bibfield  {title} {\enquote {\bibinfo {title}
  {{Rhombohedral stacking-faults in exfoliated highly oriented pyrolytic
  graphite}},}\ }\href@noop {} {\bibfield  {journal} {\bibinfo  {journal}
  {Carbon Trends}\ }\textbf {\bibinfo {volume} {15}},\ \bibinfo {pages}
  {100345} (\bibinfo {year} {2024})}\BibitemShut {NoStop}%
\bibitem [{\citenamefont {Chehab}\ \emph {et~al.}(2000)\citenamefont {Chehab},
  \citenamefont {Gu\'erin}, \citenamefont {Amiell},\ and\ \citenamefont
  {Flandrois}}]{ChehabEPJB2000}%
  \BibitemOpen
  \bibfield  {author} {\bibinfo {author} {\bibfnamefont {S.}~\bibnamefont
  {Chehab}}, \bibinfo {author} {\bibfnamefont {K.}~\bibnamefont {Gu\'erin}},
  \bibinfo {author} {\bibfnamefont {J.}~\bibnamefont {Amiell}},\ and\ \bibinfo
  {author} {\bibfnamefont {S.}~\bibnamefont {Flandrois}},\ }\bibfield  {title}
  {\enquote {\bibinfo {title} {{Magnetic properties of mixed graphite
  containing both hexagonal and rhombohedral forms}},}\ }\href@noop {}
  {\bibfield  {journal} {\bibinfo  {journal} {The European Physical Journal B}\
  }\textbf {\bibinfo {volume} {13}},\ \bibinfo {pages} {235--243} (\bibinfo
  {year} {2000})}\BibitemShut {NoStop}%
\bibitem [{\citenamefont {Murányi}\ \emph {et~al.}(2004)\citenamefont
  {Murányi}, \citenamefont {Simon}, \citenamefont {Fülöp},\ and\
  \citenamefont {Jánossy}}]{MuranyiSimon1}%
  \BibitemOpen
  \bibfield  {author} {\bibinfo {author} {\bibfnamefont {F.}~\bibnamefont
  {Murányi}}, \bibinfo {author} {\bibfnamefont {F.}~\bibnamefont {Simon}},
  \bibinfo {author} {\bibfnamefont {F.}~\bibnamefont {Fülöp}},\ and\ \bibinfo
  {author} {\bibfnamefont {A.}~\bibnamefont {Jánossy}},\ }\bibfield  {title}
  {\enquote {\bibinfo {title} {{A longitudinally detected high-field ESR
  spectrometer for the measurement of spin-lattice relaxation times}},}\
  }\href@noop {} {\bibfield  {journal} {\bibinfo  {journal} {Journal of
  Magnetic Resonance}\ }\textbf {\bibinfo {volume} {167}},\ \bibinfo {pages}
  {221--227} (\bibinfo {year} {2004})}\BibitemShut {NoStop}%
\bibitem [{\citenamefont {Simon}\ and\ \citenamefont
  {Murányi}(2005)}]{MuranyiSimon2}%
  \BibitemOpen
  \bibfield  {author} {\bibinfo {author} {\bibfnamefont {F.}~\bibnamefont
  {Simon}}\ and\ \bibinfo {author} {\bibfnamefont {F.}~\bibnamefont
  {Murányi}},\ }\bibfield  {title} {\enquote {\bibinfo {title} {{ESR
  spectrometer with a loop-gap resonator for cw and time resolved studies in a
  superconducting magnet}},}\ }\href@noop {} {\bibfield  {journal} {\bibinfo
  {journal} {Journal of Magnetic Resonance}\ }\textbf {\bibinfo {volume}
  {173}},\ \bibinfo {pages} {288--295} (\bibinfo {year} {2005})}\BibitemShut
  {NoStop}%
\bibitem [{\citenamefont {McClure}(1957)}]{McClurePR1957}%
  \BibitemOpen
  \bibfield  {author} {\bibinfo {author} {\bibfnamefont {J.~W.}\ \bibnamefont
  {McClure}},\ }\bibfield  {title} {\enquote {\bibinfo {title} {{Band Structure
  of Graphite and de Haas-van Alphen Effect}},}\ }\href@noop {} {\bibfield
  {journal} {\bibinfo  {journal} {Phys. Rev.}\ }\textbf {\bibinfo {volume}
  {108}},\ \bibinfo {pages} {612--618} (\bibinfo {year} {1957})}\BibitemShut
  {NoStop}%
\bibitem [{\citenamefont {McClure}\ and\ \citenamefont
  {Yafet}(1962)}]{McClureCoC1962}%
  \BibitemOpen
  \bibfield  {author} {\bibinfo {author} {\bibfnamefont {J.~W.}\ \bibnamefont
  {McClure}}\ and\ \bibinfo {author} {\bibfnamefont {Y.}~\bibnamefont
  {Yafet}},\ }\bibfield  {title} {\enquote {\bibinfo {title} {{Theory of the
  $g$-factor of the current carriers in graphite single crystals}},}\
  }\href@noop {} {\bibfield  {journal} {\bibinfo  {journal} {Proc. 5th
  Conference on Carbon, Pergamon Press}\ ,\ \bibinfo {pages} {22--28}}
  (\bibinfo {year} {1962})}\BibitemShut {NoStop}%
\bibitem [{\citenamefont {Möser}\ \emph {et~al.}(2017)\citenamefont {Möser},
  \citenamefont {Lips}, \citenamefont {Tseytlin}, \citenamefont {Eaton},
  \citenamefont {Eaton},\ and\ \citenamefont {Schnegg}}]{Moser2017}%
  \BibitemOpen
  \bibfield  {author} {\bibinfo {author} {\bibfnamefont {J.}~\bibnamefont
  {Möser}}, \bibinfo {author} {\bibfnamefont {K.}~\bibnamefont {Lips}},
  \bibinfo {author} {\bibfnamefont {M.}~\bibnamefont {Tseytlin}}, \bibinfo
  {author} {\bibfnamefont {G.~R.}\ \bibnamefont {Eaton}}, \bibinfo {author}
  {\bibfnamefont {S.~S.}\ \bibnamefont {Eaton}},\ and\ \bibinfo {author}
  {\bibfnamefont {A.}~\bibnamefont {Schnegg}},\ }\bibfield  {title} {\enquote
  {\bibinfo {title} {Using rapid-scan epr to improve the detection limit of
  quantitative epr by more than one order of magnitude},}\ }\href@noop {}
  {\bibfield  {journal} {\bibinfo  {journal} {Journal of Magnetic Resonance}\
  }\textbf {\bibinfo {volume} {280}},\ \bibinfo {pages} {122--133} (\bibinfo
  {year} {2017})}\BibitemShut {NoStop}%
\bibitem [{\citenamefont {Shaw}(2017)}]{DiffusionSemiconductors}%
  \BibitemOpen
  \bibfield  {author} {\bibinfo {author} {\bibfnamefont {D.}~\bibnamefont
  {Shaw}},\ }\enquote {\bibinfo {title} {{Diffusion in Semiconductors}},}\ in\
  \href@noop {} {\emph {\bibinfo {booktitle} {Springer Handbook of Electronic
  and Photonic Materials}}},\ \bibinfo {editor} {edited by\ \bibinfo {editor}
  {\bibfnamefont {S.}~\bibnamefont {Kasap}}\ and\ \bibinfo {editor}
  {\bibfnamefont {P.}~\bibnamefont {Capper}}}\ (\bibinfo  {publisher} {Springer
  International Publishing},\ \bibinfo {address} {Cham},\ \bibinfo {year}
  {2017})\ pp.\ \bibinfo {pages} {1--1}\BibitemShut {NoStop}%
\end{thebibliography}%


\appendix
\clearpage
\pagebreak
\makeatletter 
\renewcommand{\thefigure}{S\@arabic\c@figure}
\makeatother
\setcounter{figure}{0}

\section{Discussion of the non-local spin transport and Hanle spin precession experiments}
In the non-local resistance measurements, a non-equilibrium magnetization is injected on one side of the the device which decays with the spin-diffusion length $\delta_{\text{s}}$. The spatial dependence of the non-equilibrium magnetization vector reads: $\mathbf{M} (x)=\mathbf{M}_0 \exp\left(-x/\delta_{\text{s}}\right)$. This is the solution of the one-dimensional diffusion equation:
\begin{equation}
    D\frac{\mathrm{d}^2 \mathbf{M}}{\mathrm{d} x^2}-\frac{\mathbf{M}}{\tau_\text{s}}=0,
    \label{injection_diffusion}
\end{equation}
which is obtained after substitution and using that $\delta_{\text{s}}^2=D\tau_{\text{s}}$. This implies that the "surviving" concentration of spins (and thus the non-local resistance) is proportional to $\exp\left(-L/\delta_{\text{s}}\right)$, where $L$ is the distance between the electrodes. Measuring $\delta_{\text{s}}$ or $\tau_\text{s}$ from non-local resistance measurement (measuring these two quantities is essentially identical, as $D$ is usually known) is somewhat cumbersome \cite{TombrosNAT2007} as it requires to prepare a number of devices with varying electrode separation. An improved version of transport-based spin relaxation studies is based on Hanle-type spin precession \cite{TombrosPRL2008}. Then Eq. \eqref{injection_diffusion} is amended with a Larmor precession term resulting: 
\begin{equation}
    D\frac{\mathrm{d}^2 \mathbf{M}}{\mathrm{d} x^2}-\frac{\mathbf{M}}{\tau_\text{s}}+\gamma_\text{e}\left( \mathbf{M} \times \mathbf{B}\right) =0,
    \label{injection_diffusion_precession}
\end{equation}
where $\gamma_\text{e}$ is the electron gyromagnetic ratio and $\gamma/2\pi\approx 28~\text{GHz}/{\text{T}}$. This is also known as the Bloch--Torrey equation.

Assuming that $\mathbf{B}=\left(0,0,B \right)$ and that the injected magnetization at $x=0$ is $\mathbf{M}=\left(M_0,0,0\right)$ we obtain the solution of Eq. \eqref{injection_diffusion_precession} in the following form:
\begin{gather}
    M_x=M_0\exp\left(-\frac{x}{\widetilde{\delta}} \right)\cos\left(\frac{\omega_0 \widetilde{\delta}x}{2D} \right),\\
    M_y=M_0\exp\left(-\frac{x}{\widetilde{\delta}} \right)\sin\left(\frac{\omega_0\widetilde{\delta}x}{2D} \right),
\end{gather}
where we introduced the Larmor (angular) frequency as $\omega_0=\gamma B$. A substitution into Eq. \eqref{injection_diffusion_precession} gives that the value of $\widetilde{\delta}^2$ is obtained from the quadratic equation below:
\begin{gather}
    \frac{\omega_0^2}{4} \widetilde{\delta}^4+\frac{D}{\tau_\text{s}}\widetilde{\delta}^2-D^2=0.
\label{delta_tilde}
\end{gather}
Note that Eq. \eqref{delta_tilde} returns $\widetilde{\delta}=\delta_\text{s}$ when $B=0$. Given that the electrode separation and the diffusion constants are usually known, a fit to the magnetic field dependent Hanle oscillation curve yields $\tau_\text{s}$ using a \emph{single} device. 

\section{Derivation of the angular-dependent relaxation times}

Yafet proposed in his seminal paper \cite{yafet1963g} on page 60, a relatively simple approach to determine the two types of spin-relaxation ($T_1$ and $T_2$) to tackle spin-relaxation in materials. Let $z$ be the direction of the DC magnetic field of the magnetic resonance experiment and let $x$ and $y$ denote the two perpendicular directions. In addition, let $\delta B_{(x,y,z)}^2$ denote the variance of a fluctuating magnetic field along these directions. The microscopic origin of these fluctuating fields is not important for now. Then, the corresponding relaxation times are proportional to:
\begin{align}
    T_1^{-1}&\propto\delta B_x^2+\delta B_y^2,\\
    T_2^{-1}&\propto\frac{\delta B_x^2+\delta B_y^2}{2}+\delta B_z^2.
\end{align}

These expressions stem from the fact that for a given relaxation time, it is always the perpendicular fluctuating field which causes relaxation. E.g., for a $T_1$ process with the $z$-direction of the magnetic field, a perpendicular fluctuating field is along the $x$ and $y$-directions. For a $T_2$ process (decoherence in the $(x,y)$ plane), the perpendicular directions are the $z$ and in the half of the time the $x$ and $y$-directions due to the Larmor precession. This causes the factor two in the denominator in the expressions above.

For an isotropic system, the three fluctuating fields have the same magnitude, thus $T_1^{-1}=T_2^{-1}$. In an anisotropic system, these two relaxation times differ, i.e., no relationship between $T_1$ and $T_2$ can be established, except that $T_2 \leq 2 T_1$. Yafet also discussed the case of extreme anisotropy, when $T_2$ changes by a factor $2$ between the two relevant directions of the external magnetic field. In fact, this statement led us earlier to recognize that this is indeed realized in graphite such that the effect of the fluctuating magnetic fields is negligible in the graphite $(a,b)$ plane, i.e., $\delta B_{(a,b)}^2\ll \delta B_{c}^2$ which explains the experimental fact that $T_2$ is a factor two smaller when the magnetic field is in the graphite $(a,b)$ plane compared to the case when it is along the $c$ axis. 

For such a case, we previously worked out \cite{MarkusNatComm} the relaxation times for the two orientations. Let $T_{1,c}$, $T_{1,(a,b)}$ denote the longitudinal relaxation time, when the magnetic field is along the $c$ axis or in the $(a,b)$ plane, respectively. We denote similarly by $T_{2,c}$ and $T_{2,(a,b)}$ the corresponding transversal relaxation time. We then obtained:
\begin{align}
    \begin{split}
        T_{1,c}^{-1}&\propto2\delta B_{(a,b)}^2,\\
        T_{2,c}^{-1}&\propto\delta B_{c}^2+\delta B_{(a,b)}^2,\\
        T_{1,(a,b)}^{-1}&\propto\delta B_{c}^2+\delta B_{(a,b)}^2,\\
        T_{2,(a,b)}^{-1}&\propto\frac{1}{2}\delta B_{c}^2+\frac{3}{2}\delta B_{(a,b)}^2.
    \end{split}
\end{align}
These equations return the experimental situation in graphite, i.e., $T_{2,(a,b)}=\frac{1}{2}T_{2,c}$ when $\delta B_{(a,b)}^2=0$ is assumed. 

These expressions for the special orientations also led us to work out \cite{MarkusNatComm} the relaxation times for an arbitrary orientation of the magnetic field. Let $\vartheta$ denote the polar angle with respect to the crystalline $c$ axis, e.g., $\vartheta=0$ for $B \parallel c$ and $\vartheta=\pi/2$ for $B \parallel (a,b)$. We then obtained:
\begin{align}
        T_1^{-1}\left(\vartheta \right)&\propto \delta B_{(a,b)}^2 \left(1+\cos^2\vartheta \right)+\delta B_{c}^2 \sin^2\vartheta, \label{AngDepRelaxationTimes1} \\
        T_2^{-1}\left(\vartheta \right)&\propto \frac{\delta B_{(a,b)}^2 \left(2+\sin^2\vartheta \right)+\delta B_{c}^2 \left(1+\cos^2\vartheta \right)}{2}. \label{AngDepRelaxationTimes2}
\end{align}

While description with the fluctuating fields is intuitive, it appears to be more appropriate to express the angular dependence of the relaxation times with the corresponding extremal values. Let us express Eqs. \eqref{AngDepRelaxationTimes1} and \eqref{AngDepRelaxationTimes2} with the help of $T_{1,c}$ and $T_{2,(a,b)}$. The advantage is that any magnetic field orientation can be expressed with the help of the extreme values of the experimental measurables. We take $T_{1,c}^{-1}=\text{const.} \cdot 2\delta B_{(a,b)}^2 $ and $T_{2,c}^{-1}=\text{const.} \cdot \left(\delta B_{c}^2+\delta B_{(a,b)}^2 \right)$ which leads to:
\begin{align}
        T_1^{-1}\left(\vartheta \right)&= T_{1,c}^{-1}\frac{1+\cos 2\vartheta}{2}+T_{2,c}^{-1}\sin^2\vartheta \label{AngDepRelaxationTimesRewrittenSM1}\\
        T_2^{-1}\left(\vartheta \right)&=T_{1,c}^{-1}\frac{1-\cos 2\vartheta}{4}+T_{2,c}^{-1}\frac{1+\cos^2 \vartheta}{2}.
        \label{AngDepRelaxationTimesRewrittenSM2}
\end{align}

Note that the proportionality has been replaced by equality in Eqs. \eqref{AngDepRelaxationTimesRewrittenSM1} and \eqref{AngDepRelaxationTimesRewrittenSM2}. As expected, Eqs. \eqref{AngDepRelaxationTimesRewrittenSM1} and \eqref{AngDepRelaxationTimesRewrittenSM2} return $ T_1^{-1}\left(0\right)=T_{1,c}^{-1}$; $ T_2^{-1}\left(0 \right)=T_{2,c}^{-1}$; and $ T_1^{-1}\left(\pi/2\right)=T_{2,c}^{-1}$. Of these, the first two recover the definition of the respective quantities and the third equality was recognized in Ref. \onlinecite{MarkusNatComm}. It is interesting to note that $ T_{2,(a,b)}^{-1}=\frac{T_{2,c}^{-1}+T_{1,c}^{-1}}{2}$ is predicted. A delicate analysis of the $1{:}2$ linewidth anisotropy would in principle allow to determine $T_{1,c}^{-1}$. However, the latter quantity corresponds to about $4-5~\text{mT}$ in magnetic field (broadening) units, which is below the available experimental accuracy. 

We also find it intuitive to introduce the zero-field spin-relaxation times, or $\tau_{\text{s}}$, instead of the ESR measurables, $T_1$ and $T_2$. We recognize that in the zero-field limit $\tau_{\text{s},c}=T_{1,c}$ and $\tau_{\text{s},(a,b)}=T_{1,(a,b)}=T_{2,c}$. With this, we obtain the final version of the angular dependent ESR measurables, as given in the main text:
\begin{align}
        T_1^{-1}\left(\vartheta \right)&= \tau_{\text{s},c}^{-1}\frac{1+\cos 2\vartheta}{2}+\tau_{\text{s},(a,b)}^{-1}\sin^2\vartheta, \label{AngDepRelaxationTimesRewritten2SM1}\\
        T_2^{-1}\left(\vartheta \right)&=\tau_{\text{s},c}^{-1}\frac{1-\cos 2\vartheta}{4}+\tau_{\text{s},(a,b)}^{-1}\frac{1+\cos^2 \vartheta}{2}.
        \label{AngDepRelaxationTimesRewritten2SM2}
\end{align}

Eqs. \eqref{AngDepRelaxationTimesRewritten2SM1} and \eqref{AngDepRelaxationTimesRewritten2SM2} were used to simultaneously fit the relevant saturation factor and the linewidth, which yield the parameters with high accuracy. A representative fit is shown in Fig. 3. of the main manuscript.

\section{Power dependent ESR spectra, line-broadening and intensity}
\subsection{The effect of saturation}

When studying the lineshapes and relaxation times in magnetic resonance, three separate relaxation times can be distinguished : $T_1$, $T_2$, and $T_2^{\ast}$ (Refs. \onlinecite{AbragamBook, SlichterBook}). Clearly, the distinction for these relaxation times arises due to the finite magnetic field. $T_1$, known as the spin-lattice relaxation time (due to historical reasons), also known as the longitudinal relaxation time (or simply the spin-relaxation time), describes how fast a non-equilibrium spin ensemble magnetization recovers to its thermal equilibrium value, $M_0$, along the magnetic field. The $T_2$ or spin-spin relaxation time (also known as transversal relaxation time or spin decoherence time) describes how rapidly the component of the spin magnetization, which is perpendicular to the external magnetic field, relaxes to zero. As the name suggests, $T_2$ is often caused by spin-spin interactions either due to that between like-spins or between the electron spin and the nuclei \cite{AbragamBook}.

A third relaxation time, $T_2^{\ast}$, (also known as spin-dephasing time) is introduced which describes an additional line broadening. In nuclear magnetic resonance (NMR) in liquids, $T_2^{\ast}$ arises from the inhomogeneity of the external magnetic fields. In such cases, the Larmor precession frequency differs for each nuclei which gives rise to a dephasing between the nuclei. However, the effect of $T_2^{\ast}$ can be eliminated in time-resolved spin-echo based techniques, thus it is often referred to as reversible decoherence time, whereas the real $T_2$ corresponds to an entropy increasing decoherence it cannot be thus reversed \cite{AbragamBook, SlichterBook}.

In solids, both for nuclei and electron spin, the dominant source of $T_2^{\ast}$ is local magnetic field inhomogeneities due to defects, variations of the local field, e.g., due to paramagnetic ion or nuclear magnetism or even flux lattice-related magnetic field inhomogeneity in superconductors. 

In ESR, the lineshape gives a direct information about $T_2$ and $T_2^{\ast}$ only: most often the lineshape is due to $1/\gamma T_2^{\ast}$ and the much smaller $1/\gamma T_2$ remains hidden. The literature uses the concept of spin-packets \cite{AbragamBook, SlichterBook}, which corresponds to parts of the lineshape whose broadening is homogeneous, i.e., it is due to $1/\gamma T_2$. The conventional steady-state (textbook) solution of the Bloch equations is valid for a single spin-packet and the effects related to $T_2^{\ast}$ has to obtained by a convolution of the individual spin-packet lineshapes with the additional broadening due to $T_2^{\ast}$. For a lineshape, whose width is dominated by $T_2^{\ast}$ related effects, the literature uses the concept of inhomogeneous broadening. The corresponding linewidth is $\Delta B_\text{inhom}=1/\gamma T_2^{\ast}$. For a line where the true $T_2$ related broadening is observed, the concept of homogeneous broadening is used with $\Delta B_\text{hom}=1/\gamma T_2$. As we discussed in the main manuscript, a proper graphite sample made of single crystal graphite shows negligible inhomogeneous broadening and its linewidth is dominated by homogeneous relaxation effects.

Without the loss of generality, we can assume that the $T_2^{\ast}$ related broadening also results in a Lorentzian lineshape. Then, the two linewidths are simply additive, and the observed effective linewidth is $\Delta B_\text{eff}=\Delta B_\text{inhom}+\Delta B_\text{hom}$. For a more complicated case, e.g., where $T_2^{\ast}$ gives rise to a Gaussian lineshape, one has to consider the resulting Voigtian function (which is a convolution of Lorentzian and Gaussian functions) where the rules of the individual line broadening are different. 

As mentioned, in continuous-wave ESR (or cw-ESR), $T_1$ cannot be observed from the lineshape. However, its effect can be observed from saturation ESR experiments. The solution of the Bloch equations \cite{AbragamBook, SlichterBook} yields the ESR intensity, $I$, and the homogeneous linewidth changes as a function of the exciting microwave field strength, $B_1$ as:
\begin{gather}
    I(B_1)=\widetilde{I_0} \frac{B_1}{\sqrt{1+\gamma^2 B_1^2 T_1 T_2}},\label{SaturatedIntensitySM}\\
    \Delta B_\text{hom}(B_1)=\Delta B_\text{hom,0}\sqrt{1+\gamma^2 B_1^2 T_1 T_2},\label{SaturatedWidthSM}
\end{gather}
where $\widetilde{I_0}$ is the initial slope of the $B_1$ dependent ESR signal intensity and $\Delta B_\text{hom,0}$ is the linewidth in the $B_1\rightarrow 0$ limit. 

A proper treatise of the inhomogeneous broadening on the saturation ESR experiments is given in the seminal paper of Portis \cite{PortisPR1953}. It turns out that the ESR intensity drop is always present; however, a strong inhomogeneous broadening may smear out the line-broadening effect. In the case of the latter, the observed effective linewidth is:
\begin{equation}
    \Delta B_\text{eff}(B_1)=\Delta B_\text{inhom}+\Delta B_\text{hom,0}\sqrt{1+\gamma^2 B_1^2 T_1 T_2}.
\end{equation}

\subsection{Analysis of the power-dependent linewidth}

\begin{figure}[!ht]
	\begin{centering}
    \includegraphics[width=\linewidth]{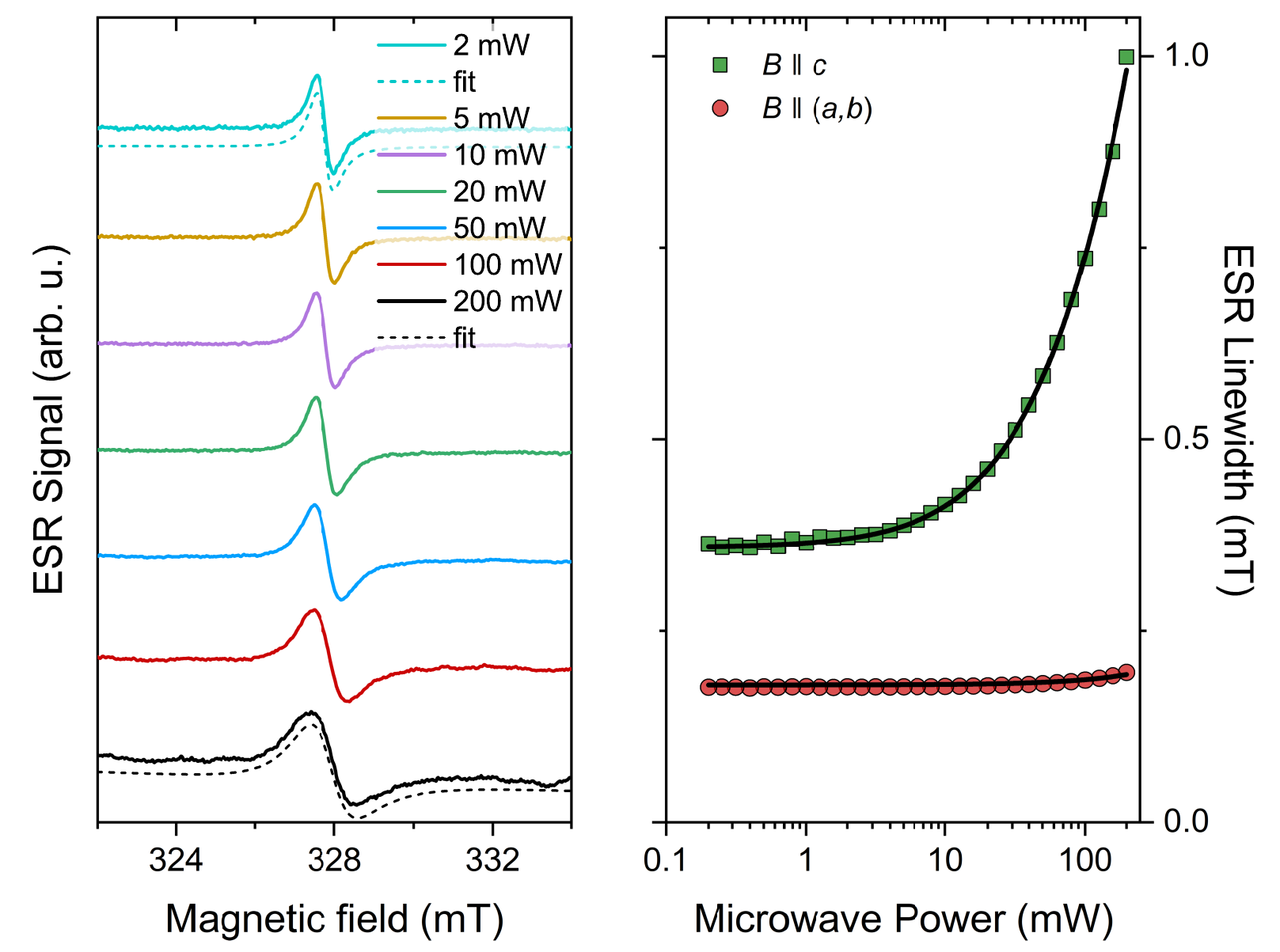}
	\caption{Left panel: The individual ESR spectra for the $B \parallel c$ configuration for a few selected powers at room temperature. The amplitude of the spectra are normalized and the spectra are offset for clarity. Dashed lines show the Lorentzian fits for the lowest and highest power spectra. Right panel: The ESR linewidth for the two orientations of the magnetic field as a function of the microwave power. A fit to the data is shown with solid line as explained in the text.}
	\label{FigSM_broadening}
	\end{centering}
\end{figure}

In Fig. \ref{FigSM_broadening}, we show the power-dependent ESR spectra when $B \parallel c$. The right panel shows the linewidths as a function of the microwave power for both orientation along with fits. To fit the data, we used the following formula:
\begin{equation}
    \Delta B\left(B_1\right)=\Delta B(p=0)\sqrt{1+\gamma^2 C\cdot p\cdot T_1 T_2}.
\end{equation}
Here, $p$ denotes the microwave power and $C$ is an instrument and microwave cavity dependent constant. The ESR instrument manufacturer gave a value of $C=0.2~\text{mT}/\sqrt{p}$ for the used "Bruker Super High Q" resonator, type ER4122SHQE for a standard quality factor of $Q=7{,}500$. The $Q$-factor of the microwave cavity was determined by conventional frequency sweep methods \cite{Klein1993, Donovan1993} and we obtained $Q=11{,}500$ (corresponding to $B_1=0.248~\text{mT}\cdot\sqrt{p[\text{W}]}$). Nevertheless, the exact magnitude of $B_1$ is somewhat uncertain, the presence or absence of a quartz insert may influence this. We in fact observed that the presence of a quartz insert in a flow-through liquid nitrogen system increases the microwave field; it focuses the microwave field, thus enhancing the local $B_1$ magnetic field around the sample \cite{chen2004microwave}. This led us to fit the $C$ constant as a free parameter in the main text, which improved the quality of the fit: the adjusted $R^2$ value increased from $0.992$ to $0.994$, while the change in $B_1$ was $10 \%$. We obtained that the power-to-microwave field conversion factor is $B_1=0.271(1)~\text{mT}\cdot\sqrt{p[\text{W}]}$.

In the fitting, we assume (following Ref. \onlinecite{MarkusNatComm}) that the linewidth is purely homogeneous (as shown below, our result strongly supports this) thus $\Delta B(p=0)=1/\gamma T_2$, which reduces the number of parameters. We introduce the notation of $\Delta B_0$ for the ESR linewidth when $B \parallel c$, i.e., $\Delta B_0=1/\gamma T_{2,c}$. Our model in Eqs. \eqref{AngDepRelaxationTimesRewritten1} and \eqref{AngDepRelaxationTimesRewritten} also gives that $1/\gamma T_{1,(a,b)}=\Delta B_0$ and $1/\gamma T_{2,(a,b)}=\frac{\Delta B_0+1/\gamma T_2}{2}$. 

Before proceeding, we discuss the magnitude of the $B_1$ in our measurement. The manufacturer specified that $B_1[\text{mT}]=0.2\sqrt{\frac{Q}{7{,}500}~p[\text{W}]}$, where $Q$ is the quality factor of the microwave cavity and $p$ is measured in Watts. The reference value of $B_1~[\text{mT}]=0.2\sqrt{p[\text{W}]}$ is set for $Q=7{,}500$. The $Q$-factor of the cavity determined by the built-in method of the instrument was $Q=9{,}800$, giving $B_1[\text{mT}]=0.2286\sqrt{ p[\text{W}]}$. However, we set the conversion factor as a free parameter in the fits and set: $B_1[\text{mT}]=C\sqrt{p[\text{W}]}$.

Altogether, we could perform a global fit (or simultaneous fit) for both magnetic-field orientations using the following formula:
\begin{align}
\begin{split}
    \Delta B(B {\parallel} c)&=\Delta B_0\sqrt{1+C\cdot p \cdot\gamma T_{1,c}/\Delta B_0},\\
    \Delta B(B {\parallel} a,b)&=\frac{\Delta B_0+\frac{1}{\gamma T_{1,c}}}{2} \sqrt{1+\frac{2 C\cdot p}{\Delta B_0\left(\Delta B_0+1/\gamma T_{1,c} \right)}},
\end{split}
\end{align}
which contains only three free parameters: $C$, $\Delta B_0$, and $T_{1,c}$. In the fits $\Delta B_0$ always turns out to be a robust parameter which is set the by zero-power linewidth and $\Delta B_0=0.358(1)~\text{mT}$. In fact $T_{1,c}$ is the desired parameter, i.e., the spin-lattice relaxation time when $B \parallel c$. In the fits, we obtain $C=0.224(20)~\text{mT}/\sqrt{p[\text{W}]}$, which is remarkably close to the manufacturer specified value but is different from the value obtained in the main text. For the spin-relaxation time, we obtain $T_{1,c}=1{,}300(250)~\text{ns}$ with the adjusted $R^2$ value being $0.9991$.

We note that the $T_{1,c}$ value corresponds to about $1/(\gamma T_{1,c})=\Delta B_{T_1}\approx4.3~\mu\text{T}$ thus it is about $100$ times smaller than $\Delta B_0$, it can therefore be even neglected in the above equations, simplifying to:
\begin{align}
\begin{split}
    \Delta B(B {\parallel} c)&=\Delta B_0\sqrt{1+C\cdot p \cdot\gamma T_1/\Delta B_0},\\
    \Delta B(B {\parallel} a,b)&=\frac{\Delta B_0}{2} \sqrt{1+\frac{2 C\cdot p}{\Delta B_0^2}}.
    \end{split}
\end{align}

\subsection{The power- and angular-dependent ESR intensity}

\begin{figure}[!ht]
	\begin{centering}
    \includegraphics[width=\linewidth]{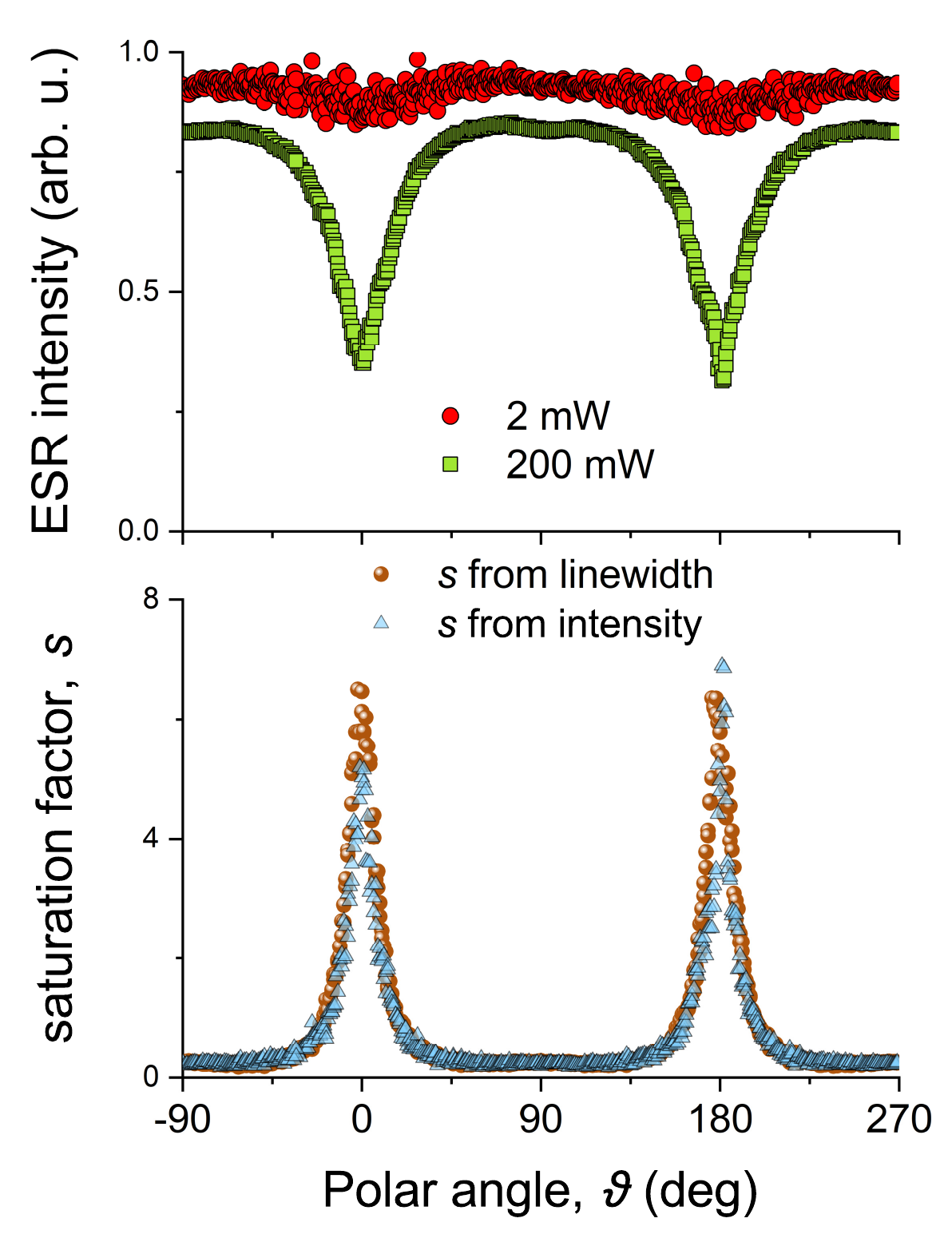}
	\caption{Top pane: the angular dependent ESR intensity in graphite for two microwave powers. The data shown is normalized by the square root of the power, i.e., by $B_1$. The effect of saturation is clearly observed when $B \parallel c$. Bottom panel: Comparison of the saturation factor, $s$, as obtained from the intensity and linewidth data, using Eqs. \eqref{SaturatedIntensitySM} and \eqref{SaturatedWidthSM}, respectively. Note the good agreement between the two types of determinations.}
	\label{FigSM_IntensitySaturation}
	\end{centering}
\end{figure}

We show the angular-dependent ESR intensity in Fig. \ref{FigSM_IntensitySaturation} (top panel) for two different power values after normalization with the square root of the power, by $B_1$. Eq. \eqref{SaturatedIntensitySM} shows that after this normalization, the saturation factor, $s$ can be obtained similarly to that from the linewidth data using Eq. \eqref{SaturatedWidthSM}. The angular-dependent ESR intensity also attests the ultralong $T_1$ when $B\parallel c$, similar to the linewidth. The comparison between the two types of data is shown in Fig. \ref{FigSM_IntensitySaturation} (bottom panel). Clearly the two types of data shows a good overall agreement even though we believe that the linewidth determination is more accurate as it is a spectral parameter rather than the intensity which depends on various factors including how well the ESR microwave bridge can be balanced, the linearity of the detecting mixer etc.

\section{The effect of mosaicity on the angular dependent ESR linewidth}

Due to the large $g$-factor anisotropy of graphite, the orientation of the planes with respect to the magnetic field plays an important role. There are two typical sources of commercial HOPG samples: SPI-1/2/3 and ZYA/ZYB/ZYC. The manufacturer provided values of the mosaicity are quote similar respectively. The quantity "mosaic spread angle" is given, however, we could not find an accurate definition. We believe that the mosaicity angle should inevitably follow a Gaussian distribution, whose standard deviation is proportional to the mosaic spread angles. The producer supplier values are as follows. For SPI-1 or ZYA: $0.4^\circ \pm 0.1^\circ$, for SPI-2 or ZYB: $0.8^\circ \pm 0.2^\circ$, and for SPI-3 or ZYC: $3.5^\circ \pm 1.5^\circ$.

The mosaicity affects the observed ESR linewidth, broadening it inhomogeneously: 
crystallites which have a slightly differing angle with respect to the external magnetic field have a resonance lying at different positions due to the $g$-factor anisotropy.

\begin{figure*}[!ht]
		\begin{centering}
        \includegraphics[width=\textwidth]{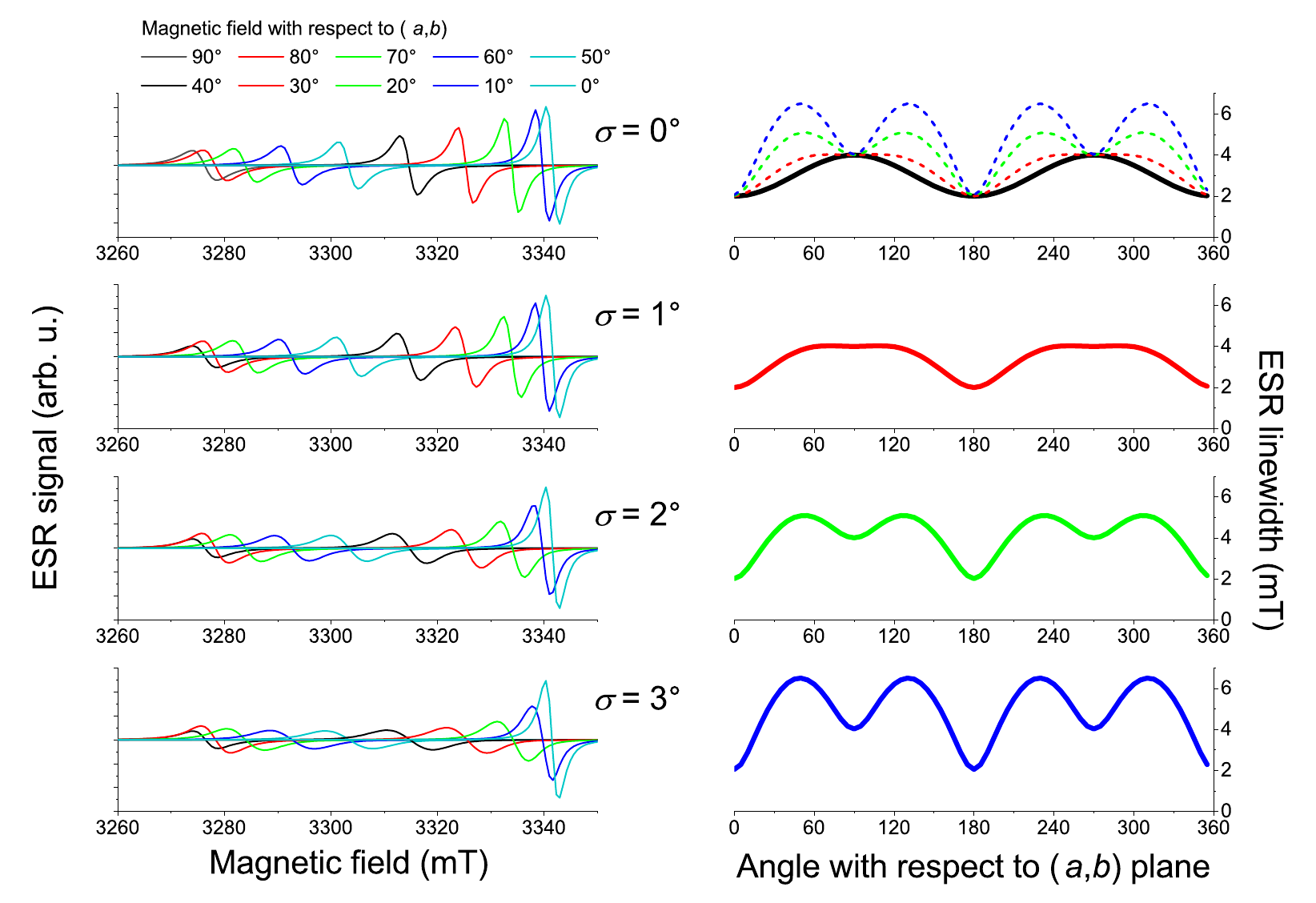}
        \caption{Effect of mosaicity on the angular dependent ESR spectra. Left panels: the individual spectra for the various levels of mosaicity. Right panels: the angular dependent linewidth as obtained from fitting to the individual spectra. The top right corner also shows all data with dashed curves for a better comparison.}
	  \label{FigSM_mosaicity}
    	\end{centering}
\end{figure*}

In Fig. \ref{FigSM_mosaicity} we show the simulated angular-dependent ESR linewidth for various levels of mosaicity. In the calculation, we assumed a sinusoidal dependence of the homogeneous (or $T_2$ related) linewidth with values $\Delta B_{(a,b)}=0.2~\text{mT}$ and $\Delta B_c=0.4~\text{mT}$ and a uniaxial $g$-factor anisotropy with $g_{(a,b)}=2.005$ and $\Delta B_c=2.045$ in agreement with Ref. \onlinecite{MarkusNatComm}. For an arbitrary angle, the effective $g$-factor reads:
\begin{equation}
    g_\text{eff}=\sqrt{g_c^2 \cos^2\vartheta+g_{a,b}^2 \sin^2\vartheta},
\end{equation}
where $\vartheta$ is the polar angle measured with respect to the $c$ axis. The mosaicity was accounted for by considering a Gaussian weight distribution of the mosaicity compared to a perfect alignment of the crystallites along the $c$ axis. (In fact, HOPG stands for Highly Oriented Pyrolytic Graphite). The standard deviation, or $\sigma$ parameter of the Gaussian distribution varies: a low value of $\sigma$ expresses a better aligned sample, i.e., a lower mosaicity.

In this approach, individual ESR spectra were simulated for the various levels of mosaicity and for a given angle of the external magnetic field with respect to the $(a,b)$ plane. It essentially involves a numerical convolution of the Gaussian distribution function with individual derivative Lorentzian functions, whose line position and linewidth are determined by the angle of the given crystallite with respect to the external magnetic field. Then, these individual spectra were fitted with derivative Lorentzian functions to obtain the linewidth. We found that for the experimentally relevant levels of mosaicity, the Lorentzian fits were appropriate, even though for high mosaicity one expects a deviation from this.

\begin{figure}[!htb]
		\begin{centering}
        \includegraphics[width=.9\linewidth]{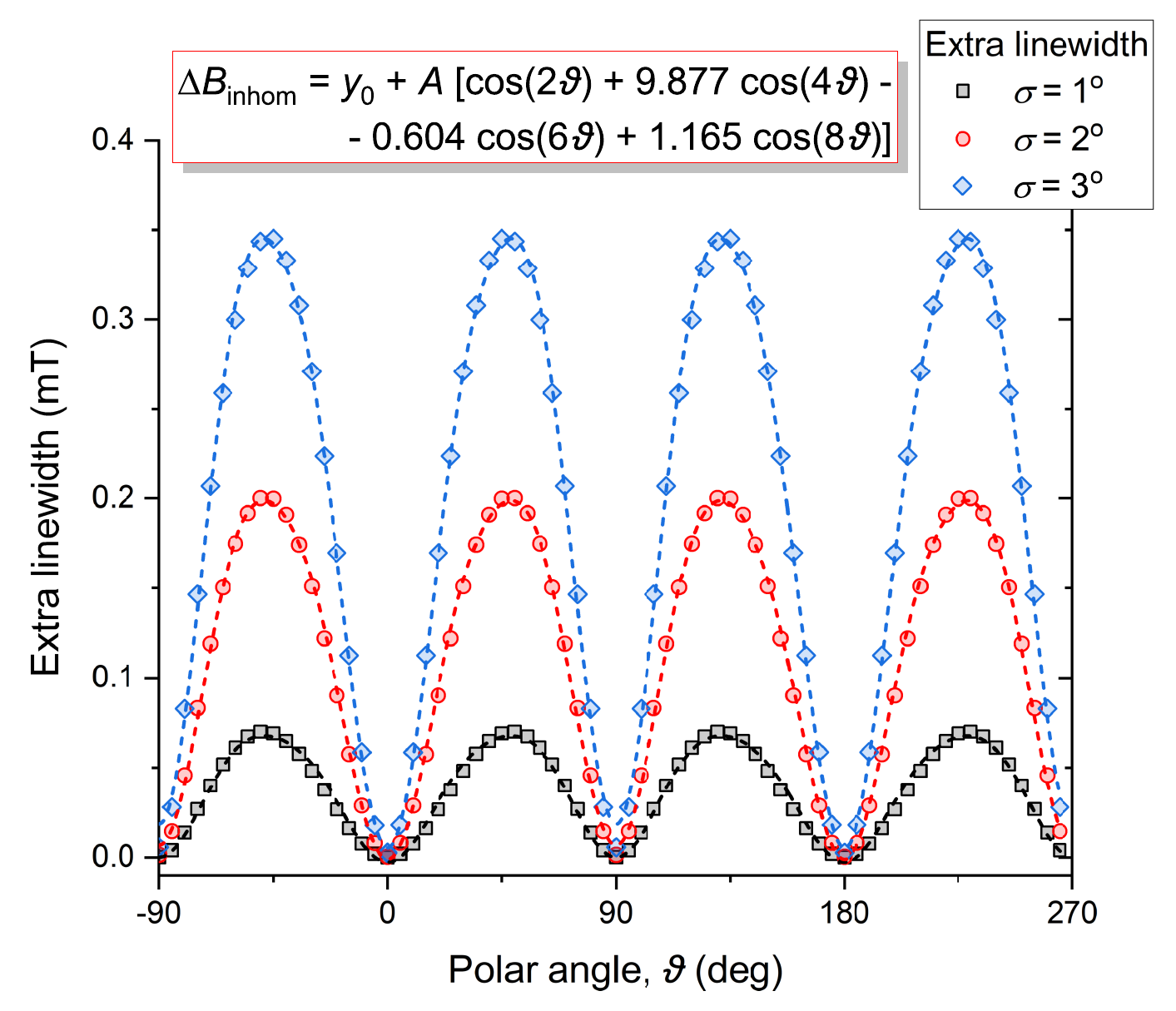}
        \caption{Extra broadening due to the mosaicity for the three considered distribution width. Dashed lines are a harmonic series fit to the data with the given expression.}
	  \label{FigSM_mosaicity_extra_broad}
    	\end{centering}
\end{figure}

We found that the mosaicity-induced broadening can be well described by a harmonic sum of components. The number of harmonics was carefully considered and it is terminated at the point where adding further components does not improve the fit. We found that this function fits the extra, inhomogeneous linewidth:
\begin{equation}
    \begin{split}
        \Delta B_\text{inhom}=y_0+A\left[\cos \left(2\vartheta \right)+9.877\cdot \cos\left(4\vartheta\right) \right. \\ 
        \left. -0.604\cdot \cos\left(6\vartheta \right)+1.165\cdot \cos\left(8\vartheta \right) \right]
    \end{split}
\end{equation}

In Fig. \ref{FigSM_mosaicity_extra_broad} we show the fitted inhomogeneous linewidth data from the mosaicity modeling with the harmonic series. For each data set, the $y_0$ and $A$ parameters are different, and these depend on the level of mosaicity as summarized in Table \ref{SM_mosaicity_table}.

\begingroup
\setlength{\tabcolsep}{8pt} 
\renewcommand{\arraystretch}{1.2} 
\begin{table}
    \begin{tabular}{c c c} 
        \hline
        $\sigma$ ($^\circ$) & $y_0$ & $A$ \\ [.5ex]
        \hline \hline
        $1$ & $0.03691$ & $-0.00345$ \\ 
        $2$ & $0.11130$ & $-0.00983$ \\
        $3$ & $0.19651$ & $-0.01673$ \\
        \hline
    \end{tabular}
    \caption{The $y_0$ and $A$ parameters as obtained from the fits for various values of the mosaicity.}
    \label{SM_mosaicity_table}
\end{table}
\endgroup

\section{Numerical solutions of the Bloch equations and simulation of phase sensitive cw ESR data}

\subsection{Scaling of the Bloch equations}
The Bloch equations read in the laboratory frame of reference as:
\begin{align}
    \frac{\mathrm{d}M_x}{\mathrm{d}t}&=\gamma\left(\mathbf{M} \times \mathbf{B} \right)_x-\frac{M_x}{T_2},\\
    \frac{\mathrm{d}M_y}{\mathrm{d}t}&=\gamma\left(\mathbf{M} \times \mathbf{B} \right)_y-\frac{M_y}{T_2},\\
    \frac{\mathrm{d}M_z}{\mathrm{d}t}&=\gamma\left(\mathbf{M} \times \mathbf{B} \right)_z-\frac{M_z-M_0}{T_1},
\end{align}
where $\gamma$ is the electron gyromagnetic ratio and is $\gamma/2\pi\approx 28.0~\text{GHz}/\text{T}$. $\mathbf{M}$ and $\mathbf{B}$ denote the vectors of the magnetization and the magnetic field, respectively. $M_0$ is the steady-state magnetization along the $z$-axis.

As it is conventional in magnetic resonance, we decompose the magnetic field to a static component with magnitude $B_0$ along the laboratory $z$-axis and an AC component with amplitude $B_1$ rotating around the $z$-axis with angular frequency $\omega$. Then we transform the Bloch equations from the $(x,y,z)$ laboratory system to the $(x',y',z')$ rotating coordinate system, which rotates together with the $B_1$. In the rotating frame of reference, let $B_1$ always point to the $x'$ direction. The Bloch equations thus read:
\begin{align}
    \frac{\mathrm{d}M_x'}{\mathrm{d}t}&=\gamma M_{y'} b_0-\frac{M_{x'}}{T_2},\\
    \frac{\mathrm{d}M_y'}{\mathrm{d}t}&=\gamma\left(M_{z'}B_1-M_{x'}b_0 \right)-\frac{M_{y'}}{T_2},\\
    \frac{\mathrm{d}M_z'}{\mathrm{d}t}&=-\gamma M_{y'} B_1 +\frac{M_0-M_{z'}}{T_1},
\end{align}
where we introduced $b_0=B_0-\omega/\gamma$ and note that the $z$ and $z'$-directions are equivalent. 

One has to avoid too large time steps for an efficient and accurate numerical solution of the Bloch equations. In addition, it is more convenient and is adapted to our particular problem to measure the magnetic field and linewidth in Gaussian (or c.g.s.) units ($1~\text{G}=0.1$ mT). In principle, it would be more precise to use Oe to characterize the magnetic field which is created in the electromagnet. In c.g.s units, Oe and G refer to the same dimensions but their use refers to a clear distinction. The Gauss unit refers to the internal magnetic fields in the presence of local dipolar fields, which are however both unknown and uncontrolled in a sample. In contrast, Oe refers to the magnetic field excited by free-flowing currents, i.e., the field generated by the coil. This terminology is properly followed by the US manufacturers, however, the manufacturer of the most often used ESR instrument (Bruker) employs G.

With all this in mind, we remind that the gyromagnetic ratio is: $\gamma=2\pi \cdot28~\text{GHz}/\text{T}$. If we measure time in microseconds, it corresponds to a $t'=10^6 \cdot t$ rescaling, where $t$ is the original time and $t'$ is time measured in microseconds. Using $\frac{\circ}{\mathrm{d}t}= \frac{\circ}{\mathrm{d}t'}\cdot \frac{\mathrm{d}t'}{\mathrm{d}t}=10^6\cdot \frac{\circ}{\mathrm{d}t'}$ we obtain for the Bloch equations which are rescaled to a numerically well-suited case:
\begin{align}
    \frac{\mathrm{d}M_x'}{\mathrm{d}t'}&=\gamma'\left[ M_{y'} b_0-\Delta B_{T_2}M_{x'}\right], \\
    \frac{\mathrm{d}M_y'}{\mathrm{d}t'}&=\gamma' \left[\left(M_{z'}B_1-M_{x'}b_0 \right)-\Delta B_{T_2} M_{y'}\right], \\
    \frac{\mathrm{d}M_z'}{\mathrm{d}t'}&=-\gamma' \left[M_{y'} B_1 +\Delta B_{T_1}\left(M_0-M_{z'}\right)\right],
\label{Bloch_numerical_rescaled}
\end{align}
where we have introduced the notations $\Delta B_{T_2}=1/\gamma T_2$ and $\Delta B_{T_1}=1/\gamma T_1$. Both quantities have dimensions of Gauss, and $\Delta B_{T_2}$ expresses the linewidth of the spin-packets \cite{SlichterBook, AbragamBook, PortisPR1953}. To our knowledge, $\Delta B_{T_1}$ has no direct and simple physical interpretation, except that it would be the observed linewidth when $T_1=T_2$. 

This transformation also means that the frequency of external modulations (e.g., magnetic field modulation of amplitude modulation of the $B_1$ field) has to be rescaled. For example, a magnetic field modulation with a frequency of $100$~kHz (a common value in ESR experiments) has to be entered into the above equations as $B_\text{m}=B_\text{m,0} \cos\left(2\pi\cdot 0.1 \cdot t' \right)$.

\subsection{Comparison of experimental and simulated phase-sensitive data}

\begin{figure}[!ht]
		\begin{centering}
        \includegraphics[width=\linewidth]{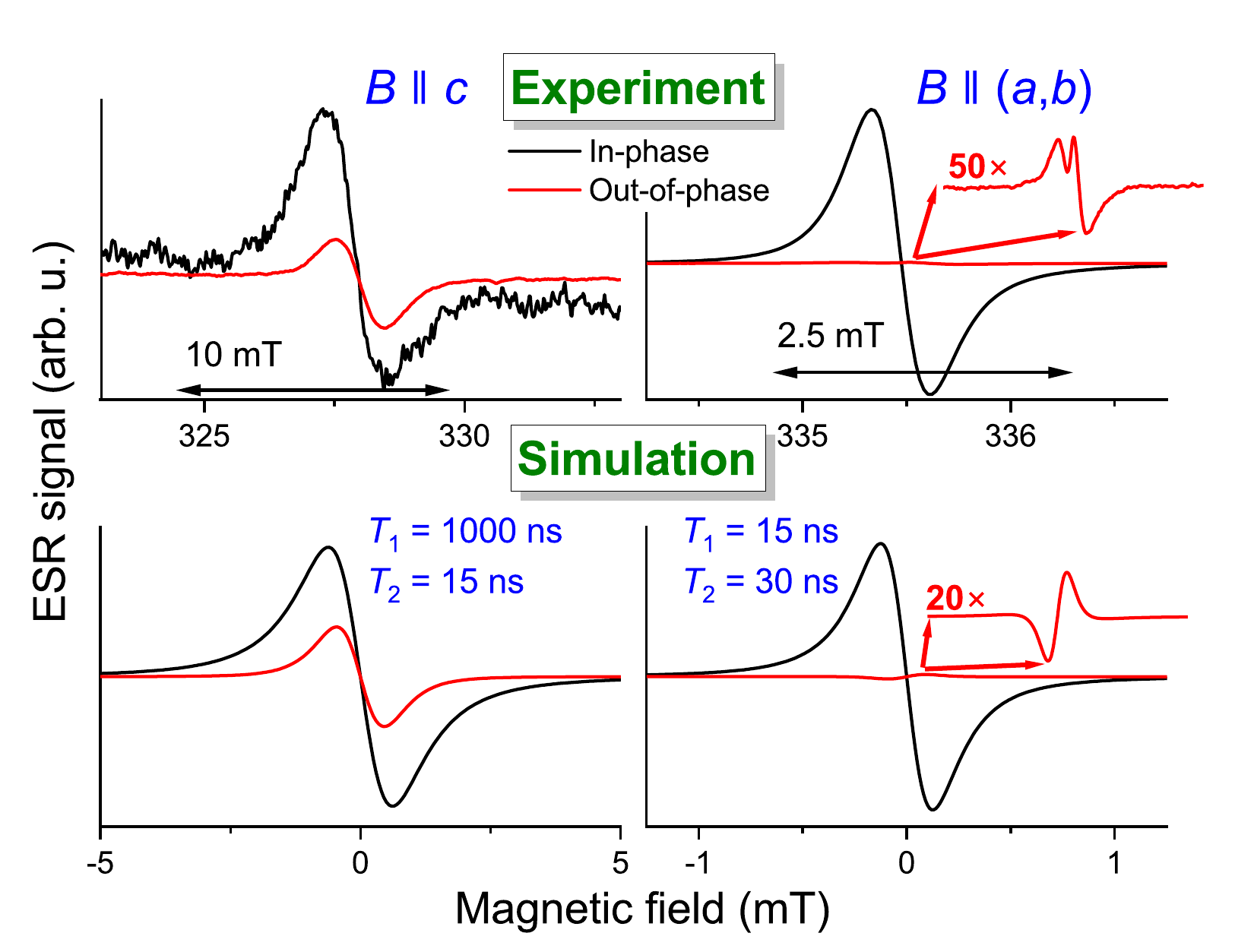}
        \caption{Phase sensitive continuous-wave ESR spectra at $200$ mW, i.e., the saturated ESR limit. Top panels: Experiment, bottom panels: Bloch quations-based simulation results. For $B\parallel c$ a significant out-of-phase component is observed, which is simulated well assuming a long $T_1$. For $B\parallel (a,b)$, the out-of-phase signal is small, which is reproduced well with a much shorter $T_1$. Note that the zoomed in data have different magnification in the experiment and simulation. }
	  \label{FigSM_PhaseInformation}
    	\end{centering}
\end{figure}

The above-described numerical solution allows us to obtain the harmonics of the modulation-detected ESR experiments in a phase sensitive manner. Here, we focus on the phase information in the first harmonic as this is the strongest. In Fig. \ref{FigSM_PhaseInformation}, we show the phase sensitive ESR spectra for the two orientations of the magnetic field using a high microwave power, $200$ mW. For $B\parallel c$, a significant out-of-phase component is observed (shown with red curve). This is reproduced well assuming a long $T_1$ as shown in the figure. In the simulated spectrum, we used $T_1=1{,}000~\text{ns}$ and $T_2=15~\text{ns}$ and employed the calculation outlined in the previous section. the other experimental parameters (magnetic field modulation amplitude and frequency) matched the respective experimental values ($0.05$ mT and $100$ kHz, respectively). For $B\parallel (a,b)$, the out-of-phase signal is much small (its amplitude is about $50$ times reduced). This negligible out-of-phase component is compatible with a short $T_1$ as shown in the simulated result. 

\begin{figure}[!ht]
		\begin{centering}
        \includegraphics[width=\linewidth]{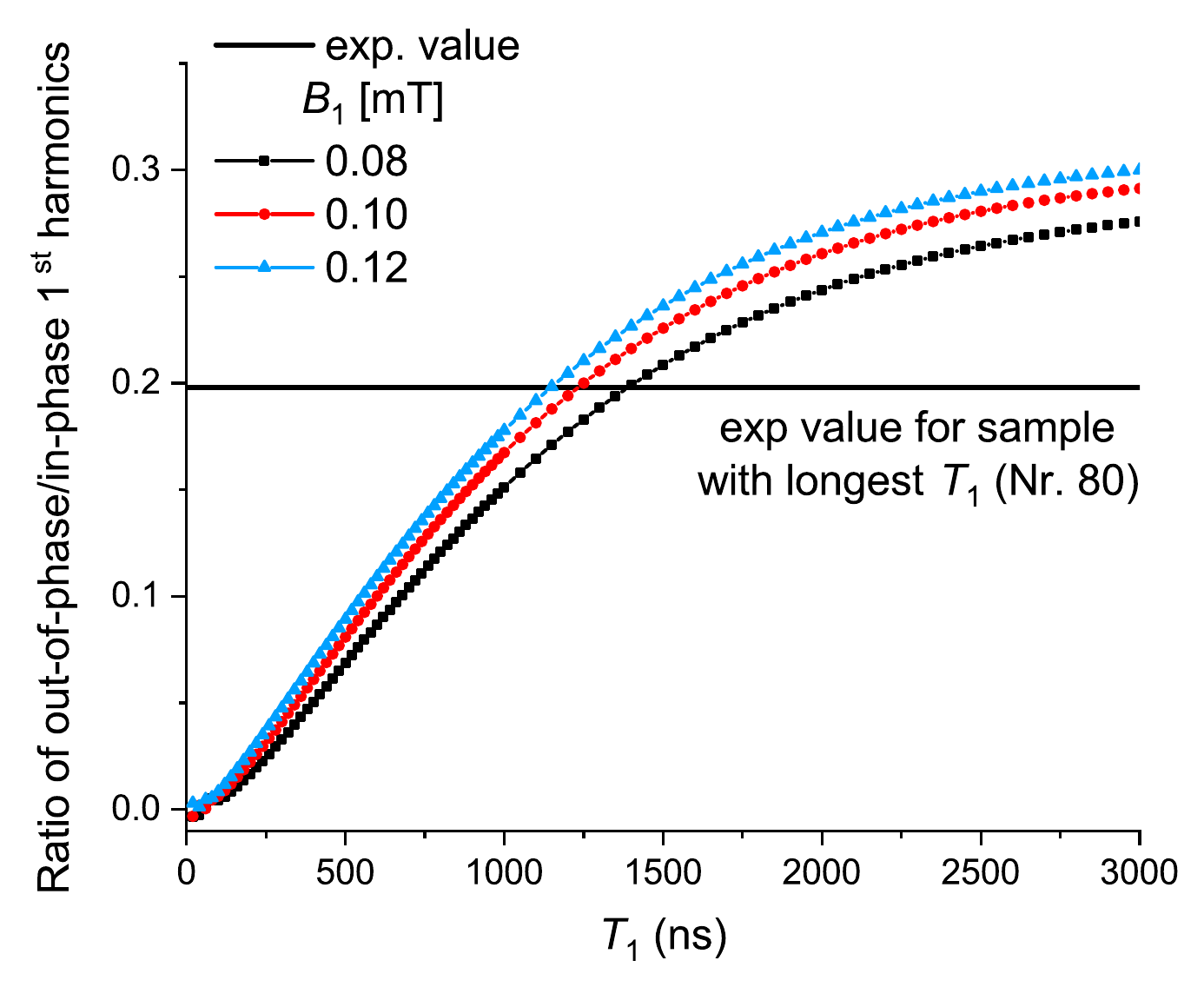}
        \caption{Calculated ratio of the out-of-phase to the in-phase cw ESR data as a function of $T_1$ for three different values of the $B_1$ amplitude. Horizontal black line is the experimental value for Sample Nr. 80. The ratio appears to be smaller than that suggested by \ref{FigSM_PhaseInformation}, however, the linewidth of the out-of-phase signal is smaller than that of the in-phase, which explains the difference. Note that the vertical scale starts from negative values as for very short $T_1$, the out-of-phase signal is small but inverted.}
	  \label{FigSM_PhaseInformation_vs_T1}
    	\end{centering}
\end{figure}

We also studied the dependence of the ratio of the two channels as a function of $T_1$ for various values of $B_1$ and the result is shown in Fig. \ref{FigSM_PhaseInformation_vs_T1}. The experimentally determined ratio is also shown with a horizontal black line. The three different values of $B_1$ correspond to different conversion factors in $B_1=C\cdot\sqrt{p[\text{W}]}$. Given that $p=0.2$~W is the highest microwave power in our case, the $B_1=0.08~\text{mT}$ corresponds to $C=0.18~\text{mT/W}$ which is about $10 \%$ lower than the factory-based conversion factor that is $C=0.2~\text{mT/W}$. The $B_1=0.12~\text{mT}$ corresponds to $C=0.27~\text{mT/W}$, which was found in the fits as discussed in the main text. 

The data shown in the figure have several consequences: first, it demonstrates that the ratio of the two channels is a sensitive function of $T_1$: it practically vanishes for $T_1\lesssim 100~\text{ns}$. The experimentally observed ratio is reproduced well with the $C=0.27~\text{mT/W}$ power-to-field conversion ratio and $T_1\approx 1000~\text{ns}$. The scaling also provides a relatively straightforward recipe for a $T_1$ determination on future samples without the need for detailed angular dependent measurement of the saturation factor.

\section{$T_1$ statistics on the samples}

\begin{figure}[!ht]
		\begin{centering}
        \includegraphics[width=\linewidth]{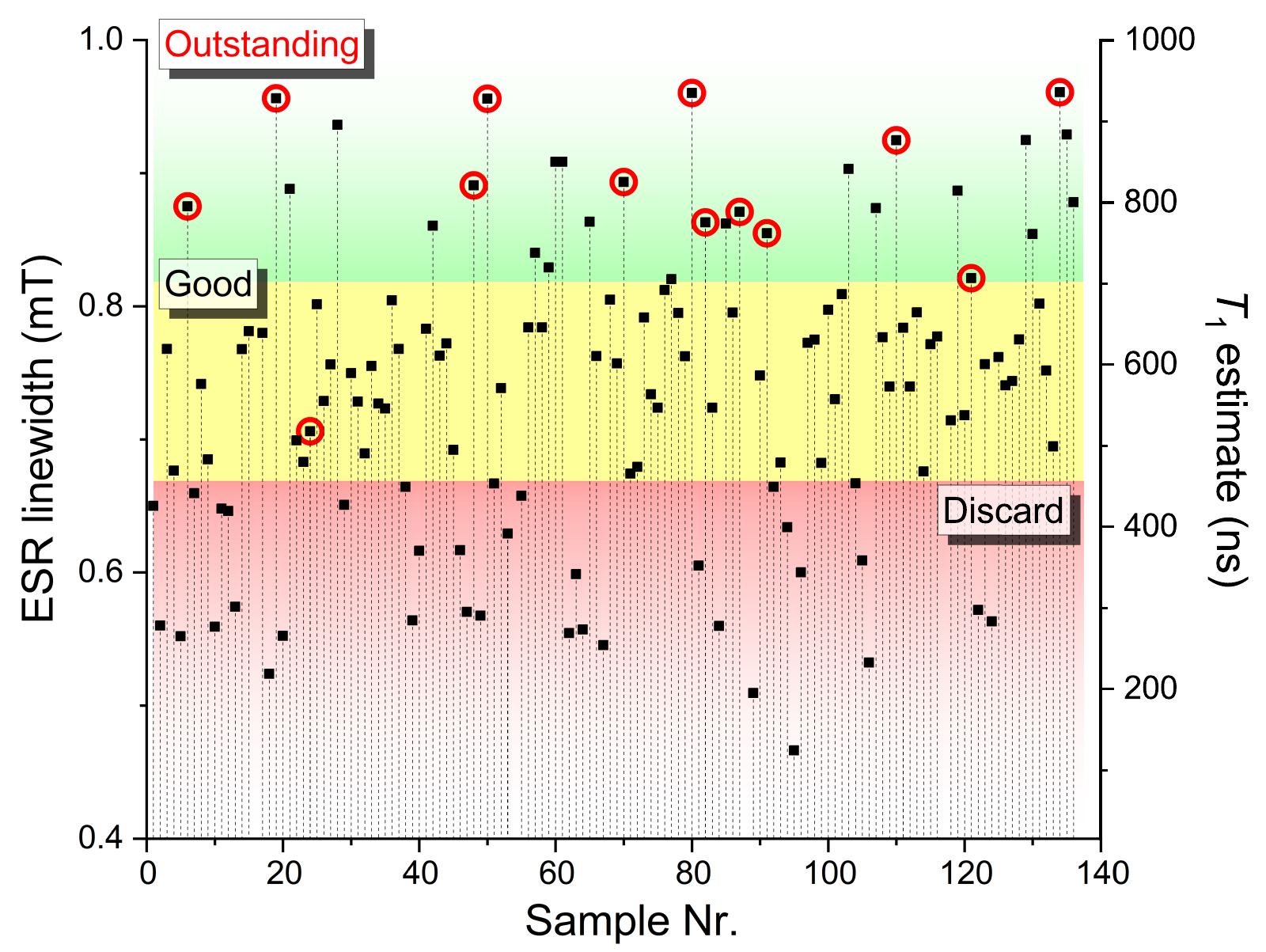}
        \caption{Statistics on the prepared samples based on the ESR linewidth as measured for $B \parallel c$ and the maximum microwave power. The samples were subjectively categorized into "Outstanding", "Good", and "Discard" quality piles according to the linewidth. The right hand scale shows an estimate on $T_1$, however it is much less accurate the studying and fitting to the full angular dependent data. The sample, which were studied in detail, are marked with a red circle.}
	  \label{FigSM_Sample_stat}
    	\end{centering}
\end{figure}

In total, we prepared more than 100 samples out of the "NGS Flaggy flakes" sample and the ESR linewidth result is shown in Fig. \ref{FigSM_Sample_stat}. The preparation method was that described in the Methods section in the main manuscript. The ESR linewidth for the samples using the highest microwave power, $0.2$ W, was studied for all of them, when $B\parallel c$. Depending on the ESR linewidth, we subjectively categorized these into "Outstanding", "Good", and "Discard" qualities. About a quarter of the samples fell into the "Outstanding" and "Discard" categories, while a half of them were "Good". The ESR linewidth also allowed us to estimate the $T_1$ values (shown on the right-hand scale in the figure). 

The ESR intensity also showed a variation, therefore we studied samples from the "Outstanding" category further which showed a reasonable linewidth. Unfortunately most "Outstanding" samples had a relatively small ESR intensity, such that signal-to-noise ratio was around $3$ for a $1$ minute ESR measurement (for a single orientation). We studied the detailed angular-dependent ESR linewidth for samples Nr. 6, 19, 24, 48, 50, 70, 80, 82, 87, 91, 110, 121, and 134. The major effect, i.e., a strongly angular dependent line broadening, indicating the presence of a long $T_1$ for $B\parallel c$ was present for all samples with small variations of the broadening value, the residual linewidth and the level of mosaicity. Most "Outstanding" and "Good" samples have a nearly symmetric ESR lineshape, indicating that the sample is thin enough that the microwave penetrates fully into these. This is analyzed in detail in the next section. 

\section{Sample thickness estimate from the lineshape}

The ESR signal lineshape is very sensitive to details of the microwave penetration depth and the sample thickness \cite{Dyson}. In fact, this dependence is so strong that it allowed to estimate the in-plane resistivity based on the ESR lineshape in the work of Walmsley and coworkers \cite{WalmsleyPRB1992, WalmsleySynthMet1989}. Graphite is characterized by a very anisotropic conductivity: the in-plane conductivity at room temperature is around $\sigma_{(a,b)}=1 \dots 3~\cdot~10^{4}~\text{S/cm}$ \cite{SouleMcClure,BrandtBook}, whereas the out-of-plane conductivity is about $10,000$ times smaller. Due to this very anisotropic conductivity, the microwave penetration depth is determined by the higher conductivity \cite{WalmsleyPRB1992,WalmsleySynthMet1989}, i.e., this enters into: $\delta_{\text{skin}}=\sqrt{\frac{2}{\sigma\omega\mu_0}}$ (it is important not to confuse $\delta_{\text{skin}}$ with the diffusion length parameters used elsewhere in this paper). Here, $\omega$ is the angular frequency of the radiation ($\omega=2\pi\cdot 9.4~\text{GHz}$) and $\mu_0\approx 1.256\cdot 10^{-6}~\text{Vs/Am}$ is the permeability of the vacuum. This gives $\lambda \approx 3~\mu\text{m}$. 

The typical size of our samples were $1\dots3\times1 \dots3~{\text{mm}\times\text{mm}}$, with varying combinations of the two lateral dimensions, and thickness below about $50~\mu\text{m}$ but in any case these were. The thickness was estimated with the help of profilometry and photography and also from the fact that the HOPG disk samples have a manufacturer specified thickness of $70~\mu\text{m}$ which represented the thick limit as discussed below. 

Dyson discusses in his seminal paper \cite{Dyson} that the relevant parameters are: $T_2$, $T_{\text{diff}}$, and the $\lambda=d/\delta_{\text{skin}}$ ratio. Samples, where $\lambda\gg1$ are usually considered as "thick" and where $\lambda\ll1$ as "thin". $T_{\text{D}}$ denotes the time it takes for the electrons to diffuse through the skin-depth. In our case, the relatively small $D_c$ diffusion constant enters in the calculation of $T_{\text{diff}}$, namely: 
$\delta_{\text{skin}}=\sqrt{D_c T_{\text{diff}}}$. Using the mean value for $D_c=0.06~\text{cm}^2/\text{s}$, we obtain $T_{\text{diff}}=1.5~\mu\text{s}$. In fact, Dyson introduced the parameter $R=\sqrt{\frac{T_{\text{diff}}}{T_2}}$, systems where $R\gg1$ are usually considered as having "slow" spin diffusion (this is also known as the "NMR-limit", as nuclei are fully stationary) and where $R\ll1$, as having a "rapid" spin diffusion.

With the above values of the skin-depth diffusion time and $T_2$, we obtain for our case $R\approx 7$. Therefore graphite clearly belongs to the case of slow spin diffusion from the point of view of the Dysonian theory. The phase in the Dysonian signal in this limit arises due to the changing phase of the exciting electromagnetic wave across the skin depth. 

It is worth discussing why $T_2$ is the relevant timescale and not $T_1$; $T_2$ is the timescale on which the ESR signal, i.e., the $x,y$ plane magnetization in the Bloch sphere decays to zero without external excitation. Therefore, electrons, which diffuse into a solid, lose their magnetization on this timescale without further external excitation. 

We follow the result given in Ref. \onlinecite{Dyson} to establish the connection between the observed lineshapes and the sample properties. Dyson solves the problem of spin diffusion for a flat plate in the normal skin limit. The result is also given for the anomalous skin limit, which is realized in ultra-clean samples at high frequencies, i.e., when the electron mean-free path becomes larger than the skin depth. The absorption component of the magnetic field derivative ESR signal then reads:
\begin{equation}
    \frac{\mathrm{d}\chi'}{\mathrm{d}B}=A \cdot \Re\left(F(\lambda)^2 \frac{\mathrm{d}G(\lambda,R,B_0,B,\Delta B)}{\mathrm{d}B}  \right),
\end{equation}
where $A$ is a normalizing constant, which does not influence the lineshape. The appearing functions, $F$ and $G$, are:
\begin{align}
\begin{split}
    F&=-u \tan\left( u \right), \\
    G(B)&=\frac{\mathrm{i}}{\left(w^2-u^2\right)^2}\left[\frac{2u^2}{w\tan(w)}+\frac{w^2-3u^2}{u\tan(u)}+\frac{w^2-u^2}{\sin^2(u)} \right],
\end{split}
\end{align}
with
\begin{align}
\begin{split}
    u&=\frac{\lambda}{2}\left(1+\mathrm{i} \right),\\
    w&=\frac{\lambda R}{2}\left(\xi+\mathrm{i}\eta \right),\\
    \xi&=\sgn(x)\sqrt{\sqrt{1+x^2}-1},\\
    \eta&=\sqrt{\sqrt{1+x^2}+1},\\
    x&=\frac{B_0-B}{\Delta B},
\end{split}
\end{align}
where $B_0$ is the resonance field and $\Delta B$ is the ESR linewidth. We note that there is a minus sign in the equation for $x$ as compared to the usual $B-B_0$ usage as Dyson expressed his result as $x=(\omega-\omega_0)T_2$. 

\begin{figure}[!ht]
		\begin{centering}
        \includegraphics[width=.91\linewidth]{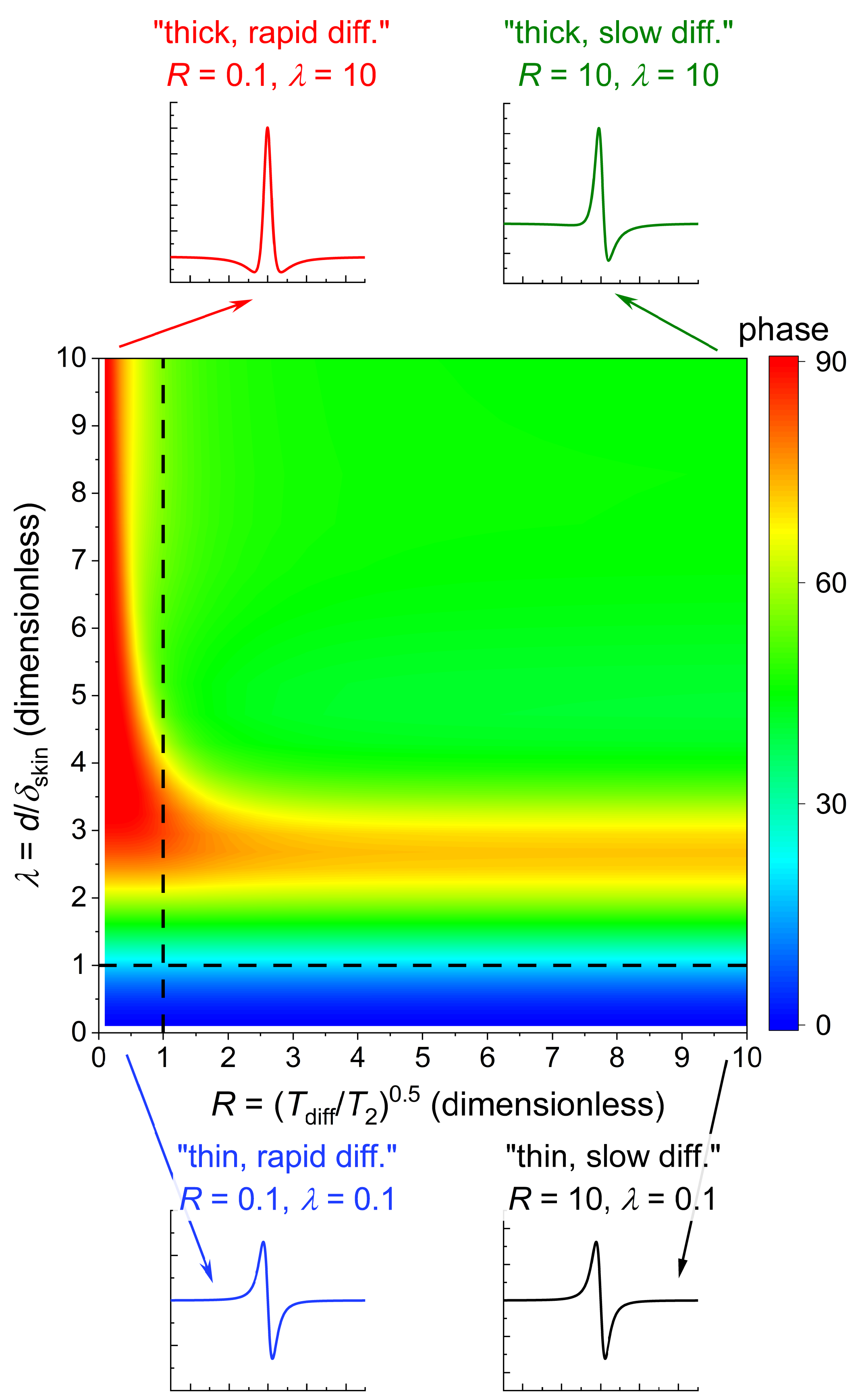}
        \caption{The phase of the Dysonian lineshape when fitted with a mixture of Lorentzian absorption-dispersion. Individual spectra are shown for a few particular values of $R$ and $\lambda$. Dashed vertical and horizontal lines are guides to the eye when $R=1$ and $\lambda=1$, respectively.}
	  \label{FigSM_Dyson}
	\end{centering}
\end{figure}

We generated the Dysonian lineshapes for various values of $R$ and $\lambda$ and fitted these with a mixture of derivative Lorentzian absorption-dispersion as follows:
\begin{align}
    \text{ESR-sig.}&=\cos\varphi\cdot\text{L-abs}_{\text{der}}(B)+\sin\varphi \cdot\text{L-disp}_{\text{der}}(B), \nonumber \\
    \text{L-abs}_{\text{der}}(B)&=\frac{-2}{\pi}\frac{b\Delta B}{\left(b^2+\Delta B^2 \right)^2}, \nonumber \\
    \text{L-disp}_{\text{der}}(B)&=\frac{-1}{\pi}\frac{b^2-\Delta B^2}{\left(b^2+\Delta B^2 \right)^2}, \nonumber \\
    b&=B-B_0.
\end{align}

A value of $\varphi$ close to $0$ means a nearly absorption Lorentzian derivative (realized when $\lambda \ll 1$), while $\varphi$ close to $90^\circ$ means a dispersion lineshape (e.g., for $R=0.1$ and $\lambda=10$). The result is shown in Fig. \ref{FigSM_Dyson} along with spectra for some particular combinations of $R$ and $\lambda$. Clearly, the $\varphi<20^\circ\dots~30^\circ$ is only realized when $\lambda$ is smaller than $1$.

\begin{figure}[!htb]
		\begin{centering}
        \includegraphics[width=.91\linewidth]{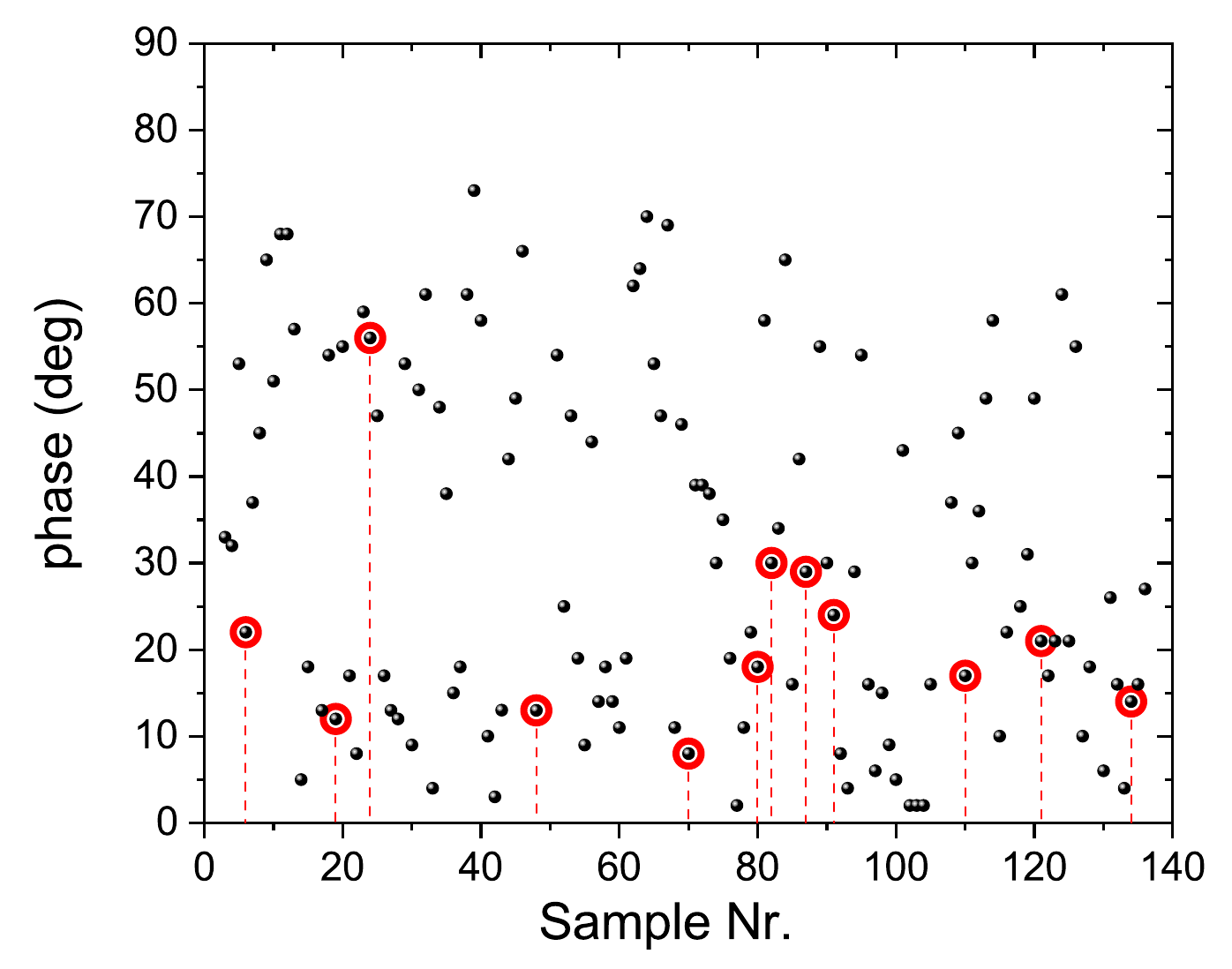}
        \caption{The fitted phase for the prepared samples. Red circles indicate those which were studied in detail.}
	  \label{FigSM_PhaseStat}
	\end{centering}
\end{figure}

In Fig. \ref{FigSM_PhaseStat}, we show the fitted phase values for all of the prepared samples. Red circles mark those which were studied in more detail. The presented data in the main manuscript was taken on Samples Nr. 80 (the angular-dependent linewidth values, $\varphi=18^\circ$), Nr. 121 and Nr 134 ($\varphi=21^\circ$ and $\varphi=14^\circ$, respectively). We can therefore conclude that these samples had a $\lambda$ parameter around $1$ and thus a thickness, $d$ of about $3$ microns. With the van der Waals layer-to-layer thickness of graphite being $0.336$ nm, we can establish that our samples contain about $5{,}000\dots 10{,}000$ graphene layers.

\section{Temperature-dependent studies, details and reproducibility of the linewidth measurement}

\begin{figure}[!ht]
		\begin{centering}
        \includegraphics[width=\linewidth]{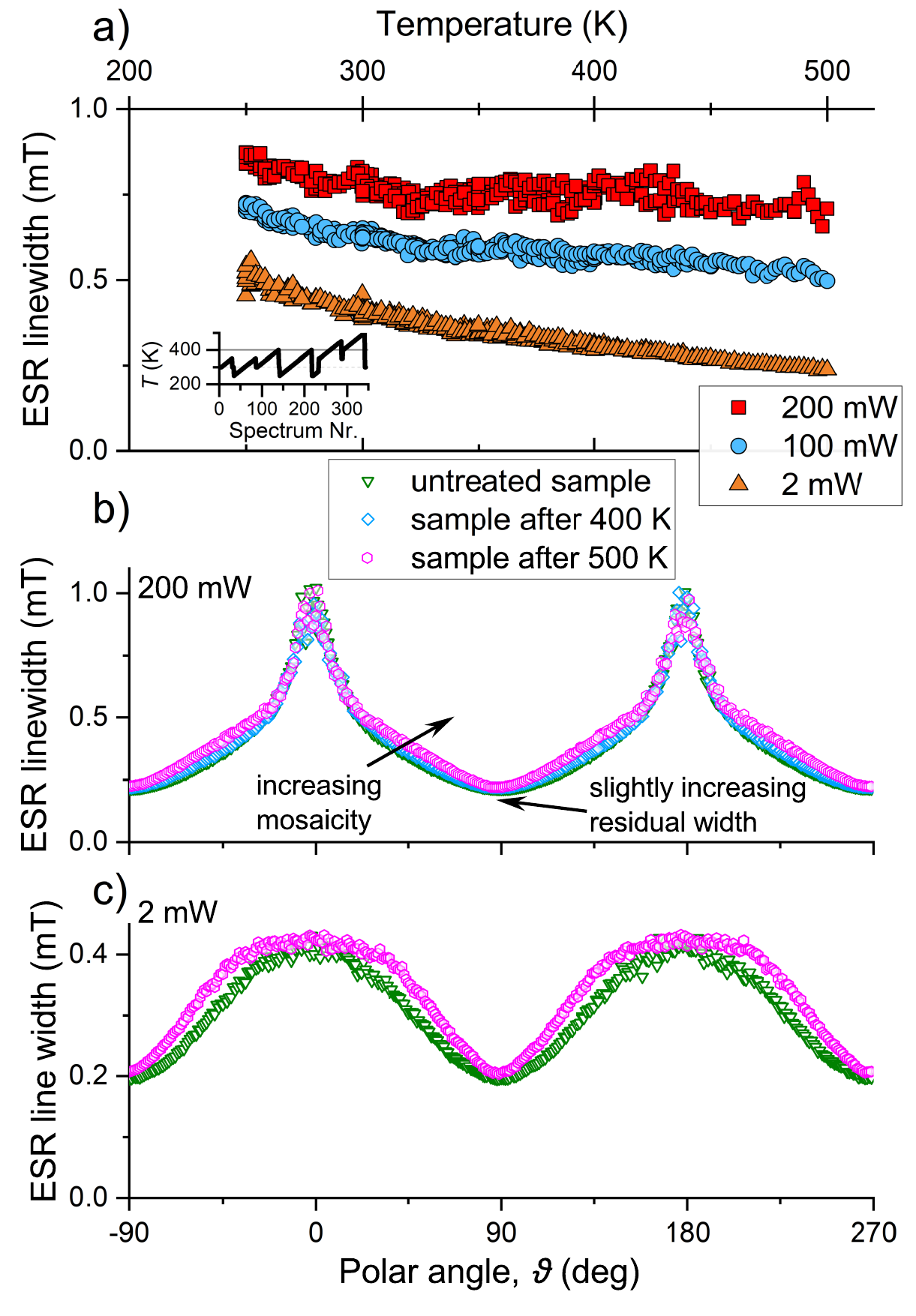}
        \caption{a) The raw temperature-dependent linewidth data for a polar angle offset of $7^{\circ}$ for three different values of the microwave power. The temperature dependent $T_1$ data in the main text were directly obtained from these data. Inset shows the actual temperature protocol, we stepped the temperature by $2$ K in each step and thermalization was about $20$ seconds. b-c) The detailed angular dependent data taken with $200$ mW and $2$ mW for the untreated sample, after the first $400$ K treatment, and after the $500$ K treatment as per the inset of the top panel (the intermediate data is now shown for the $2$ mW measurement).}
	  \label{FigSM_thermal_history}
    	\end{centering}
\end{figure}

As mentioned in the main text, the quality of the graphite crystals is very important in our study. We found that breaking a high-quality flake results in a lower $T_1$. Similarly, temperature dependence can induce irreversible changes to the sample. We therefore carefully examined the reproducibility of the ESR linewidth during the heat treatment steps. We remind that the samples are between two Scotch tapes which is then fixed onto the suprasil rod of the goniometer with vacuum grease (Dow Corning high vacuum grease). The freezing of the grease and the unequal thermal contraction/expansion of the two types of materials probably also leads to sample breaking and also the Scotch tape cannot withstand elevated temperatures. We found that cooling below the freezing point of the grease ($233$ K) and heating above $500$ K induces irreversible changes. These are evident as a drop in the $200$ mW linewidth from a maximum value of $1$ mT to $0.6$ mT. 

In Fig. \ref{FigSM_thermal_history}, we show the temperature-dependent linewidth data for a polar angle offset of $7^{\circ}$ for a few microwave power values. In this experiment, we measured the three power values automatically while staying on a given temperature. The temperature-dependent $T_1$ data, shown in Fig. 4. in the main text were directly obtained from these data. The inset shows the temperature protocol as a function of the individual spectrum number. The temperature protocol was as follows: the sample was controllably cooled/warmed to the starting temperature value, avoiding a significant (not more than $5$ K) overshoot. We then changed the temperature by $2$ K steps and thermalization was about $20$ seconds between two measurement. Fig. \ref{FigSM_thermal_history} also shows the angular dependent linewidth data at $200$ mW for the untreated sample, after the first $400$ K treatment, and after the $500$ K treatment as per the inset of the top panel. 

The most important observation is that the maximum value of the linewidth is unaffected by the treatments. This means that the $T_1$ value is unchanged. However, we do observe minor changes to the sample quality as indicated by a slight increase of the ESR linewidth for the shoulders of the linewidth data around $45^\circ$ and also a small change in the width for the $B\parallel (a,b)$. The earlier indicates a slight increase in the mosaicity of the sample and the latter hints at a small added spin scattering. 

\section{Impact of mechanical deformation on the relaxation times}

\begin{figure}[!ht]
		\begin{centering}
        \includegraphics[width=\linewidth]{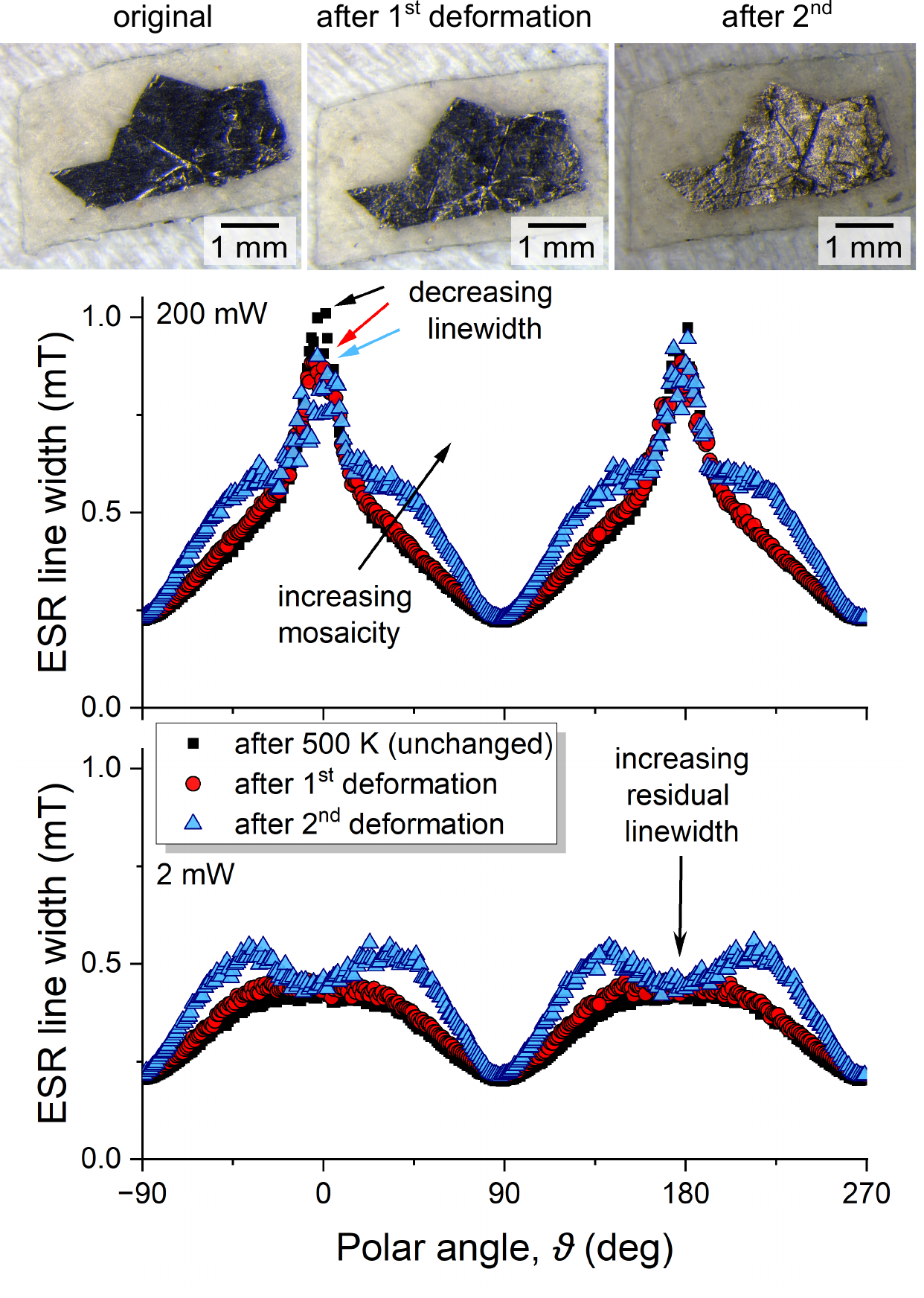}
        \caption{Effect of mechanical deformation on the spin-relaxation lifetime. Arrows indicate the increasing mosaicity of the sample during the mechanical deformation and also a drop in the maximum linewidth, which indicates a lowered $T_1$.}
	  \label{FigSM_mechanical_deformation}
    	\end{centering}
\end{figure}

As mentioned in the main text, we hypothesized that the sample quality, its perfect flat structure and the size of the individual flakes, play an important role on the spin-lattice relaxation time. To test it, we subjected an "Outstanding" sample (Nr. 134) to repeated mechanical deformation. This samples had been previously investigated in detail: it was thermally cycled between $250$ K and $500$ K with no observable change in the sample properties as shown in Fig. \ref{FigSM_thermal_history}. The sample had an approximate lateral size of $2\times 2$ mm, held between two Scotch tapes of about $3\times 3$ millimeters. It was subjected to bending with tweezers, such that the two ends touch and held in this position for $5$ seconds. Following this, it was bent along the opposite direction until the Scotch tapes were flat again. This procedure was repeated $5$ times (denoted as "1$^{\text{st}}$ deformation" in Fig. \ref{FigSM_mechanical_deformation}). Then the microwave power-induced broadening factor was determined. Following this, the same procedure was repeated ("2$^{\text{nd}}$ deformation") and the sample was measured again. 

The result is shown in Fig. \ref{FigSM_mechanical_deformation}. Three trends are clear during the deformations: i) the maximum linewidth in the $200$ mW experiment drops, which indicates a lowered value of $T_1$, ii) the mosaicity increases in the sample, iii) the residual linewidth when $B\parallel (a,b)$ increases in the $2$ mW experiment. These changes indicate that the observed $T_1$ drops by a factor or $2$.

Moreover, a recent paper discusses that micromechanical preparation also introduces rhombohedral stacking faults in Bernal graphite\cite{BoiCT2024}, which has somewhat different ESR properties\cite{ChehabEPJB2000} and shorten the $T_2$ relaxation time (and possibly $T_1$, too) with increasing concentration.


\section{Time-domain detectability of the spin-relaxation time in graphite using magnetic resonance}

\paragraph{Spin-echo ESR}
The most common technique to detect relaxation times in ESR is the use of spin-echo detected techniques \cite{PooleBook, SlichterBook}. However, a typical pulsed ESR instrument is limited to $T_2^{\ast}$, $T_2$, and $T_1$ times above $20-50$~ns \cite{NafradiNatComm}. In our case, for the $B \parallel c$ orientation we estimated that $T_2=15$~ns and for $B\parallel(a,b)$ our result gave $T_2\approx 30$~ns and $T_1\approx 15$~ns, which clearly limits the observability using the pulsed ESR method.

\paragraph{Longitudinally detected ESR}
The longitudinally detected (LOD) ESR method \cite{Martinelly_LOD, Atsarkin, Schweiger1, Schweiger2, MuranyiSimon1, MuranyiSimon2} is in principle capable of detecting $T_1$ values as low as a few nanoseconds. The method relies on the detection of the longitudinal, or $M_z$ component of the magnetization using a pick-up coil which is parallel to the DC magnetic field. During the experiment, the microwave power is strongly modulated or chopped. Due to this direct detection scheme, LOD experiments require a large sample amount to yield a large magnetization. we estimate that the relatively small sized graphite crystals and the low density of states in graphite \cite{McClurePR1957, McClureCoC1962} prevent such an experiment at $0.3$ T, however, we cannot rule out that a future, specialized ESR instrument may be capable of detecting the reported $T_1$ values. 

We in fact attempted an LOD study on our samples with a specially fabricated coil, which surrounds the sample while leaving the cavity quality factor unchanged, but no signal was detected even though we had a previous success with the method on larger susceptibility materials including KC$_{60}$ and the superconducting MgB$_2$ (Refs. \onlinecite{MuranyiSimon1, MuranyiSimon2}).

\paragraph{Saturation recovery ESR}
Saturation recovery ESR (SR-ESR) is a method developed for the study of $T_1$ times in the range of $100 \dots 1{,}000$ ns\cite{EatonsRSI1992, EatonsAMR2017, EatonsRSI2019} even when the $T_2$ is short, which prevents the detection of a spin echo. The method is based on saturating the ESR signal with intensive microwave irradiation, which is then turned off and during the recovery, the conventional cw ESR is detected. The major difference between this method and that of the spin-echo, is that SR-ESR works even when $T_2$ is short (below $100$ ns), which prevent measurement with pulsed methods, as the echo decay is comparable to instrumental dead times. 

During the saturation recovery, the system is continuously irradiated with a low power $B_1$. Therefore even when $T_2$ is short (thus the transversal magnetization in the rotating frame of reference decays rapidly to zero) the excitation continuously pumps a finite transversal magnetization, which is then detected. The SR-ESR requires specialized ESR bridges with a number of isolating/protecting/blanking switches between the excitation and detection, which was not available to us. We also mention that the so-called rapid passage method is not capable of studying $T_1$ but instead it induces oscillation, which are characteristic for the $T_2$ relaxation time \cite{Moser2017}.

\section{Diffusion limited transport}
In general, the diffusion length, for a time period of $t$, is described by $\delta=\sqrt{D t}$, where $D$ is the diffusion constant. In a simple diffusion model, which considers the electrons in the quasiparticle approximation, we obtain in dimension $d$: $D=\frac{1}{d}v_\text{F}^2\tau$. Here $v_\text{F}$ is the Fermi velocity and $\tau$ is the momentum relaxation time. This yields the final result as $\delta=\frac{1}{\sqrt{d}}v_\text{F}\sqrt{\tau t}$. This can be used generically for the diffusion of any non-equilibrium physical quantity in a solid. E.g., for the diffusion of excess-charge carriers in semiconductors with lifetime $\tau_\text{c}$, we obtain the charge-carrier diffusion length: $\delta_\text{c}=\frac{1}{\sqrt{d}}v_\text{F}\sqrt{\tau \tau_\text{c}}$ or for an injected or excited non-equilibrium spin concentration with lifetime $\tau_{\text{s}}$: $\delta_\text{s}=\frac{1}{\sqrt{d}}v_\text{F}\sqrt{\tau \tau_\text{s}}$.

It is a well-known effect in semiconductors \cite{pierret2002advanced, schroder2006semiconductor} that the effective charge-carrier lifetime is shortened by finite size effects. This occurs when the relevant dimension, $l$ of a semiconductor (e.g., the thickness of a wafer) is smaller than the calculated $\delta_\text{c}$. For such, so-called, diffusion-limited charge-carrier lifetime \cite{DiffusionSemiconductors}, the effective $\tau_\text{c,eff}$ is obtained from $l=\frac{1}{\sqrt{d}}v_\text{F}\sqrt{\tau \tau_\text{c}}$. Although this effect has not been observed for spins, we argue that this situation is encountered in graphite. 

A key fingerprint of diffusion-limited charge-carrier lifetime is that $\tau_\text{c,eff}$ usually lengthens with increasing temperature as it reduces the charge-carrier mobility (thus $\tau$, too). We find that the same effect is observed in graphite. 

\end{document}